\documentclass[10pt,aps,pra,twocolumn,floatfix,nofootinbib,superscriptaddress,longbibliography]{revtex4-2}
\usepackage[utf8]{inputenc}	
\usepackage[english]{babel}
\usepackage[dvipsnames]{xcolor}
\usepackage{bm,graphicx,mathrsfs,amsmath,amssymb,mathtools,makecell,bbm, amsthm,dsfont,times,txfonts,nicefrac,framed,enumitem,tikz,physics,wrapfig,amsfonts,lipsum,nicematrix,wrapfig,halloweenmath,tabularray}
\usepackage[colorlinks=true,linkcolor=equationcolor,citecolor=NavyBlue]{hyperref}
\usepackage[most]{tcolorbox}
\usepackage{pifont}
\usepackage{etoolbox}
\usepackage{xargs}

\definecolor{changescolor}{rgb}{0, 0, 0.7}

\usepackage[colorinlistoftodos,prependcaption,textsize=tiny]{todonotes}
\newcommandx{\unsure}[2][1=]{\todo[linecolor=red,backgroundcolor=red!25,bordercolor=red,#1]{#2}}
\newcommandx{\change}[2][1=]{\todo[linecolor=blue,backgroundcolor=blue!25,bordercolor=blue,#1]{#2}}
\newcommandx{\info}[2][1=]{\todo[linecolor=OliveGreen,backgroundcolor=OliveGreen!25,bordercolor=OliveGreen,#1]{#2}}
\newcommandx{\improvement}[2][1=]{\todo[linecolor=Plum,backgroundcolor=Plum!25,bordercolor=Plum,#1]{#2}}
\newcommandx{\thiswillnotshow}[2][1=]{\todo[disable,#1]{#2}}

\hypersetup{urlcolor=NavyBlue}
\definecolor{equationcolor}{RGB}{222,94,100}
\definecolor{alecolor}{RGB}{198,113,190}

\newcommand{\ms}[1]{\textsf{#1}}
\usetikzlibrary{tikzmark}
\renewcommand{\v}[1]{\ensuremath{\boldsymbol #1}}

\newcommand{\example}[1]{
  \begin{center}
    \makebox[\linewidth]{\hrulefill\quad \emph{\large{\ding{46}} \small #1} \quad\hrulefill}
  \end{center}
}

\newcommand{\examplend}{
  \noindent
  \makebox[\linewidth]{\rule{0.4\linewidth}{0.4pt}%
  \hfill\large{\ding{46}}\hfill%
  \rule{0.4\linewidth}{0.4pt}}
}

\newenvironment{coolexample}[1][]{
  \begin{center}
    \makebox[\linewidth]{\hrulefill\quad \emph{\large{\ding{46}} \small #1} \quad\hrulefill}
  \end{center}
  }
  {
    \noindent
  \makebox[\linewidth]{\rule{0.4\linewidth}{0.4pt}%
  \hfill\large{\ding{46}}\hfill%
  \rule{0.4\linewidth}{0.4pt}}
}

\newcounter{globalboxcounter} 
\renewcommand{\theglobalboxcounter}{\arabic{globalboxcounter}}

\newtcolorbox[auto counter]{mybox}[3][]{%
    breakable,
    enhanced,
    sharp corners,
    colback=white,
    colframe=black!50!,  
    fonttitle=, 
    colbacktitle=white, 
    coltitle=black, 
    before skip=-6mm, 
    after skip=5mm,  
    boxrule=0.25mm, 
    arc=0mm, 
    width=\linewidth, 
    boxsep=0mm, 
    left=2mm, right=2mm, top=2mm, bottom=2mm, 
    before title={%
        \stepcounter{globalboxcounter} 
    },
    title={\centering \strut Experimental Box \theglobalboxcounter: #2}, 
}

\begin{document}

\title{A friendly guide to exorcising Maxwell's demon}

\author{A. de Oliveira Junior}
\author{Jonatan Bohr Brask}
	\affiliation{Center for Macroscopic Quantum States bigQ, Department of Physics,
Technical University of Denmark, Fysikvej 307, 2800 Kgs. Lyngby, Denmark}
\author{Rafael Chaves}
\affiliation{International Institute of Physics, Federal University of Rio Grande do Norte, 59078-970, Natal, RN, Brazil}
\affiliation{School of Science and Technology, Federal University of Rio Grande do Norte, Natal, Brazil}

\date{\today}

\begin{abstract}

The birth, life, and death of Maxwell's demon provoked a profound discussion about the interplay between thermodynamics, computation, and information. Even after its resolution, the demon continues to inspire a multidisciplinary field. This tutorial offers a comprehensive overview of Maxwell's demon and its enduring influence, bridging classical concepts with modern insights in thermodynamics, information theory, and quantum mechanics.

\end{abstract}

\maketitle

\tableofcontents

\newpage

\section{Introduction}

\setlength\intextsep{0pt}
 \setlength{\columnsep}{0pt}%
\begin{wrapfigure}{l}{0.07\textwidth}
     \vspace{-0.04cm}
     \includegraphics[width=0.07\textwidth]{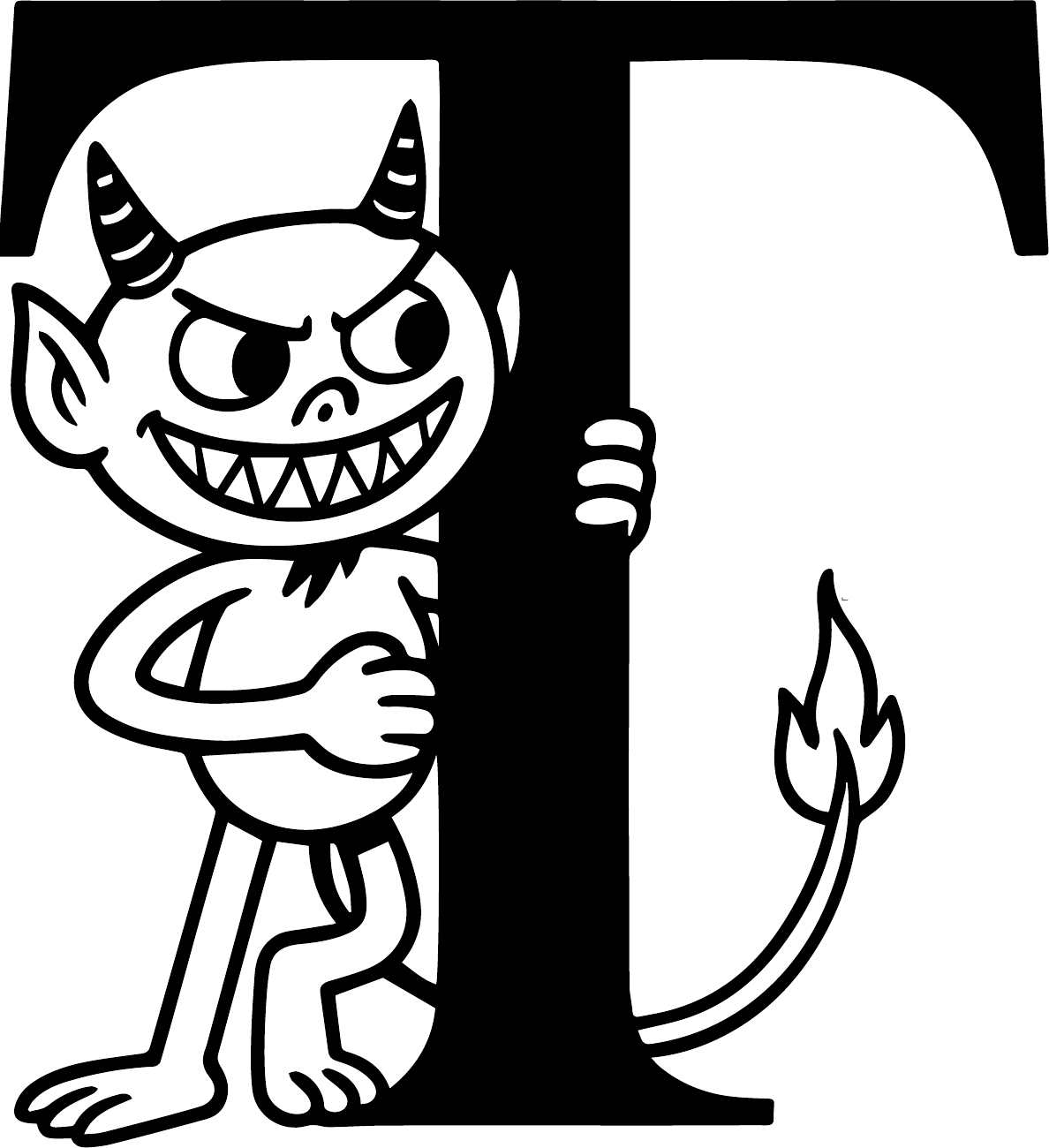}
 \end{wrapfigure}
\setlength\intextsep{0pt} 
\noindent 
hought experiments in physics are the birthplace of new paradigms. Einstein's relativity defied Newtonian notions of space and time~\cite{Einstein1905,Einstein1914} and led him to have what he later called his ``happiest thought,'' when he conceived the equivalence principle~\cite{Einstein1907,Einstein1912}. Meanwhile, Schrödinger’s cat~\cite{Erwin1926} illustrates the strangeness of quantum mechanics. Many of these \emph{mental exercises} have evolved from theoretical explorations to experimental realisations. When Einstein questioned whether quantum mechanics offered a complete description of reality~\cite{Einstein1935}, John Bell provided an elegant answer decades later with his groundbreaking theorem~\cite{Bell1964}. Not long after, experiments by Alain Aspect, Jean Dalibard, and Gérard Roger confirmed Bell's predictions~\cite{Aspect1982}—bringing such thought experiments to life as tangible realities. Yet, some thought experiments remain unresolved, such as Wigner's friend~\cite{Wigner1995}, which highlights the need for a deeper interpretation of quantum mechanics (see Ref.~\cite{Nurgalieva2020} for a state-of-the-art overview related to the topic).

This tutorial discusses a thought experiment that took more than a century of scientific development to be resolved. By the 19th century, it was commonly understood that thermodynamic fluctuations at the particle level could lead to apparent violations of the second law of thermodynamics. To illustrate how these fluctuations could be systematically exploited to seemingly break the second law, Maxwell proposed a thought experiment, now famously known as Maxwell's demon~\cite{maxwell1911letter,maxwell2001theory}. 

\begin{figure}[t]
    \centering
    \includegraphics{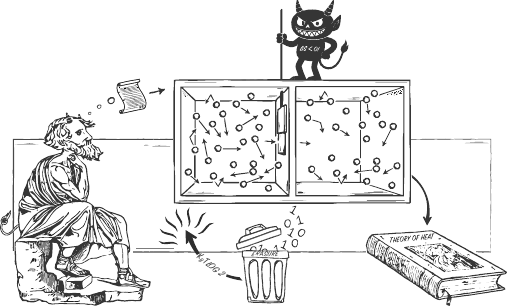}
    \caption{\emph{Maxwell's Gedanken experiment}. In 1867 James Clerk Maxwell addressed a letter to Peter Guthier Tait with the provocative idea that the second law is a statistical principle that only holds on average~\cite{knott1911life}. There, he introduced his famous demon -- an intelligent being capable of sorting gas molecules by their velocities. By creating a temperature difference, the demon apparently defies the second law of thermodynamics, decreasing the system’s entropy without expending energy. Later on, this apparent paradox appeared in his book ``Theory of Heat''~\cite{maxwell2001theory}. }
    \label{F-Maxwell-demon-intro}
\end{figure}

Maxwell envisioned a scenario in which, by knowing the positions and velocities of gas particles, one could reduce the entropy of the system without performing any work, thus appearing to defy the second law. More specifically, the thought experiment involves a demon controlling a trapdoor between two gas chambers, allowing fast particles to pass into one chamber and slow particles into the other. Using \emph{information} about the velocity of the particles, the demon creates a temperature difference, reducing the entropy without expending energy~(see Fig.~\ref{F-Maxwell-demon-intro} for a pictorial representation and historical context).

The demon’s ability to reduce entropy without performing work hinges entirely on the information it possesses about the gas particles. This suggests that information somehow plays an important role in characterising thermodynamic processes. This interplay was tightened with the Szilárd engine~\cite{szilard1929entropieverminderung,Szilard1964} -- a more operational version of Maxwell’s demon, based on a device that trades information for work. The setup consists of a chamber containing a single gas molecule in thermal equilibrium with a heat bath at temperature $T$\footnote{Throughout this tutorial, we will use both the inverse temperature, $\beta = 1 / k_B T$ (where $k_B$ is the Boltzmann constant), and the temperature $T$, depending on the context.}. Based on information about which half the chamber the molecule is in, a thin wall attached to a weight is inserted. The one-molecule gas then pushes the wall, expanding the chamber back to its initial size and extracting $k_B T \log 2$ of work from the thermal reservoir (approximately $3 \times 10^{-21}  \text{J}$ at room temperature). Like Maxwell's demon, the Szilárd engine appears to violate the second law of thermodynamics, as the heat extracted from the environment is entirely converted into mechanical work. Although Szilárd acknowledged the need for an entropy increase to close the thermodynamic cycle, he did not specify its source--whether the increase arose from measuring the particle's position, storing the collected information, inserting and removing the partition, or erasing the recorded data. 

In the following years, information and computation theories were developed~\cite{shannon1938symbolic,shannon1948mathematical,von1993first}. Although they were not yet tied to physics, Shannon’s theory~\cite{shannon1948mathematical} inspired Brillouin~\cite{brillouin1951maxwell} to attempt to save the second law. He proposed a mathematical theory that linked the acquisition of information and the increase in entropy needed to restore the second law. Two years after Brillouin's paper, Landauer~\cite{landauer1988dissipation,landauer1991information} asked whether thermodynamics imposes physical limits on information processing. The answer to such a simple question led to the conclusion that erasing information is inherently linked to a heat cost. Finally, this intriguing connection between thermodynamics and computation was later strengthened by Roger Penrose~\cite{PENROSE1970} and Charles Bennett~\cite{bennett1973logical}, which ultimately led to the resolution of Maxwell's demon~\cite{bennett1982thermodynamics}.

What seemed to be the end of the demon's journey turned out to be just the beginning. Both stochastic thermodynamics~\cite{seifert2008stochastic} and quantum thermodynamics~\cite{Gemmer2004,Kosloff2013,Goold2016,Vinjanampathy2016,Alicki2018,Binder2018,Deffner2019} have reinterpreted and extended the concept of Maxwell's demon beyond its original scenario. Beautifully, not only has theoretical progress been made, but numerous experiments on these two topics have also been reported (a summary of experiments in the field of stochastic thermodynamics can be found in Ref.~\cite{Ciliberto2017} and an overview of the proposed and realised quantum thermodynamic devices is presented in Ref.~\cite{Myers2022}). 

\begin{center}
    \small{\emph{How to use this guide?}}
\end{center}

This tutorial revisits Maxwell's demon, from its early phase to modern approaches, while providing historical context and connections to state-of-the-art experiments. It offers an introduction for quantum scientists, exploring the intersection of thermodynamics, information theory, and quantum mechanics. We do not intend to provide a detailed review of the literature here; for that, we recommend both the reviews~\cite{plenio2001physics, Vedral2009,Lutz2015} and the book by Leff \& Rex~\cite{leff1990maxwell}. In each section, we discuss problems (along with their solutions) relevant to the topic. These problems often introduce results that will be useful in subsequent discussions. To highlight these key results, they are marked with the symbol \ding{46} and are enclosed between horizontal separators. Starting from Sec.~\ref{Sec:picking-a-hole}, we present experimental realisations related to the discussed topic. These realisations are independent and can be skipped without compromising the understanding of the main text. The experiments are placed in boxes, and the discussion is kept short. Our goal is to show that what might seem theoretical has already been demonstrated experimentally. The experiments were selected based on the physical platforms rather than their chronological order of appearance in the literature.

This tutorial is organised as follows. We start with a brief recap of basic thermodynamics concepts in Sec.~\ref{Sec:thermodynamic-nutshell}, followed by a concise summary of relevant quantum mechanics notions in Sec.~\ref{Sec:quantum-mechanics}. The central problem of this tutorial is presented in Sec.~\ref{Sec:picking-a-hole}, where Maxwell's thought experiment is discussed [\hyperref[Subsec:maxwells-demon]{IV-A}], along with Szilard's reinterpretation [\hyperref[Subsec:szilard-engine]{IV-B}] and attempts to resolve the apparent violation of the second law of thermodynamics [\hyperref[Subsec:efforts-to-patch]{IV-C}]. Next, Sec.~\ref{Sec:information-theory} starts with a discussion of what information is, introducing Shannon’s theory and some mathematical results [\hyperref[Subsection:classical-info-processing]{V-A}]. Then we briefly touch on some basic ideas and historical facts about computation [\hyperref[Subsection:computation-nutshel]{V-B}]. One of the most important points of this tutorial is then presented in Sec.~\ref{Sec:thermodynamics-of-computation}, where Landauer's principle is introduced~[\hyperref[Subsec:Landauer]{VI-A}], rigorously proven~[\hyperref[Subsec:Landauer-proof]{VI-B}], and its first experimental verification is discussed [\hyperref[Subsec:Landauer-experimental]{VI-C}]. To finish this section, the notion of reversible computation is briefly explained [\hyperref[subsec:reversible-computation]{VI-D}]. The resolution of Maxwell's demon is presented in Sec.~\ref{Sec:exorcisim} and three different approaches to solving this ``paradox'' are discussed. Finally, in Sec.~\ref{Sec:whatsnext}, we briefly discuss recent progress related to Maxwell's demon, as well as topics inspired by it.

\section{Thermodynamics in a nutshell}\label{Sec:thermodynamic-nutshell}

This introductory section revisits some fundamental formulations of thermodynamics, introduces the notation used, and briefly highlights recent developments in the field. A few problems, which will later connect to other sections, are also discussed. As this section is a warm-up, readers already familiar with thermodynamics are welcome to skip it.

Thermodynamics emerged from the desire to understand how macroscopic systems exchange energy, particularly the interconversion between heat and work~\cite{fermi,callen}. It offers a framework for describing the transformations of complex systems made up of many particles without the need to take into account their microscopic details. The key insight is that macroscopic systems in equilibrium can be very well described by just a few variables, like temperature, pressure, or magnetisation, depending on the specific physical system being studied. This allows us to make accurate predictions about the system behaviour without having to track every individual particle. While this might seem like an oversimplification, the universality of thermodynamic principles led to the formulation of four fundamental laws\footnote{Modern thermodynamic approaches propose a ``fifth Law'', related to Onsager’s theorem~\cite{Onsager1931,Onsager1931b}, which describes symmetry and reciprocity in transport coefficients for systems near local equilibrium fundamental laws.}:
\begin{enumerate}[start=0]
     \item \textbf{Zeroth law.} \emph{If two systems are each in thermal equilibrium with a third system, they are also in thermal equilibrium with one another.} 
     \item \textbf{First law.} \emph{Energy is conserved. For a system interacting with an environment, energy transfer can be divided into two contributions: work and heat.}
     \item \textbf{Second law.} \emph{The entropy of an isolated system undergoing a physical process can never decrease.}
     \item \textbf{Third law.} \emph{It is impossible to attain absolute zero in a finite number of thermodynamic operations. At absolute zero, systems with degenerate ground states may retain nonzero entropy.}
\end{enumerate}

The central focus of this tutorial is on the second law of thermodynamics, which can be formulated in various ways, with all formulations being fundamentally equivalent. In its most general form, the second law asserts that the entropy of the universe tends to increase $\Delta S_{\ms{univ}} \geq 0$. The change in the system's entropy, known as entropy production, is denoted by $\Sigma := \Delta S_{\ms{univ}}$. Thus, a succinct way to express the second law is through the non-negativity of the entropy production. This quantity also allows us to classify processes as reversible, where $\Sigma = 0$, or irreversible, with $\Sigma > 0$. For a detailed discussion of entropy production in both classical and quantum systems, see Ref.~\cite{landi2021irreversible}.

Throughout this tutorial, a main system interacts with its environment, resulting in changes to both the energy and the entropy of each. The total entropy of the universe is divided into the entropy of the system and that of the environment: $S_{\text{univ}} = S_{\ms{S}} + S_{\ms{B}}$. This leads to the second law being written as $\Delta S_{\text{univ}} = \Delta S_{\ms{S}} + \Delta S_{\ms{B}} \geq 0$. Moreover, we assume that the environment consists of a work source, which exchanges only energy with the system, leaving its entropy unchanged, and a heat bath, which exchanges both energy and entropy with the system. Thus, $\Delta S_{\ms{B}}$ simplifies to the change in entropy of the heat bath, which is modelled as being in equilibrium at temperature $T$. The change in entropy of the heat bath is then given by $\Delta S_{\ms{B}} = \int \textrm{d}Q / T$, where $\textrm{d}Q$ represents an infinitesimal heat flow into the system. The total heat exchanged is $Q = \int \textrm{d}Q$. Here, we adopt the convention that heat is positive when it flows into the system from the environment. Work, however, is performed by an external controlling agent—an entity that manipulates macroscopic parameters (e.g., volume, magnetic fields) to exchange energy with the system. Crucially, the agent’s ability to extract or inject work depends on what they know about the system’s state. In typical thermodynamics, the agent possesses only macroscopic information (temperature, volume, pressure, etc). However, a hypothetical all-knowing agent--like the star of this tutorial, the ``famous'' Maxwell’s demon--would be privy to microscopic information about each molecule. Such an agent could, in principle, exploit thermal fluctuations to convert heat entirely into work, seemingly violating the second law. As we will discuss in detail later (via Landauer’s principle), this paradox is resolved when one accounts for the entropy production associated with the demon’s measurement process and the erasure of its memory, which compensates for any apparent entropy decrease in the system. For now, we retain the classical assumption of macroscopic control and minimal knowledge, leading us to the second law in its traditional form, Clausius’s inequality~\cite{fermi}:
\begin{equation}\label{Eq:clausius-inequality}
    \Sigma = \Delta S_{\ms S} + \int \frac{\textrm{d}Q}{T} \geq 0.
\end{equation}
If the heat bath is only slightly perturbed, meaning it is large enough so that its temperature remains effectively constant, the second term in Eq.~\eqref{Eq:clausius-inequality} simplifies to $Q/T$. From this point onwards, we assume that the bath remains at its initial temperature at all times and that the system does not significantly perturb it. In the presence of multiple heat baths Eq.~\eqref{Eq:clausius-inequality} generalises naturally to $\Sigma = \Delta S_{\ms S} + \sum_i Q_i/T_i \geq 0$, where the index $i$ refers to the $i$th reservoir. 

Traditionally, classical thermodynamics describes quasi-static (reversible) transformations, in which the system and every reservoir remain in equilibrium at each instant. In that equilibrium limit, the entropy-production rate vanishes, $\dot\Sigma = 0$, and, because no heat or particle currents flow, the entropy flux--defined as $\dot{\Phi} := \sum_i \dot{Q}_i / T_i$-- is also zero $\dot\Phi = 0$, so that $\mathrm dS_{\ms S}/\textrm{d}t = 0$ (where $t$ is time). However, many processes occur over finite times and lead to entropy production. A scenario that can emerge after finite-time transients is one where there is a continuous flow of energy or matter, with certain system properties becoming constant in time despite the absence of thermal equilibrium. After this transient period has elapsed, the condition $ \dot S_{\ms S}=0 $ together with the balance relation $ \dot S_{\ms S}= \dot{\Sigma}-\dot{\Phi} $ implies $ \dot{\Sigma} = \dot{\Phi}>0 $, which means that entropy is continually generated within the system and is transferred entirely to the reservoirs. This defines a non-equilibrium steady state~\cite{Lax1960}, in which the entropy-production rate remains constant.

The concepts presented so far can be illustrated by considering a thermodynamic system interacting with heat baths at different temperatures and a work reservoir. The notion of entropy production allows us to incorporate time into the analysis and describe the continuous interaction between the system and its environment. Consequently, we express the first and second laws in terms of rates, as follows:
\begin{align}
    \frac{\textrm{d} E_{\ms S}}{\textrm{d}t} &= \dot{W}- \sum_i\dot{Q}_i, \label{Eq:first-law-finite}\\
    \frac{\textrm{d}S_{\ms S}}{\textrm{d}t} &= \dot{\Sigma} - \sum_i \frac{\dot{Q}_i}{T_i}. \label{Eq:second-law-finite}
\end{align}
where $\dot{E}_{\ms S}$ represents the rate of change of the system’s internal energy, $\dot{Q}_i$ denote the heat exchanged with $i$th bath, respectively, and $\dot{W}$ is the rate of work done on the system, which is assumed to be positive in the case of work being supplied to the system.

Let us discuss three paradigmatic examples (presented in Ref.~\cite{landi2021irreversible}), in which we analyse a system that has operated for a sufficiently long time and has reached a steady state. This implies that both Eq.~\eqref{Eq:first-law-finite} and Eq.~\eqref{Eq:second-law-finite} are equal to zero, and the steady state is therefore characterised by a continuous conversion of heat into work, accompanied by a constant production of entropy.

Suppose that there is no work involved in the process and only heat exchange occurs between the two thermal reservoirs at temperatures $T_c$ and $T_h$ with $T_c < T_h$. The heat exchanged with the hotter and colder reservoirs are denoted by $\dot{Q}_h$ and $\dot{Q}_c$, respectively. As a result, the first law simplifies to $\dot{Q}_h = - \dot{Q}_c$, which, when substituted into Eq.~\eqref{Eq:second-law-finite}, gives:
\begin{equation}
    \dot{\Sigma} = \qty(\frac{1}{T_c}-\frac{1}{T_h})\dot{Q}_c \geq 0.
\end{equation}
Since the above equation must be non-negative and $T_c < T_h$, it follows directly that $\dot{Q_c} \geq 0$, which according to our convention means that heat must flow from the hot to the cold reservoir. This is precisely Clausius's statement of the second law:
\newline
\begin{wrapfigure}{l}{0.09\textwidth}
     \includegraphics[width=0.08\textwidth]{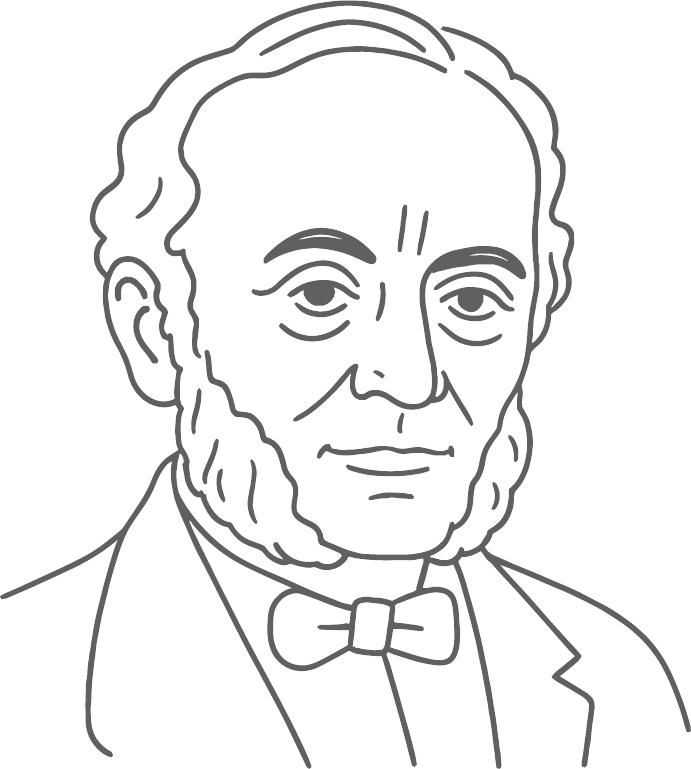}
 \end{wrapfigure}
 \noindent\emph{``A transformation whose only final result is to transfer heat from a body at a given temperature to a body at a higher temperature is impossible.''~\cite{clausius1854veranderte}}
\newline

Now, consider the case where the system is connected only to the hot reservoir. In this case, the first law becomes $\dot{W} = \dot{Q}_h$, and the second law is expressed as: 
\begin{equation} \dot{\Sigma} = \frac{\dot{Q}_h}{T_h} = \frac{\dot{W}}{T_h} \geq 0. 
\end{equation} 
By definition, positive work implies that work is done on the system rather than extracted from it. Therefore, work cannot be extracted from a single heat reservoir, which directly corresponds to Kelvin's statement:
\newline
\begin{wrapfigure}{l}{0.09\textwidth}
     \includegraphics[width=0.079\textwidth]{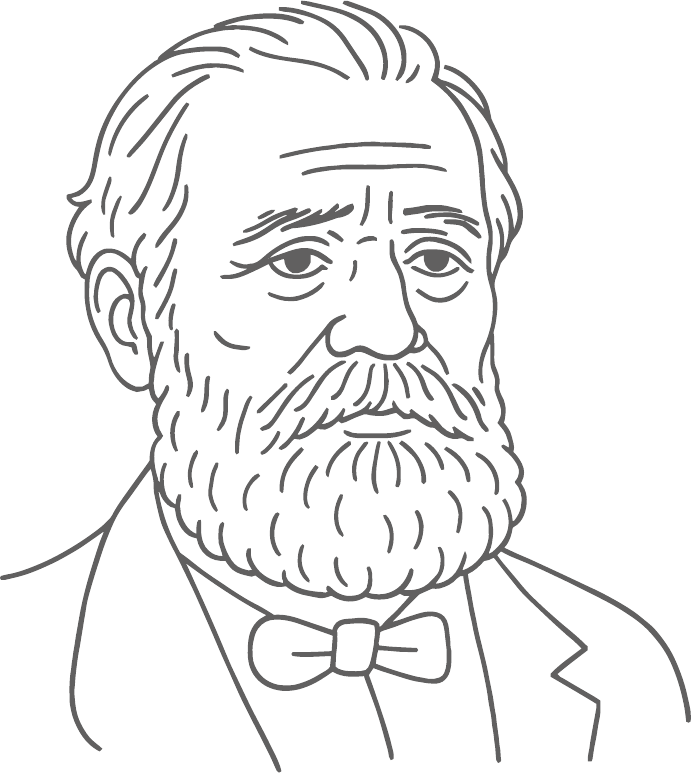}
 \end{wrapfigure}
\noindent\emph{``A transformation whose only final result is to transform into work heat extracted from a source which is at the same temperature throughout is impossible.''~\cite{thomson1853xv}}
\newline

As a final example, we now consider a machine comprised of two heat baths at temperatures $T_c$ and $T_h$, and a work reservoir. Assuming that this machine has reached a steady state allows us to rewrite Eqs.~\eqref{Eq:first-law-finite} and \eqref{Eq:second-law-finite} as $\dot{W} = \dot{Q}_h + \dot{Q}_c$ and $\dot{\Sigma} = \dot{Q}_h/T_h + \dot{Q}_c/T_c$. Computing the efficiency of the engine (i.e.\ the ratio of output work to input heat), we arrive at
\begin{equation}\label{Eq:efficiency}
    \eta = \frac{\dot{W}}{\dot{Q_h}} = 1+\frac{\dot{Q}_c}{\dot{Q}_h} = \underbrace{\qty(1-\frac{T_c}{T_h})}_{:=\eta_C}+\frac{T_c}{\dot{Q}_h}\dot{\Sigma}.
\end{equation}
The term in brackets represents the well-known Carnot efficiency $\eta_{C}$. Since $\dot{Q}_h < 0$, as the heat flows from the hot reservoir to the system, the second term is strictly negative. Therefore, the efficiency of a heat engine operating in finite time is given by Carnot's efficiency with a negative correction proportional to the entropy production. This result leads to Carnot's statement of the second law: 
\newline
\begin{wrapfigure}{l}{0.09\textwidth}
     \includegraphics[width=0.07\textwidth]{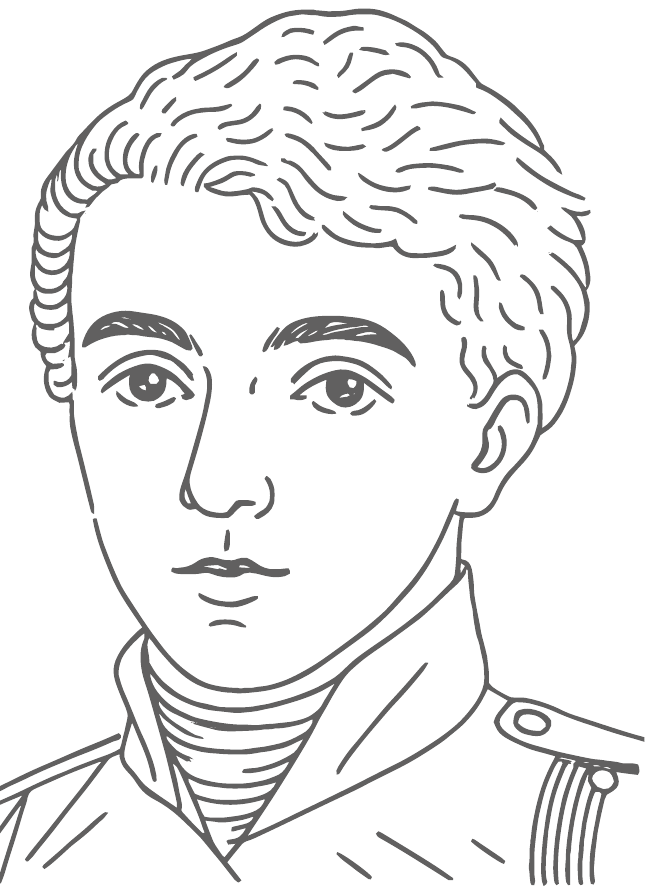}
 \end{wrapfigure}
\noindent\emph{The efficiency of any reversible engine operating between two heat reservoirs depends only on the reservoir temperatures and is equal to the Carnot efficiency. No heat engine operating between the same two temperatures can be more efficient.~\cite{Carnot}}
\newline

The previous example shows that incorporating the concept of entropy production provides a deeper understanding of Carnot's theorem. Specifically, the efficiency of a heat engine operating in finite time can always be expressed as the Carnot efficiency minus a correction term, directly related to the irreversibility of the process.

So far, we have made general statements about thermodynamic processes, mostly for systems in thermal equilibrium, but also considering non-equilibrium states. However, we have remained completely agnostic about the ``form" of the system's state. Since the Gibbs state is crucial for much of what follows, we will now simply present its expression without justifying its form—readers are instead referred to the books by Landau \& Lifshitz~\cite{landaubook}, Pathria~\cite{pathria2016statistical}, or Callen~\cite{callen} for a more detailed explanation. To do so, we consider a classical continuous system\footnote{The system does not need to be classical. We first introduce classical systems for convenience, since the next section will address quantum systems.} in contact with a heat bath at temperature $T$. The system is described by a Hamiltonian $H(\v z,\lambda)$, where $\v z = (q,p)$ represents a point in the phase space, and $\lambda$ is a real parameter (e.g., volume, magnetic field, or any other externally controllable parameter). We then define a thermal state as a probability distribution over phase space
\begin{equation}\label{Eq:thermal-state-classical}
    \rho_{\ms E}(\v z, \lambda):= \frac{e^{-\beta H(\v z, \lambda)}}{Z(\lambda)},
\end{equation}
where $Z(\lambda) = \int \textrm{d}\v z\, e^{-\beta H(\v z, \lambda)}$ is the partition function. 

\begin{coolexample}[Average thermodynamic quantities]
    Using Eq.~\eqref{Eq:thermal-state-classical}, one can calculate important average quantities for the system. For example, the average equilibrium energy is given by:
\begin{align}
     E_{\ms S}(\lambda):&= \int \textrm{d}\v z \frac{e^{-\beta H(\v z, \lambda)}}{Z(\lambda)}H(\v z, \lambda) = \frac{\frac{-\partial}{\partial \beta}\int \textrm{d}\v z e^{-\beta H(\v z,\lambda)}}{Z(\lambda)} \nonumber\\&= -\frac{\partial}{\partial \beta}\log Z(\lambda).\label{Eq:average-energy-classical}
\end{align}
Similarly, the Gibbs entropy takes the form (again, yet without justifying the reasoning behind this equation):
\begin{align}
  \frac{S(\lambda)}{\beta} &=-\frac{1}{\beta}\int \textrm{d}\v z \frac{e^{-\beta H(\v z, \lambda)}}{Z(\lambda)} \log \qty(\frac{e^{-\beta H(\v z, \lambda)}}{Z(\lambda)}) \nonumber \\ &= E_{\ms S}(\lambda) + \frac{\log Z(\lambda)}{\beta}. \label{Eq:Gibbs-entropy-classical}
\end{align}
\end{coolexample}

Another important formulation of the second law is given in terms of free energy $F$. When thermodynamic transformations are described between equilibrium states at constant temperature (isothermal processes), this quantity provides a bound on the work that can be extracted in the process. Namely, 
\begin{equation}\label{Eq:free-energy-bound}
    \langle W\rangle \geq \Delta F.
\end{equation}
Here, $F := E - TS$, where $E$ and $S$ denote the system’s internal energy and thermodynamic entropy, respectively. To simplify notation, we omit the index $\ms S$ when the context is clear and include it only when needed to avoid confusion. Note that from Eq.~\eqref{Eq:Gibbs-entropy-classical}, we directly get $F(\lambda)=-\frac{\partial}{\partial \beta} Z(\lambda).$ Moreover, we can verify Eq.~\eqref{Eq:free-energy-bound} considering a simple protocol as in the example below:

\begin{coolexample}[Free energy bound]
Consider a classical system whose microstate is $\v z=(q,p)$ and whose Hamiltonian depends on a control parameter $\lambda\in[0,1]$. The protocol lasts a time $\tau$ and we take a constant switching rate $\dot\lambda=1/\tau$. For a trajectory that begins in $\v z_0$ and follows Hamiltonian dynamics, the mechanical work is
    \begin{equation}\label{Eq:work-trajectory}
        W_{\v z_0}= \int_0^\tau \! \textrm{d}t\,\dot\lambda
        \frac{\partial H(\v z(t),\lambda)}{\partial\lambda}
        =\int_0^1 \! \mathrm d\lambda\,
        \frac{\partial H(\v z(\lambda),\lambda)}{\partial\lambda}.
    \end{equation}
    Averaging \eqref{Eq:work-trajectory} over initial conditions drawn from $\rho(\v z_0)$ gives
    \begin{equation}
        \langle W\rangle = \int \mathrm d\v z_0\,\rho(\v z_0)\,W_{\v z_0}.
    \end{equation}
 where $\rho(\v z_0)$ is the probability distribution of the initial state, which can be in or out of equilibrium. For an equilibrium system, this reduces to the Boltzmann distribution, as given in Eq.~\eqref{Eq:thermal-state-classical}, leading to:
    \begin{equation}\label{Eq:average-work-equil}
        \langle W\rangle
        =\int_0^{1}\! \mathrm d\lambda
        \int \mathrm d\v z\,
        \frac{e^{-\beta H(\v z,\lambda)}}{Z(\lambda)}
        \frac{\partial H(\v z,\lambda)}{\partial\lambda}.
    \end{equation}
In the limit $\tau\!\to\!\infty$, the system remains in instantaneous equilibrium, and \eqref{Eq:average-work-equil} reduces to
    \begin{align}
        \langle W\rangle
            &= -\frac{1}{\beta}\int_0^{1}
               \frac{\mathrm d}{\mathrm d\lambda}\ln Z(\lambda)\,
               \mathrm d\lambda
               = -\frac{1}{\beta}\bigl[\ln Z(1)-\ln Z(0)\bigr] \nonumber\\
            &= F(1)-F(0),
    \end{align}
    where $F(\lambda)=-\beta^{-1}\ln Z(\lambda)$ was noted below Eq.~\eqref{Eq:Gibbs-entropy-classical}. Therefore, a quasistatic isothermal transformation extracts an average work $\langle W\rangle\!=\!\Delta F$ saturating the free-energy bound~\eqref{Eq:free-energy-bound}. Any finite-time drive ($\tau\!<\!\infty$) produces additional dissipation, so that $\langle W\rangle\!>\!\Delta F$.
\end{coolexample}

The last example shows that, for an infinitely slow process, the work done on the system is exactly equal to the free energy difference between the final and initial states. Conversely, finite-time processes lead to irreversibility, and the thermodynamic work required to transform a system, in contact with an environment, can be expressed as $W = \Delta F + W_{\text{diss}}$, where the dissipated work is directly proportional to the entropy production~\cite{Jarzynski2012,landi2021irreversible}.

Surprisingly, the inequality corresponding to the second law as given in Eq.~\eqref{Eq:free-energy-bound} can be expressed as an equality~\cite{Jarzynski1997,Liphardt2002}, namely\footnote{The second law, as expressed in Eq.~\eqref{Eq:free-energy-bound}, can be directly derived using Jensen's inequality~\cite{Jensen1906} applied to the Jarzynski equality. Specifically, for any convex function $f$ and a random variable $X$, Jensen's inequality states that $f(\langle X\rangle \leq \langle f(X)\rangle.$},
\begin{equation}\label{Eq:Jarzynski-relation}
    \langle e^{-\beta W} \rangle = e^{-\beta \Delta F},
\end{equation}
where $\langle \cdot \rangle$ denotes an average over multiple realisations of the given protocol, and $\beta = 1/k_B T$. This expression, known as the Jarzynski equality, holds for arbitrary processes, even those far from equilibrium. It is remarkable because it connects equilibrium and nonequilibrium quantities: while the left-hand side represents the averaged work over many realisations of a process, the right-hand side corresponds to the free energy difference between two equilibrium states. Its applicability extends beyond physics~\cite{Jarzynski2012}.

Understanding the thermodynamics of systems far from equilibrium, where fluctuations cannot be neglected, is anything but simple. However, if the system evolves under well-defined external protocols—for instance, via external driving that deterministically changes the parameters describing the system—thermodynamic quantities such as work, heat, or entropy production can be treated as random variables, each characterised by a probability distribution constructed over many realisations of the protocol. Remarkably, fluctuations in these quantities satisfy universal constraints known as fluctuation theorems~\cite{Crooks1999,Campisi2011}, with the Jarzynski equality being a prominent example of an integral fluctuation theorem. While we do not explore the full details of fluctuation theorems in this tutorial, it is important to note that many of the concepts and illustrations we discuss here are based on and derived from these foundational relations.

Originally, free energy was defined only for states in thermal equilibrium. However, given its operational meaning, the definition can be extended to non-equilibrium states~\cite{Esposito2009, Parrondo2015}. Specifically, in the next section, we will naively replace the thermodynamic entropy with a different notion of entropy. This will allow us to define a \emph{non-equilibrium free energy}. In this broader context, the system interacts with a thermal reservoir at temperature $T$, although the system itself may not have a well-defined temperature. While we will not explicitly demonstrate it here, non-equilibrium free energy quantifies the maximum average work that can be extracted from a system in an out-of-equilibrium state~\cite{Gour2015,lostaglio2019introductory}.

\section{Quantum mechanics in a nutshell}\label{Sec:quantum-mechanics}

\vspace{-0.4cm}
\begin{wrapfigure}{l}{0.08\textwidth}
     \includegraphics[width=0.07\textwidth]{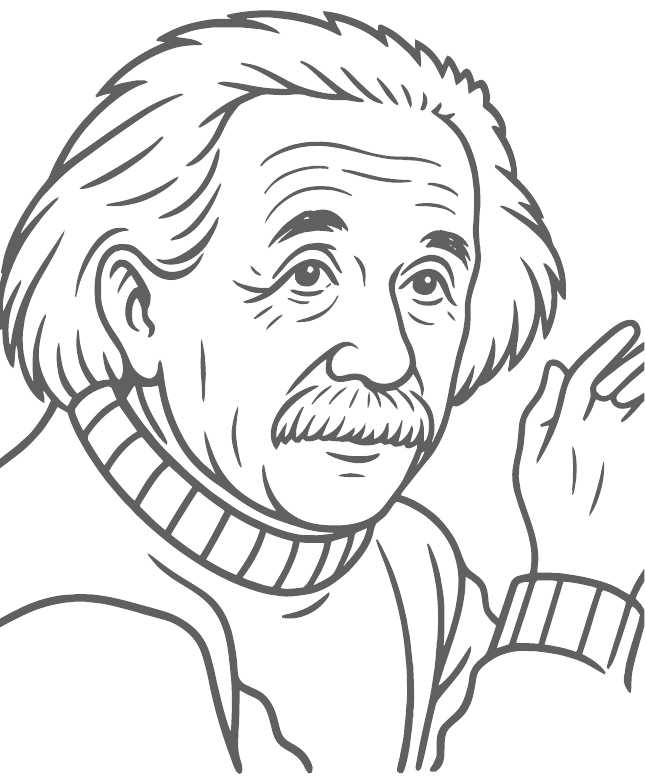}
 \end{wrapfigure}
\emph{\phantom{This text will be invisible This text will be invisible This text will be invisible This text will be invisible This text will be invisible This text will be invisible} 
``Quantum mechanics: Real Black Magic Calculus.''~\cite{ponomarev2021quantum}}

\phantom{This text will be invisible This text will be invisible This text will be invisible This text will be invisibleThis text will be invisible This text will be invisible This text will be invisible This text will be invisible }

For completeness and to make this tutorial as self-contained as possible, in this section, we will cover the basic tools needed to describe quantum systems, focussing on what is useful to discuss Maxwell's problem and information erasure later on. As a result, we avoid going into too much detail and are less rigorous in both notation and justification -- proving Einstein’s words to be surprisingly accurate. For those who want a comprehensive and detailed material, we recommend the books by Sakurai~\cite{Sakurai2017}, Peres~\cite{peres1995quantum}, Nielsen and Chuang~\cite{nielsen2010quantum}, or Bertlmann and Friis~\cite{bertlmann2023modern}. Readers already familiar with quantum mechanics and quantum information theory are welcome to jump ahead — it will not make resolving Maxwell's demon any more difficult.

To simplify the mathematical description, we limit our discussion to finite-dimensional systems, as many experimentally relevant physical systems can be well approximated this way. To keep the discussion general, we avoid focussing on specific physical systems while still aiming to keep the explanations as simple as possible. This approach allows us to focus on the core ideas without getting bogged down in unnecessary complexity.

In a classical system with a configuration space of size $d < \infty$, the system can exist only in one of the $d$ distinct configurations. However, in quantum theory, describing a system with $d$ configurations means associating it with a finite-dimensional Hilbert space $\mathcal{H} \in \mathbb{C}^d$, where each configuration corresponds to a vector in an orthonormal basis. But what exactly is a Hilbert space and how do we construct one?

For a system with $d$ distinct configurations labelled $\{1, \dots, d\}$, the associated Hilbert space $\mathcal{H}$ is a complex vector space equipped with an inner product. Each configuration corresponds to a vector $\ket{i} \in \mathcal{H}$, and the set $\{\ket{1}, \ket{2}, \dots, \ket{d}\}$ forms an orthonormal basis for $\mathcal{H}$. Our workhorse in this tutorial is the quantum version of a classical bit, the qubit, which can be described by two configurations, $0$ and $1$. The Hilbert space of the qubit is $\mathcal{H} = \mathbb{C}^2$, and the two classical configurations correspond to the quantum states $\ket{0}$ and $\ket{1}$ in $\mathcal{H}$. Examples of such systems include a photon of light with two orthogonal polarisations, an electron with two possible spin states, or a two-level atom with two possible energy states, among many others.

We can move on to the postulates of quantum mechanics, which will guide us in: \emph{(i)} defining the possible states of a quantum system, \emph{(ii)} understanding and calculating measurement outcomes, and \emph{(iii)} describing how the system evolves over time.

\begin{enumerate}[label=\textbf{I}.]
    \item A pure state of a quantum system is associated with a vector in the Hilbert space, represented by \mbox{$\ket{\psi} \in \mathcal{H}$}. 
\end{enumerate}
In some cases, we can only associate a vector $\ket{\psi_i}$ with a classical probability $p_i$. For example, imagine placing qubits randomly in a box, sometimes in the state $\ket{0}$ and sometimes in the state $\ket{1}$. If we blindly pick a qubit from the box, the state will be $\ket{0}$ or $\ket{1}$, each with a probability of $\frac{1}{2}$. These states are called mixed states and are mathematically described by an operator $\rho = \sum_i p_i \ketbra{\psi_i}{\psi_i}$. As a result, the previous postulate can be refined as follows:
\begin{enumerate}[label=\textbf{I'}.]
    \item To any state of a system, we associate a density operator $\rho$, which \emph{(i)} is Hermitian ($\rho = \rho^\dagger$), \emph{(ii)} is positive semidefinite ($\rho \geq 0$), and \emph{(iii)} has a trace equal to one $[\tr(\rho) = 1]$.
\end{enumerate}

\begin{enumerate}[label=\textbf{II}.]
    \item An observable is associated with a Hermitian operator $O = O^{\dagger}$ that acts on the Hilbert space of the system.
\end{enumerate}
Given an observable $O$, we can always express its spectral decomposition as $O = \sum_i o_i \mathbbm{P}_i$, where $o_i$ are real numbers, and $\mathbbm{P}_i$ are orthogonal projection operators onto the eigenspace corresponding to the eigenvalue $o_i$. These projectors satisfy $\mathbbm{P}_i \mathbbm{P}_{i'} = \delta_{i,i'} \mathbbm{P}_i$ and $\sum_i \mathbbm{P}_i = \mathbbm{1}$.
\begin{enumerate}[label=\textbf{III}.]
    \item Given a state $\rho$, when we measure an observable $O$, the outcome will be one of its eigenvalues $o_i$. The probability $p_i$ of obtaining this outcome and the state $\rho_i$ after the measurement are given by $p_i = \tr(\mathbbm{P}_i \rho)$ and $\rho_i = \mathbbm{P}_i \rho \mathbbm{P}_i / p_i$, respectively.
\end{enumerate}
Note that in quantum measurements, the act of measuring generally changes the state of the system, typically collapsing it into one of the eigenstates of the observable~\cite{Bassi2013}. This postulate
also tells us that it is possible for the same experiment, repeated
under identical conditions, to give different outcomes.

\begin{enumerate}[label=\textbf{IV}.]
    \item The dynamics of a closed system is defined by a unitary operator $U$ that satisfies $U U^{\dagger} = U^{\dagger} U = \mathbbm{1}$ as
    \begin{align}
        \ket{\psi} \to \ket{\psi'} &= U\ket{\psi} \quad \text{and} \quad  \rho \to \rho' = U\rho U^{\dagger}.
    \end{align}
\end{enumerate}
In many cases of interest, we can express the time evolution operator as $U = e^{-i Ht}$, where $H$ is the Hamiltonian, $t$ is the elapsed time, and we set $\hbar = 1$. Importantly, unitary evolution occurs when the system is completely isolated, which we refer to as a closed system. However, systems are rarely fully isolated, as they typically interact with other degrees of freedom, known as the environment. Accounting for the environment introduces the possibility of more general dynamics~\cite{kossakowski1972quantum,gorini1976completely,lindblad1976generators}, which will also not be directly discussed in this tutorial, although many examples discussed here are described in open dynamics.

Let us now briefly comment on a fundamental quantity that appears in von Neumann's seminal work from 1932~\cite{von2018mathematical}, which today bears his name. Motivated by a desire to better understand the nature of the measurement process, he devised a thought experiment\footnote{von Neumann considers an ideal gas where mechanical properties follow classical mechanics, while quantum states act as informational labels. This separation of physical behaviour and information allows him to use semipermeable membranes and the second law of thermodynamics to derive Eq.~\eqref{Eq:von-Neumann-entropy}. See Ref.~\cite{Minagawa2022} for a rigorous formulation of this thought experiment in a purely operational way.} (this time without challenging the second law) and ended up deriving what we now call the von Neumann entropy:
\begin{equation}\label{Eq:von-Neumann-entropy}
    S(\rho):=-\tr(\rho \log \rho) = -\sum_i p_i \log p_i,
\end{equation}
where $p_i$ are the eigenvalues of $\rho$. The equation above quantifies the amount of mixedness or lack of knowledge about the exact state of a system when it is described by a density matrix $\rho$. Therefore, we often say that the von Neumann entropy is a measure of the system's mixedness. Consequently, the entropy of a pure state is zero because there is no ambiguity about the system's state. Mathematically, this is evident since a pure state has one eigenvalue $\lambda_1 = 1$, with all the others zero. However, if the system is in a mixed state, its entropy is larger than zero, which captures our incomplete knowledge about the system. In the extreme case, where we have no information about the system's state (i.e., when it is maximally mixed), its entropy reaches the maximum value of $\log d$, where $d$ is the dimension of the system.

The von Neumann entropy has many interesting properties. In the upcoming discussion, we will focus on three relevant ones. Although these can be proven formally, we will skip the proofs and instead rely on their physical interpretations to understand the results. The first property is that von Neumann entropy is additive for uncorrelated systems: given $\rho_{\ms A}$ and $\rho_{\ms B}$, describing independent systems $\ms A$ and $\ms B$, we have 
\begin{equation}\label{Eq:entropy-additive}
    S(\rho_{\ms A} \otimes \rho_{\ms B}) = S(\rho_{\ms A}) + S(\rho_{\ms B}).
\end{equation}
This property simply tells us that in the absence of any correlation between $\ms {A}$ and $\ms{B}$, the total uncertainty is just the sum of the uncertainties of the individual systems. The second property is that the von Neumann entropy is invariant under unitary operations
\begin{equation}\label{Eq:entropy-invariance}
    S(\rho) = S(U\rho U^{\dagger}).
\end{equation}
Unitary evolution corresponds to a change of basis in the Hilbert space. Since no information is lost and unitary processes are reversible, it is expected that the von Neumann entropy remains unchanged in such cases. Finally, the von Neumann entropy is subadditive. Given two subsystems $\ms{A}$ and $\ms{B}$, the entropy of the composite system satisfies the following inequality
\begin{equation}\label{Eq:entropy-subadditive}
    S(\rho_{\ms{AB}}) \leq S(\rho_{\ms A}) + S(\rho_{\ms B}).
\end{equation}
This inequality captures the fact that there may be correlations between the two subsystems, which can reduce the total uncertainty about the joint system. Consequently, it is saturated only when the two systems are uncorrelated. Thus, we can use this as a measure of the total degree of correlations by defining the mutual information:
\begin{equation}\label{Eq:mutual-information-quantum}
   I(\ms{A}:\ms{B})_{\rho_{\ms{AB}}} := S(\rho_{\ms A}) + S(\rho_{\ms B})-S(\rho_{\ms{AB}}).
\end{equation}
This quantity represents the amount of information stored in $\ms{AB}$ that is not contained in $\ms A$ and $\ms B$ individually. However, it is important to note that mutual information quantifies the total degree of correlations without distinguishing between classical and quantum contributions. Another important quantity that we will come across in the next sections is the quantum relative entropy, or the Kullback-Leibler divergence~\cite{kullback1951information}. Given two density matrices $\sigma$ and $\rho$, it is defined as:
\begin{equation}\label{Eq:relative-entropy}
    S(\sigma \|\rho) := \tr[\sigma(\log \rho - \log \sigma)] = -S(\sigma) + \tr(\sigma \log \rho).
\end{equation}
This quantity is important to us because it satisfies the Klein inequality $S(\sigma \|\rho) \geq 0$, and $S(\sigma \|\rho) = 0$ if and only if $\sigma = \rho$. 

In Sec.~\ref{Sec:information-theory}, we will introduce another entropy, which is mathematically very similar to the von Neumann entropy and can be interpreted as its classical counterpart. Importantly, while von Neumann and thermodynamic entropy frequently coincide—particularly for equilibrium states described by Gibbs ensembles—this is not guaranteed in general. A simple example illustrating this discrepancy is a single isolated quantum system prepared in a pure energy eigenstate. Suppose a system with Hamiltonian $H$ is prepared in a pure state $\ket{E}$ of energy $E$. Its density operator is $\rho = \ketbra{E}{E}$, which has zero von Neumann entropy. However, if the system were in thermodynamic equilibrium at the same energy $E$, we would describe it using a canonical or microcanonical ensemble, each typically possessing nonzero thermodynamic entropy. This contrast demonstrates that the two notions of entropy cannot, in general, be identified—similar differences can arise in other contexts, such as strong coupling~\cite{Gallego2014,Strasberg2016,Perarnau2018} or non-equilibrium scenarios~\cite{Talkner2009,Sone2023}. We will return to this discussion at the end of Sec.~\ref{Sec:information-theory}.

\begin{coolexample}[Entropy of a system at thermal equilibrium]

Consider a system, described by a Hamiltonian $H$, and prepared in a thermal Gibbs state at inverse temperature $\beta$:
\begin{equation}\label{Eq:Gibbs-state}
    \gamma = \frac{e^{-\beta H}}{Z} \quad \text{where} \quad Z:= \tr(e^{-\beta H}).
\end{equation}
Substituting Eq.~\eqref{Eq:Gibbs-state} into Eq.~\eqref{Eq:von-Neumann-entropy}, we obtain 
\begin{align}\label{Eq:entropy-thermal-state}
    S(\gamma) &\!=\! -\tr(\gamma \log \gamma) \!=\! -\tr(\gamma \log \frac{e^{-\beta H}}{Z}) \!=\! -\tr[\gamma\qty(-\beta H - \mathbbm{1}\log Z)] \nonumber
    \\&= \beta\tr(\gamma H)+\log Z.
\end{align}
\end{coolexample}

By defining the average energy as $E := \tr(\gamma H)$ and identifying the last term as the free energy $F := -\log Z/\beta$ in the canonical ensemble~\cite{pathria2016statistical}, where $Z = \tr(e^{-\beta H})$, the free energy of the system can be expressed in terms of the von Neumann entropy as $F = E - \beta^{-1} S(\gamma)$.

Motivated by its operational role as the maximum average extractable work, the
free energy can be extended from equilibrium to arbitrary quantum states. Consider a finite–dimensional system with Hamiltonian $H$ in an arbitrary state $\rho$, whose average energy is $E(\rho):=\tr(\rho H)$. If the system exchanges energy with a heat bath at inverse temperature
$\beta$, we define the nonequilibrium free energy~\cite{Parrondo2015,Esposito2009}
\begin{align}
    \mathcal{F}(\rho):&=E(\rho) - \frac{1}{\beta}S(\rho)  ,\nonumber\\ &\overset{\eqref{Eq:relative-entropy}}{=} \frac{1}{\beta}[S(\rho\|\gamma) - \log Z], \label{Eq:noneq-F}
\end{align}
where $S(\rho)$ is the von Neumann entropy given by Eq.~\eqref{Eq:von-Neumann-entropy}, $\gamma$ is thermal Gibbs state defined in Eq~\eqref{Eq:thermal-state} and $Z$ is the partition function. Because $S(\rho\|\gamma)\geq 0$ with the equality if and only if $\rho = \gamma$, the Gibbs state uniquely minimises $\mathcal F(\rho)$. Equivalently, writing the equilibrium free energy $F=-\beta^{-1}\log Z$, Eq.~\eqref{Eq:noneq-F} can be recast as $\mathcal{F} = \beta^{-1}S(\rho\|\gamma) + F$.

\begin{coolexample}[Recording quantum information]
As a warm-up example for what follows, imagine that we have a particle in a box, where its exact position is unknown, but we know it can be in one of two possible states: either left (L) or right (R). We will demonstrate that by using an additional subsystem, a \emph{memory}, we can store the position of the particle in the memory. To describe this scenario, we first assume the particle is in a mixed state while the memory is in a blank state, so the composite system is represented by the following density operator:
\begin{equation}
    \rho_{\textsf{SM}} = \frac{(\ketbra{L}{L}_{\ms{S}}+\ketbra{R}{R}_{\ms{S}})}{2}\otimes \ketbra{0}_{\textsf{M}}.
\end{equation}
We now apply a controlled-NOT (CNOT) gate in the composite system. This is a unitary operation that acts as follows: $U_{\text{CNOT}} = \ketbra{L}{L}_{\ms{S}}\otimes\mathbbm{1} + \ketbra{R}{R}_{\ms{S}}\otimes \sigma_x$, where $\sigma_x$ is the Pauli-$x$ matrix. Its effect can be understood by observing how it transforms the states: 
\begin{align}
\begin{split}
\ket{L}_\ms{S}\otimes \ket{0}_\ms{M} &\to \ket{L}_\ms{S}\otimes\ket{0}_\ms{M},\\
\ket{R}_\ms{S}\otimes\ket{0}_\ms{M} &\to \ket{R}_\ms{S}\otimes\ket{1}_\ms{M}.\\
\end{split}
\end{align}
As we can see, this unitary operation changes the state of the second qubit conditioned on the state of the first one. Hence, after the unitary, the composite state of the system is now given by
\begin{align}\label{Eq:output-state}
    \sigma_{\ms{SM}}&= U_{\text{CNOT}}\rho_{\textsf{SM}}U_{\text{CNOT}}^{\dagger} = \frac{1}{2}(\ketbra{L,0}{L,0} + \ketbra{R,1}{R,1}).
\end{align}
We observe that if the particle is on the left, the memory remains in the blank state $\ket{0}_{\ms M}$, but if the particle is on the right, the memory is in the state $\ket{1}_{\ms M}$. Once we know the particle's position, we can proceed to use this information for a thermodynamic protocol or any other information-processing task that we wish. However, note that if we now look at the state of the memory [which corresponds to tracing out the main system in Eq.~\eqref{Eq:output-state}], we find that the memory is in a mixed state. However, if we measure the state of the memory, we will obtain a pure state for the system, conditioned on the outcome of the measurement. This means that the memory is entangled with the system, and the memory can be used to extract information about the system. In practice, if we discard the measurement result (i.e., ignore the outcome), the memory is in a mixed state. In this case, the memory can no longer be reused for the same process unless it is reset to its initial state or replaced with a new one.
\end{coolexample}

Let us summarise this example. We begin with the system in an unknown state, while the memory is in a blank known state. After the operation, the memory becomes correlated with the system: it is in the state $\ket{0}_{\ms M}$ if the particle is on the left side of the box $\ket{L}_{\ms S}$, or in the state $\ket{1}_{\ms M}$ if the particle is on the right side $\ket{R}_{\ms S}$. This information can then be used to perform a specific protocol, such that by the end of the transformation, the system returns to its initial state, but the memory does not—it ends up in a mixed state between blank and non-blank. We will revisit this problem in Sec.~\ref{Sec:exorcisim} and reinterpret it in light of its connection to Maxwell’s demon.

\begin{coolexample}[Simple model for thermodynamic work in quantum mechanics ]

Consider a closed system described by a Hamiltonian $H$, satisfying the eigenproblem $H\ket{\psi_n} = \epsilon_n\ket{\psi_n}$, where $\ket{\psi_n}$ and $\epsilon_n$ denote the $n$th eigenstate and eigenenergy, respectively. The average energy of the system is given by $E = \sum_n p_n \epsilon_n$, where $p_n$ represents the occupation probability of the $n$th eigenstate. Note that if the system is in thermal equilibrium, $p_n$ is determined by the Gibbs distribution [Eq.~\eqref{Eq:thermal-state}]. The differential of the internal energy $E$ is
\begin{equation}\label{Eq:dU-decomposition}
    \textrm{d}E = \sum_n (\underbrace{\epsilon_n \textrm{d}p_n}_{\textrm{d}Q}+\underbrace{p_n \textrm{d}\epsilon_n}_{-\textrm{d}W}).
\end{equation}
From the first law of thermodynamics for an isothermal process, we have $T\textrm{d}S = \textrm{d}E + \textrm{d}W$. Consequently, we can identify that the first term in Eq.~\eqref{Eq:dU-decomposition} is the heat exchanged, and the second term is the negative of the work done by the system. This decomposition is not universal; rather, it is motivated by scenarios where the system’s Hamiltonian depends on an external parameter $\lambda$ that can be modified, thus changing $\epsilon_n$ - for example, by applying external fields. In this context, the second term should be associated with $T \textrm{d}S$, as the entropy is given by Eq.~\eqref{Eq:von-Neumann-entropy}. Let us now derive an expression for thermodynamic work in an isothermal process when the system starts at thermal equilibrium at inverse temperature $\beta$ and the external parameter $\lambda$ is varied from $\lambda_1$ to $\lambda_2$ at constant temperature. First, we note the following identity:
\begin{equation}\label{Eq:pn-identity}
   \!\!\!\!\! -\frac{1}{\beta}\frac{\partial \log Z}{\partial \epsilon_m} \!=\! -\frac{1}{\beta Z}\frac{\partial Z}{\partial \epsilon_m} \!=\! \frac{1}{Z}\qty[\frac{\partial}{\partial \epsilon_m}\sum_n e^{-\beta \epsilon_n}] = \frac{e^{-\beta \epsilon_n}}{Z} =  p_n.
\end{equation}
Since $\epsilon_n$ depends on $\lambda$, substituting Eq.~\eqref{Eq:pn-identity} in the expression for $W$, we obtain 
\begin{align}\label{Eq:simple-work}
    W&=\frac{1}{\beta}\sum_n\int_{\lambda_1}^{\lambda_2}  \qty(\frac{\partial \log Z }{\partial \epsilon_n})\qty(\frac{\partial \epsilon_n}{\partial \lambda}) \,\textrm{d}\lambda \nonumber \\ &=\frac{1}{\beta} [\log Z(\lambda_2) - \log Z(\lambda_1)].
\end{align}
This expression relates the work done during an isothermal process, as the external parameter $\lambda$ varies from $\lambda_1$ to $\lambda_2$.
\end{coolexample}

In the previous example, we introduced a simple model for thermodynamic work in quantum mechanics. In general, the definition of work in quantum thermodynamics depends on the specific context (see Ref.~\cite{Binder2018} for a detailed discussion), and \emph{no} universal definition exists. Here, we adopt a straightforward model to illustrate, in the next section, how energetic costs can arise in the operation of Maxwell’s demon.

\section{Picking a hole}
\label{Sec:picking-a-hole}

Maxwell's demon was conceived by James Clerk Maxwell in 1871, as described in his book \emph{Theory of Heat}~\cite{maxwell2001theory}. This concept was introduced toward the end of the book, in a section discussing the limitations of the second law of thermodynamics. However, the idea originally appeared in a letter written by Maxwell in 1867, where he boldly proposed \emph{``...picking a hole''} in the second law~\cite{knott1911life}. The famous term ``demon'' for this imaginary intelligent being was coined by Lord Kelvin 3 years after Maxwell's book. Although one might assume that this name was chosen to evoke a malicious creature, it was intended to highlight the idea of a supernatural being—a concept rooted in Greek mythology~\cite{delahunty2008foreign,cresswell2014origins}.
\begin{figure}[t]
    \centering
    \includegraphics{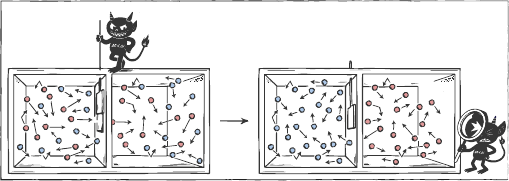}
    \caption{\emph{Temperature demon}. An intelligent being, a demon, controls a door between two gas chambers. As gas molecules approach the door, the demon selectively opens and closes it, allowing only fast molecules to pass one way and slow molecules the other. Since the kinetic temperature of a gas is directly linked to the velocities of its constituent molecules, the demon's strategy results in one chamber warming up while the other chamber cools down. This reduces the overall entropy of the system without requiring any work, thereby violating the second law of thermodynamics. }
    \label{F-Maxwell-temperature-demon}
\end{figure}

In this section, we introduce the ideas of Maxwell's demon and the Szilard engine. Rather than presenting the solution straight away, we take a more gradual, historical approach by first laying out the problem and leaving it open. Later, we bridge this discussion with tools from information theory, which allows us to resolve the apparent paradox. However, when discussing the Szilard engine, we also introduce new insights and results that provide a deeper understanding of this thought experiment.

\subsection{Maxwell's demon \label{Subsec:maxwells-demon}}

We start by analysing what is known as the \emph{temperature demon}. The setup depicted in Fig.~\ref{F-Maxwell-temperature-demon} illustrates the core concept of Maxwell's thought experiment: a tiny creature has access to a gas of particles that are initially in thermal equilibrium. The demon controls a trapdoor that it can freely open or close without friction. The creature's task is to generate a temperature difference in the gas without performing any work on it. Essentially, this scenario is analogous to producing a heat flow from a lower to a higher temperature without any other effect, which contradicts Clausius’s formulation.. 

The kinetic theory of gases tells us that temperature is related to the average kinetic energy of the particles, and according to the Maxwell-Boltzmann distribution, some particles are faster or slower than the average~\cite{boltzmann1868studien}. Thus, the demon must be able to distinguish between the velocities of the particles that make up the gas. In other words, it is capable of measuring each particle, ``remembering'' what it measures, and then opening or closing the trapdoor accordingly. This implies that the \emph{information} collected from the measurement is recorded in some sort of device, which we will refer to as \emph{memory}.

If we assume that the demon itself is a physical system that is subject to the laws of thermodynamics, what do the first and second laws tell us about this setup? Since both the demon and the gas are thermally isolated, the demon’s energy remains unchanged during the sorting process. According to the second law, however, the demon’s entropy must increase by at least as much as the gas's entropy decreases. Therefore, the demon’s entropy must increase while its energy remains constant.
\begin{figure}[t]
    \centering
    \includegraphics{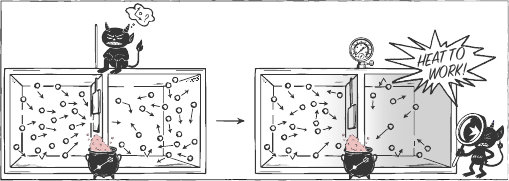}
    \caption{\emph{Pressure demon}. A less intelligent demon controls a door between two gas chambers connected by a constant temperature reservoir. As gas molecules approach the door, the demon selectively opens and closes it, allowing only right-moving molecules to pass in one direction and only left-moving molecules to pass in the opposite direction. This strategy increases the pressure on one side of the system while decreasing it on the other. The demon's actions convert heat from the reservoir into work, seemingly in violation of Kelvin's statement of the second law of thermodynamics }
    \label{F-Maxwell-pressure-demon}
\end{figure}

Now, consider another demon that differs from the previous one by not being as smart --  the \emph{pressure demon} (see Fig.~\ref{F-Maxwell-pressure-demon} for a pictorial representation). Instead of knowing the precise position and velocity of each particle, it only knows the direction in which each particle is moving. In this case, the demon allows all particles moving in one direction to pass through while stopping all those moving in the opposite direction. Effectively, this creates a pressure difference. It operates by making the gas interact with a heat bath at a constant temperature after generating a pressure inequality. Note that the sole effect of this process is the conversion of heat transferred from the heat bath into work, which constitutes a direct violation of Kelvin's statement of the second law. As we shall see in the next section, this type of demon was carefully discussed and analysed by Leo Szilard~\cite{szilard1929entropieverminderung,Szilard1964}, who, for simplicity, considered a single particle instead of a gas.

Is it possible to design an autonomous machine that exploits statistical fluctuations to convert heat into work without a demon operating it? Smoluchowski~\cite{smoluchowski1912experimentell,smoluchowski1967gultigkeitsgrenzen} provided a negative answer to this question by demonstrating that a purely mechanical version of Maxwell’s demon is impossible. A microscopic trapdoor designed to let only fast molecules pass would be affected by its own Brownian motion, causing it to move randomly. These thermal fluctuations would result in molecules passing between chambers in both directions, preventing any net energy transfer and making such a machine unreliable. Later, Feynman expanded on Smoluchowski’s thought experiment to show that a Brownian ratchet, much like a microscopic trapdoor, would slip both forward and backward, preventing net work extraction~\cite{Feynman1964}\footnote{While purely mechanical devices cannot harness thermal fluctuations to perform useful work, supplying external energy and information to the system makes it possible. This principle was demonstrated with a molecule called rotaxane~\cite{Serreli2007}. By adding light as an energy source, random Brownian motion can be overcome to produce a controlled movement that performs useful work. Importantly, this process does not contradict the second law of thermodynamics, as the external energy input drives the system away from equilibrium.}.

\vspace{1cm}
\begin{mybox}{Maxwell's demon at work~\cite{Cottet2017}}{Maxwell's demon at work}

This experiment realises a Maxwell's demon while tracking all involved thermodynamic quantities, allowing full characterisation of the demon's memory and its role in the thermodynamic process.

\vspace{0.2cm}

The demon is modelled as a microwave cavity, while the system is a superconducting qubit. They interact via the following Hamiltonian:
\begin{equation}\label{Eq:Hamiltonian-system-demon}
    H = \hbar \omega_s \ketbra{e}{e}+\hbar \omega_D d^{\dagger} d - h\chi d^{\dagger}d \ketbra{e}{e},
\end{equation}
where $\ket{g}$ and $\ket{e}$ represent the ground and excited states of the qubit, with an energy difference $h \omega_s$, $d$ is the annihilation operator for a photon in the cavity, with resonant frequency $\omega_D$ and $h$ is Planck's constant. From this Hamiltonian, we observe that no direct energy exchange occurs between the qubit and the cavity. However, the interaction induces a frequency shift of $-\chi$ in the cavity when the qubit is excited. If $N$ photons are present in the cavity, this results in a total frequency shift of $-N\chi$.

\vspace{0.2cm}
The system and the demon's memory are initially prepared in thermal and vacuum states, respectively. A pulse at frequency $\omega_D$ is then applied to record the state of the system in the demon's memory. As described by the Hamiltonian, the cavity is excited only if the qubit is in the ground state. Work is extracted by applying a pulse at frequency $\omega_s$, which acts on the qubit, functioning as a battery system. As indicated in Eq.~\eqref{Eq:Hamiltonian-system-demon}, the demon controls the energy transfer by preventing the qubit from absorbing or emitting energy in an uncontrolled manner. If the correlation between the demon and the qubit is perfect, the energy flows from the qubit to the battery, ensuring that work is extracted. In this process, the system's entropy decreases while the demon's memory records the change. The demon's memory is then reset, completing the cycle.
\end{mybox}

A purely mechanical version of Maxwell’s demon cannot work, but this does not rule out the possibility that such a violation occurs when an intelligent being, with information about the system, is present. In other words, what would happen to the second law of thermodynamics if we consider a mechanical demon equipped with a device, such as a computer, that gathers and processes information?

\subsection{Szilard engine \label{Subsec:szilard-engine}}

The interplay between thermodynamics and information came up 62 years after Maxwell's demon was conceived. The Szilárd engine~\cite{szilard1929entropieverminderung,Szilard1964}, a machine with a one-molecule working fluid, is a simple device that implements a cyclic process and extracts energy as work from a thermal reservoir. As with Maxwell's demon, the Szilárd engine can apparently violate the second law of thermodynamics whenever some information about the state of the system is available.

\begin{figure}[t]
    \centering
    \includegraphics{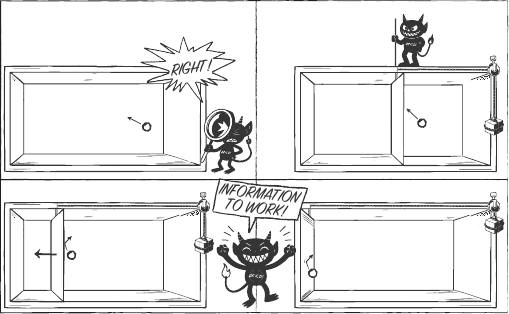}
    \caption{\emph{Szilárd engine}. A chamber containing a single particle (atom or molecule) is initially in an unknown position. A demon measures the particle's position and inserts a movable, massless wall with an attached mass. As the gas in the chamber expands at a constant temperature, the attached mass is raised, resulting in work being performed. This mechanism appears to enable the conversion of information into useful energy, seemingly contradicting the second law of thermodynamics, as the demon extracts heat from the heat bath and converts it fully into mechanical work.}
    \label{F-Szilard-engine}
\end{figure}

Szilard’s engine is a detailed version of Maxwell's pressure demon. Its work protocol is illustrated in Fig.~\ref{F-Szilard-engine}. The engine consists of a chamber with volume $V$ and a single gas molecule, in thermal equilibrium with a reservoir at temperature $T$, undergoing a three-step process. First, the demon measures the position of the particle, determining whether it is on the left or right side of the chamber. Second, based on the particle's position, it inserts a thin, massless, adiabatic partition into the chamber, dividing it into two halves. Importantly, a specific mass load is attached to the movable partition on the side where the molecule is located. Third, by maintaining the chamber at a constant temperature, the gas undergoes a quasistatic isothermal expansion, with the partition acting as a piston. When the partition reaches the end of the chamber, the gas returns to its original state, occupying the entire volume $V$. During this expansion, the heat extracted from the environment is completely converted into mechanical work.

The work extracted in the process can be calculated considering the following points: $\emph{(i)}$ the gas undergoes an isothermal expansion; $\emph{(ii)}$ the insertion and removal of the thin partition are performed reversibly, without expending energy; and $\emph{(iii)}$ the only energy contribution comes from the reversible expansion of the particle against the piston from $V/2$ to $V$. It should also be noted that the load must be continuously varied to match the gas pressure during expansion, ensuring that the process remains quasistatic and reversible. This condition allows us to express the pressure as $p = k_B T/V$. Thus, the amount of extracted work is given by:
\begin{equation}\label{Eq:Szilard-work}
    W = -\int_{V_i}^{V_f} pdV= k_B T \int_{\frac{V}{2}}^{V}\frac{dV}{V} = -k_BT \log 2.
\end{equation}
This simple protocol shows a direct conversion of heat into work, contradicts Kelvin's statement, and thus seemingly violates the second law of thermodynamics. However, it is worth noting that if the system is modified to include more partitions, the volume increase would be larger, leading to the potential for extracting more work. In this case, the work extracted would scale with the number of partitions, enhancing the total energy conversion.

\vspace{1cm}
\begin{mybox}{Single-electron box as a Szilard engine~\cite{koski2014experimental}}{Single-electron box as a Szilard engine}

This experiment shows the extraction of $k_B T \log 2$ of work from a heat bath per bit of information, with a fidelity of $75\%$.

\vspace{0.2cm}

The setup consists of two metallic islands connected by a tunnel junction. These islands contain an electron gas, and an additional ``extra'' electron is introduced by adjusting the gate charge $n_g$. The difference in chemical potential between the islands is controlled by a gate voltage $V_g$, applied to one of them. Initially, $V_g$ is set so that the extra electron is equally likely to be on either island. Next, the extra electron is detected using a single-electron transistor and is captured on one of the islands. To trap the electron, the gate charge is rapidly changed, resulting in an energy increase that prevents the electron from tunnelling out. Finally, $n_g$ slowly returns to its original condition, extracting energy from the heat bath in the process, and completing the cycle.

\vspace{0.2cm}

Unlike the classical Szilard engine, this realisation involves a gas of electrons with an additional ``extra'' electron, instead of a single particle. After measuring the electron, no partition is installed. Instead, the charge configuration is changed, and the extracted work is not related to the volume but rather to the energy associated with changes in the gate charge.

\end{mybox}

If Szilard's conclusions are indeed correct, the system’s entropy should decrease, making us wonder why that could not actually be the case. Interestingly, Szilard, in his writings, assumes that any measurement procedure is inherently associated with a certain amount of entropy production. When this entropy production is taken into account, it restores the validity with the second law. Moreover, he argues that the amount of entropy produced could be greater than, but not less than, the value given by Eq.~\eqref{Eq:Szilard-work}. This is a remarkable observation, as we will see in Sec.~\ref{Sec:thermodynamics-of-computation}, , where we shall connect it with a fundamental bound on computation.

Although not entirely convinced, Szilard suspected that the act of measuring and recording information in a memory could eventually explain such a possible violation. Motivated by this, he considered two additional models that involve memory. However, it remained unclear whether the thermodynamic cost arose from the measurement, the act of recording, or the process of forgetting. Nevertheless, memory was regarded to play a crucial role in the demon's operation. Importantly, Szilard's reinterpretation of Maxwell's demon contributed to the concept of a bit of information. Years later, this notion would become a central concept in information and computing theories.

However, we may question the conclusions of Szilárd's engine, given that the entire analysis relies on a single molecule, where fluctuations cannot be neglected, and traditional thermodynamic methods may not apply. This objection can be overruled by noting that linking multiple Szilárd engines together minimises fluctuations, making it possible to apply thermodynamics. Yet, even if we accept the single-molecule system, the gas behaves in a way that seems to violate the ideal gas law. When the partition is inserted into the chamber, the gas is suddenly confined to half of its original volume without any change in temperature or expenditure of energy.

The second objection is more difficult to refute than the first. Szilárd’s position holds that, although the molecule is on one side of the partition after it is inserted, the key point is that we do not know which side it is on until a measurement is made. This uncertainty is critical because, from an informational perspective, the gas as a system still effectively occupies both sides of the partition. However, one might question whether this argument is subjective, as it depends on the observer's information (see Ref.~\cite{jauch1972entropy} for an extended discussion). Until we measure which side of the partition the molecule is on, we cannot fully define the system's state or extract any useful work. This blurs the line between an objective thermodynamic process—one that should operate independently of the observer—and a subjective one, where the observer's knowledge plays a role. The apparent subjectivity is resolved by appealing to stochastic thermodynamics~\cite{seifert2008stochastic,Strasberg2022}. Before any measurement, the molecule is described by an objective probability distribution over the two half-boxes, and the associated free energy already accounts for the cost of inserting the partition. A fully quantum treatment by Zurek \cite{Zurek1986,Zurek2018} reaches the same conclusion and, by invoking spatial superposition and decoherence, makes the observer-independence of the argument explicit at the microscopic level.

A third objection concerns the assumption that the insertion and removal of the partition do not incur any energy cost~\cite{Kim2011}. While this assumption is valid in the classical case, it does not hold when the problem is approached quantum mechanically. This is evident if one considers Szilard's setup as a single particle of mass $m$ confined within a one-dimensional square-well potential of length $L$. Solving the time-independent Schrödinger equation yields the following energy spectrum for the particle $\epsilon_n = \hbar^2 \pi^2 n^2 / 2m L^2$. Consequently, when the partition is introduced — a process that can be modelled as a potential barrier erected in the middle of the box — the boundary conditions change. As a result, the energy spectrum is altered to $E'_n = \hbar^2 \pi^2 n^2 / 2m (L/2)^2$. Using the decomposition provided by Eq.~\eqref{Eq:dU-decomposition}, one observes that $\textrm{d}\epsilon_n$ is non-zero due to the shifting energy levels. This implies that $\textrm{d}W$ is also non-zero, according to Eq.~\eqref{Eq:simple-work}, which quantifies the work performed during this process. The next question is whether this process can be assumed to be adiabatic. Using again Eq.~\eqref{Eq:dU-decomposition}, this would imply that $\textrm{d}p_n = 0$, meaning that the occupation probabilities remain constant during the process. However, as discussed, the energy levels shift non-uniformly when the partition is inserted. Consequently, the ratio $p_n/p_m$ no longer satisfies the Boltzmann distribution, indicating that the system is no longer in thermal equilibrium. However, since the system interacts with a heat bath, thermalisation will occur, allowing the entire process to be described as adiabatic and isothermal.

\begin{figure}[t]
    \centering
    \includegraphics{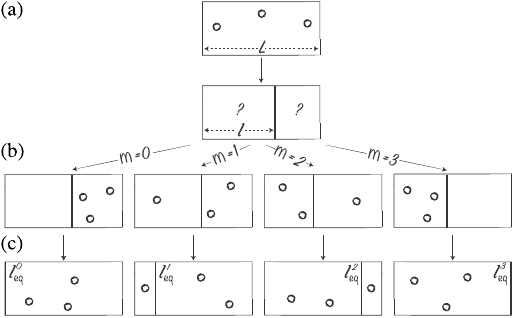}
    \caption{\emph{Quantum Szilárd engine}. Three quantum particles are confined in a box of size $L$. (a) A partition is installed at location $l$, and (b) a measurement is made revealing the number $m$ of particles on the left. (c) Then, the particles expand isothermally until they reach equilibrium at $l^m_{\rm eq}$, and the partition is removed.}
    \label{F-quantum-Szilard-engine}
\end{figure}

\example{Quantum Szilard engine }
Consider $N$ identical particles prepared in a potential well of size $L$~(see Fig.~\ref{F-quantum-Szilard-engine}) and the four-step protocol: \emph{(i)} partition insertion, \emph{(ii)} measurement, \emph{(iii)} expansion and \emph{(iv)} partition removal. Unlike the classical Szilárd engine, we insert the partition before making any measurements. This is justified because, before measurement, the particles are delocalised across the entire box, with a probability amplitude describing their likelihood of being found at any position. Following Ref.~\cite{Kim2011}, we calculate the thermodynamic cost associated with each step.

\emph{(i)} A wall is isothermally inserted in a position $l$. At this moment, the partition function $Z(l)$ includes all accessible microstates. Since the system is in a superposition of states with different numbers of particles $m$ on the left and $N-m$ on the right, the system's partition function is $Z(l) = \sum_{m=0}^N Z_m(l)$. The work required for the insertion process is then
\begin{equation}
W_{\rm ins} =\frac{1}{\beta} \left[ \ln Z(l) - \ln Z(L) \right].
\end{equation}

\emph{(ii)} We perform a measurement, collapsing the superposition and projecting the system into a specific state with a definite number of particles $m$ on the left and $N-m$ on the right. The measurement is assumed to be perfect and does not require any work. Consequently, the system can be in any of the possible states characterised by different $m$ with probability $p_m = Z_m(l)/Z(l)$. 

\emph{(iii)} The system undergoes an isothermal expansion. The wall moves until it reaches an equilibrium position $l^m_{\rm eq}$ determined by the balance of forces, $F^{\rm left}+F^{\rm right}=0$, where the generalised force $F$ is defined as $\sum_n P_n ({\partial \epsilon_n}/{\partial X})$. The average work extracted during the expansion is calculated by finding the work for each possible $m$ weight it by its probability $p_m$, namely
\begin{equation}\label{Eq:W-exp}
W_{\rm exp} =\frac{1}{\beta} \sum_{m=0}^N p_m\left[ \ln Z_m(l^m_{\rm eq}) - \ln Z_m(l) \right],
\end{equation}

\emph{(iv)} The wall is removed isothermally and quasistatically\footnote{In reality, the wall has a finite potential height $\lambda_\infty$, ensuring during expansion that the tunnelling time $\tau_t$ is much longer than the process time $\tau$, keeping $m$ well-defined. During removal, as $\tau_t$ decreases to match $\tau$ in $\lambda_0$, the eigenstates delocalise and the partition function changes from $Z_m(l^m_{\rm eq})$ to $Z(l^m_{\rm eq}) = \sum_{n=0}^N Z_n(l^m_{\rm eq})$. The integral appearing in Eq.~\eqref{Eq:simple-work}, 
for each $m$ splits into $ \int_{\lambda_\infty}^{\lambda_0} \frac{\partial \ln Z_m(l^m_{\rm eq})}{\partial X} dX $ and $ \int_{\lambda_0}^{0} \frac{\partial \ln Z(l^m_{\rm eq})}{\partial X} dX $, where the first term vanishes in the quasi-static limit ($\tau \to \infty$).}. 

The average work in the removal process is:
\begin{equation}\label{Eq:W-rem}
W_{\rm rem} =\frac{1}{\beta} \sum_{m=0}^N p_m\left[ \ln Z(L) - \ln Z(l^m_{\rm eq}) \right].
\end{equation}
Observe that the second term of Eq.~\eqref{Eq:W-rem} accounts for the partition function $Z(l_{\rm eq}^m) = \sum_{m=0}^N Z_m(l_{\rm eq})$ as after removal, all particles are once again delocalised [compare this with.~\ref{Eq:W-exp}]. 

Combining all the contributions, the total work during the cycle is given by
\begin{equation}\label{Eq:total-work}
W_{\rm tot}= W_{\rm ins} + W_{\rm exp} + W_{\rm rem} = -\frac{1}{\beta} \sum_{m=0}^N p_m \ln \left(\frac{p_m}{p^*_m}\right),
\end{equation}
where $p^*_m = Z_m(l^m_{\rm eq})/Z(l^m_{\rm eq})$. Let us analyse Eq.~\eqref{Eq:total-work}. First, note that the classical scenario is recovered by setting $N=1$, $l=L/2$. This implies that $p_0 = p_1 = 1/2$ and since $m=0,1$, the wall reaches the end of the box so that $Z(l_{\rm eq}^m) = Z_m(l_{\rm eq}^m)$ and $p^*_m = 1$ are always true. Consequently, Eq.~\eqref{Eq:total-work} reads $W_{\rm tot} = \beta^{-1} \log 2$. We can also compute each term independently to show that $W_{\rm ins} = \beta^{-1} \ln 2 - \Delta$, where $\Delta = \log [ z(L) / z(L/2) ]$ with $z(l) = \sum_{n=1}^\infty e^{ -\beta \epsilon_n(l) }$ and $\epsilon_n(l) = \hbar^2 \pi^2 n^2 / (2 m l^2)$; and similarly, $W_{\rm exp} = \Delta$ and $W_{\rm rem} = 0$. As temperature increases, the classical results of individual steps are recovered, i.e., $W_{\rm ins} \to 0$, $W_{\rm exp} \to k_{\rm B} T \log 2$, and $W_{\rm rem} = 0$, since $\Delta$ approaches $k_{\rm B} T \log 2$ in this limit.

Another interesting case to analyse is when $N=2$ and $l=L/2$. In this setup, $p^*_0 = p^*_2 = 1$, and for $m=1$, the wall does not move during the expansion process, as $l = l^1_{\rm eq}$ and $f_1 = p^*_1$. This leads to $W_{\rm tot} = -2\beta^{-1}p_0 \log p_0$. Ignoring particle spin and considering the extreme cases of temperature, we find that for $T \to 0$: $W_{\rm tot} = 2/3 \log 3$ for bosons, as they can occupy the same state, and $W_{\rm tot} = 0$ for fermions, due to the Pauli exclusion principle. The other limit, $T \to \infty$, both bosons and fermions yield $W_{\rm tot} = \log 2$, as the particles behave classically at high temperatures.

\examplend

The quantum Szilard engine, as described in the example above, sparked considerable discussion regarding the role of particle statistics and the relationship between information gain and extractable work. The assumptions and conclusions of Ref.~\cite{Kim2011} were later critiqued in Ref.~\cite{Plesch2013}, which questioned the handling of the partition removal stage and suggested that the original protocol could lose potential work due to tunnelling effects. The critique proposed an alternative approach, emphasising the role of generalised forces during partition removal and suggesting that the extracted work should always be non-negative. In their reply, the original authors defended their approach, arguing that the critique misunderstood the role of partition function discontinuities and incorrectly characterised their protocol as non-optimal~\cite{KimReply2013}. They clarified that under their protocol, optimal conditions ensure positive work extraction, and any negative results stem from misinterpreting the non-optimal scenarios.

The discussion culminated in a subsequent work~\cite{Plesch2014}, which reframed the debate by connecting the extraction of work with the gain in information during the measurement phase. This work showed that: \emph{(i)} The total extractable work, is determined solely by the mutual information gained from the measurement, regardless of particle type or statistics. \emph{(ii)} Quantum effects influence intermediate stages (e.g., expansion and removal), but when averaged, total work depends only on the number of measurement outcomes.

\subsection{Efforts to patch the hole \label{Subsec:efforts-to-patch}}

The measurement aspect raised by Szilard was further explored in the 1950s by Leon Brillouin~\cite{brillouin1951maxwell} and Dennis Gabor~\cite{gabor1969progress}. Independently, both demonstrated that the information obtained by performing a measurement results in an increase in entropy. To understand the main argument behind this claim, imagine that in Szilard's engine, we want to determine the position of the gas particle. The chamber containing this single gas particle is at a constant temperature, and, optically speaking, the radiation inside is that of a black body. According to quantum theory, this electromagnetic radiation, composed of photons, has a well-defined energy distribution~\cite{planck1900theory}. Hence, a high-temperature lamp is sufficient to provide light signals (photons) distinguishable from the existing black-body radiation, enabling the detection of the gas particle's position. 

Brillouin then concluded that introducing a light source to observe the particle is accompanied by an increase in entropy, which is sufficient to save the second law of thermodynamics. More precisely, the measurement process would dissipate energy on the order of $k_B T$. A bold comment we might make here is that Brillouin's reasoning is based on a vicious circle: if we assume that the chamber in which the demon operates is a blackbody, then it follows Kirchoff's law~\cite{kirchhoff1978verhaltnis}—however, this law is derived using the second law of thermodynamics. Consequently, it is no surprise that we can prove the second law to be satisfied.

Interestingly, Brillouin tackled this problem mathematically and, inspired by Shannon's recent theory of information (discussed in the next section), treated information and thermodynamics on equal footing. This fusion provided a mathematical basis for proving the increase in entropy. It is amusing to compare this to Maxwell's original idea—Brillouin actually narrowed the demon's capabilities, as the demon now requires a physical means to obtain information about the particle's position. Ironically, in their attempt to resolve Maxwell's puzzle, Brillouin and Gabor discarded the role played by the demon's memory. 


\section{Information and computation interlude}
\label{Sec:information-theory}

Szilard's engine planted the seeds of modern information theory two decades before the formal mathematical framework emerged. This setup can be seen as a device that acquires information and trades it for work extraction, seemingly in contradiction to the second law. Brillouin, however, mathematically argued that information gathering is inherently linked to an increase in entropy. What is more, he treated \emph{informational entropy} and thermodynamic entropy as fundamentally equivalent. But what exactly is informational entropy, and how does it relate to thermodynamics? 

From another angle, we can break down Szilard’s engine into a simple three-step process: the demon gets some information, makes a decision based on the acquired information, and then performs a protocol. So, could we loosely argue that Szilard’s engine is a thermodynamic model of computation? After all, it is essentially processing information (where the molecule is), making a logical decision (which side to open), and carrying out an action (allowing the gas to expand and do work). The whole process feels very much like a computation happening through a physical system. 

This section builds on these points to strengthen the connection between Szilard's engine, information theory, and computation. We begin with a recap of Claude Shannon's mathematical theory of information~\cite{shannon1938symbolic,shannon1948mathematical}, followed by brief remarks on the concept and evolution of computation.

\subsection{Classical information processing \label{Subsection:classical-info-processing}}

\begin{figure*}
    \centering
    \includegraphics{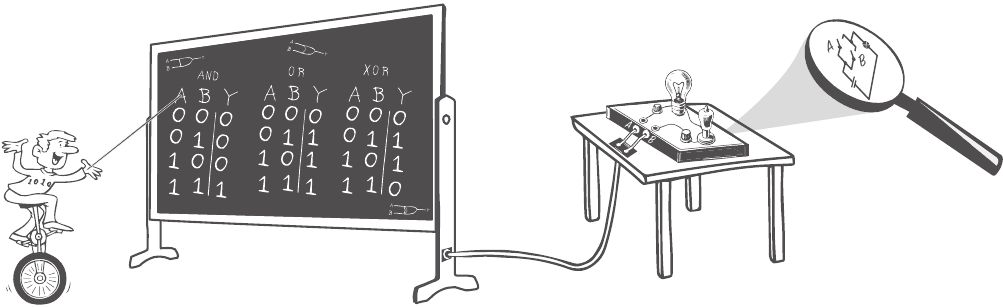}
    \caption{\emph{A Symbolic analysis of relay and switching circuit}. Shannon, famously known for riding a unicycle around the halls of Bell Labs, demonstrated that Boolean algebra could be implemented using relay and switching circuits. Boolean algebra consists of three elementary operations: \textbf{AND}, \textbf{OR}, and \textbf{NOT}. The rules for these operations are outlined in the truth table sketched in Shannon’s blackboard. For example, the implementation of an \textbf{OR} gate can be represented as a simple electrical circuit. In this case, switches A and B are arranged in parallel, meaning the circuit will be completed (and the lightbulb will turn on) if either switch A, switch B, or both are closed.}
    \label{F-shannon-scheme}
\end{figure*}

When we think about information, it is easy to imagine an endless stream of zeros and ones, just like in The Matrix ~\cite{wachowski1999matrix}. But beyond the cool visuals, this collection of binary digits obeys its own set of rules, known as Boolean algebra~\cite{boole1854investigation}. Developed in the mid-19th century, this toolkit provided a new theoretical playground for logic and philosophy. Most importantly, operations such as \textbf{AND}, \textbf{OR}, and \textbf{NOT} make it possible to manipulate bits of data in surprisingly simple ways~(see Fig.~\ref{F-shannon-scheme}). However, applying these concepts to real-world processes was still a distant reality. It took 84 years for Boolean algebra to find its practical application in the design of digital circuits, thanks to Shannon’s groundbreaking master’s thesis~\cite{shannon1938symbolic}.
Shannon bridged the gap between Boolean logic and electrical circuits by demonstrating that the rules of Boolean algebra could be used to design switching circuits. By treating the \textbf{ON} and \textbf{OFF} states in electrical circuits as 1 (true) and 0 (false), he showed that Boolean logic operations could be directly implemented using relays and switches. In other words, computation, once primarily understood as the manipulation of numbers and symbols to solve mathematical problems or to reason logically, is now seen as more physically and practically—linked to the manipulation of binary information in electrical circuits. This shift made computation synonymous with the automatic and efficient processing of data by digital computers. With this in mind, one might now ask a deeper question: what exactly is information?

To illustrate the modern concept of information, consider two different coins: one always lands heads, while the other lands randomly, sometimes heads, sometimes tails. Before flipping the first coin, we already know the result, so we learn nothing from flipping it. In contrast, before flipping the second coin, our uncertainty is at its maximum, and by observing the result, we gain new information. This example shows that information can be understood as both the uncertainty we have before observing an event and the knowledge we gain afterward. Information relates to uncertainty, and since uncertainty is linked to disorder, this establishes an initial connection between information and entropy, albeit in a rudimentary form. To formalise this intuition about information, in what follows we will adopt an axiomatic approach similar to that of Shannon in his seminal 1948 article~\cite{shannon1948mathematical}, which introduced a mathematical theory of communication. 

If a given event has a probability $p$, we expect that the amount of information contained in such an event should satisfy certain properties. First, it should be label-independent, meaning that the information content of an event is determined entirely by how likely or unlikely the event is, regardless of what the event is called or how it is labelled, as long as the mapping between the labels is a bijection. Thus, it depends only on the probability $p$ of the event. Second, the amount of information $i(p)$ for a given event should be greater the less likely the event is to occur -- less likely an event is to occur, the more information it provides when it happens. Mathematically, the amount of information $i(p)$ should be a continuous and monotonically decreasing function of $p$ as the probability of an event increases (making the event more likely), the information content decreases. This reflects the idea that more likely events are less surprising and, therefore, carry less information. Importantly, in the extreme cases where an event is completely certain (with probability $p=1$), the amount of information is zero because no surprise occurs when the event happens, and in the case of an impossible event (with probability $p=0$), the amount of information should be infinite, as you would be infinitely surprised to observe such an event. This can also be understood through a continuity argument, where as $p$ approaches zero, the information content grows without bound. Third, the information from two independent events with probabilities $p$ and $q$ should sum up to $ i(p) + i(q)$. This additivity captures the idea that when two independent events occur, the total information gained is simply the sum of the information from each event individually. Since the probability of two independent events occurring is the product of their individual probabilities, $pq$, the additivity condition implies that the information from such events should satisfy $i(pq) = i(p) + i(q)$. The only function that satisfies this property is the logarithm. This implies that, up to a constant factor $k$, the information content of an event with probability $p$ should be given by:
\begin{equation}\label{Eq:information-content}
    i(p) = -k \log p.
\end{equation}
The constant factor $k$ can be understood as a scaling factor that adjusts the units in which we measure information, often chosen to make the mathematical expressions more convenient, such as when using logarithms with base 2, 10 or $e$. Observe that the function in Eq.~\eqref{Eq:information-content} is the only one satisfying the three properties~\cite{Thomas2001}. The logarithm depends solely on the probability $p$ and is independent of any specific labels or characteristics of the event. It decreases as $p$ increases, and the factor $-k$ ensures that $i(p)$ remains positive and decreasing. Finally, the logarithmic function naturally exhibits additivity, which means that the information content from two independent events sums appropriately, making it the ideal candidate for quantifying information.

Consider a random variable $X$ described by a probability distribution $p(X = x) := p_X(x)$. The information content of a specific outcome $x$, given by $-k \log p_X(x)$, measures how much information is gained when that outcome occurs. To capture the overall uncertainty or the average information content associated with the entire distribution of $X$, we sum up this quantity across all possible outcomes, weighted by their probabilities. Setting $k=1$ in the expression for information content, we obtain the famous Shannon entropy (or information entropy):
\begin{equation}\label{Eq:shannon-entropy}
    S(X) := - \sum_{x} p_X(x) \log p_X(x).
\end{equation}
We slightly abuse the notation by representing thermodynamic, Shannon, and von Neumann entropies with the same letter. While this choice is justified in the case of Shannon and von Neumann entropies, as any probability distribution naturally corresponds to a diagonal quantum state, we also use it for the Gibbs entropy because, in the scenarios of interest, these notions coincide (see discussion below).

Even though they come from different contexts, the Shannon and von Neumann entropies share a striking similarity—they both measure uncertainty or ``missing information." In fact, von Neumann entropy can actually be thought of as a natural extension of Shannon entropy.  This becomes clear if we imagine a source that prepares messages composed of $n$ letters, each letter chosen from an ensemble of quantum states represented by $\rho$. The probability of any measurement outcome for a letter from this ensemble—assuming the observer has no knowledge of the specific letter prepared—can be fully characterised by the density operator $\rho$. When we select an orthonormal basis that diagonalises $\rho$, the vector of eigenvalues forms a probability distribution, and the von Neumann entropy then corresponds to the Shannon entropy of this distribution. However, it is important to emphasise that von Neumann entropy serves multiple roles. It not only quantifies the quantum (and classical) information content per letter of a pure state ensemble but also measures the entanglement of a bipartite pure state, among various other applications in quantum information theory~(for a detailed discussion, see the books by Bengtsson \& Życzkowski~\cite{BengtssonZyczkowski_2006}, Wilde~\cite{Wilde2013}, as well as the lecture notes by Preskill~\cite{preskill2015lecture}). 

Interestingly, as mentioned in Sec.~\ref{Sec:quantum-mechanics}, von Neumann entropy was introduced almost 20 years before Shannon’s. The funny thing is that Shannon was not sure what to call his function at first. He thought about calling it by either ``information" or ``uncertainty", but both terms were already overused in the literature. That is when von Neumann stepped in and gave him this advice~\cite{tribus1971energy}:

\vspace{0.2cm}
\begin{wrapfigure}{l}{0.07\textwidth}
     \includegraphics[width=0.06\textwidth]{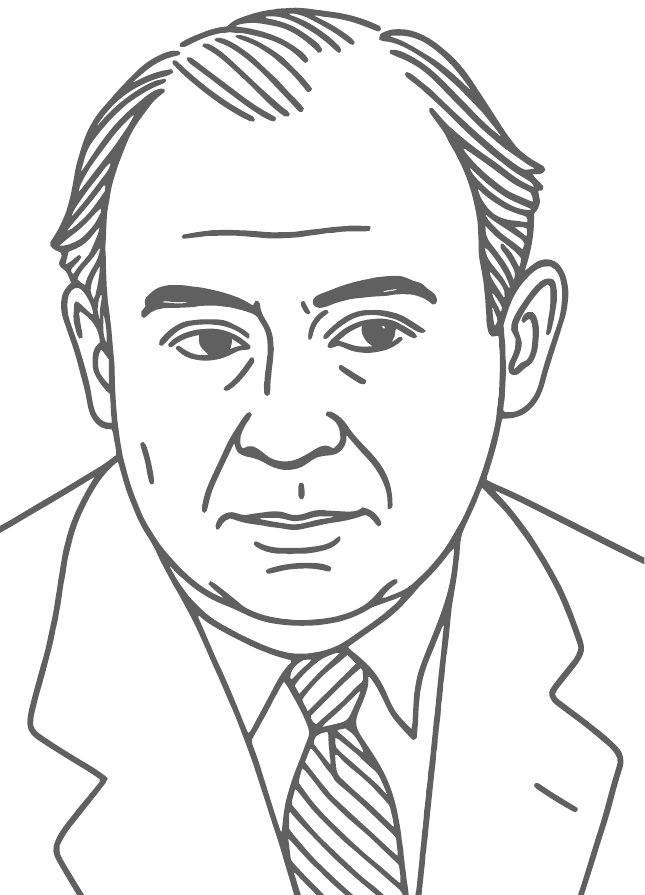}
 \end{wrapfigure}

\emph{``You should call it entropy, for two reasons. In the first place, your uncertainty function has been used in statistical mechanics under that name, so it already has a name. In the second place, and more important, no one really knows what entropy really is, so in a debate you will always have the advantage''.}

While this anecdote is often cited, its historical accuracy is anecdotal rather than rigorously documented. More importantly, the mathematical and conceptual connection between information-theoretic entropy and statistical mechanics (e.g., Gibbs entropy) plays a meaningful role in modern thermodynamics—a connection we discuss at the end of this section.

For continuous random variables, the sum in Eq.~\eqref{Eq:shannon-entropy} is replaced by an integral. It follows directly from Eq.~\eqref{Eq:shannon-entropy} that for an identically distributed binary variable, where $p(x=0)=p(x=1)=1/2$, $S(X)=1$ (using a base-2 logarithm), representing the unit of information known as a bit. Note that Eq.~\eqref{Eq:shannon-entropy} bears a mathematical resemblance to the entropy in the canonical ensemble, $S = k_B \log \Omega$, where $\Omega$ represents the number of possible microscopic arrangements of the atoms or molecules in a thermodynamic system. However, these two expressions are not merely similar; they coincide (up to a constant factor) within the assumptions of the microcanonical ensemble. Specifically, when there are $\Omega$ equiprobable microstates, the corresponding probability is $p_x = \frac{1}{\Omega}$, leading to the same form as Eq.~\eqref{Eq:shannon-entropy}. The derivation and interpretation of these two types of entropy arise from very different contexts. Recall from the previous section that Brillouin boldly postulated a direct connection between the two\footnote{Brillouin proposed that acquiring information reduces a system’s thermodynamic entropy, treating thermodynamic and information-theoretic entropy as equivalent and introducing the notion of negentropy~\cite{Rothstein1951,brillouin1953negentropy}. While this contributed to the formulation of the thermodynamics of information, it has been criticised for equating observer-dependent knowledge with objective thermodynamic quantities~\cite{jauch1972entropy}. Today, Brillouin’s view is used mainly as a heuristic, not as a general physical law.}.

\begin{figure}[t]
    \centering
    \includegraphics{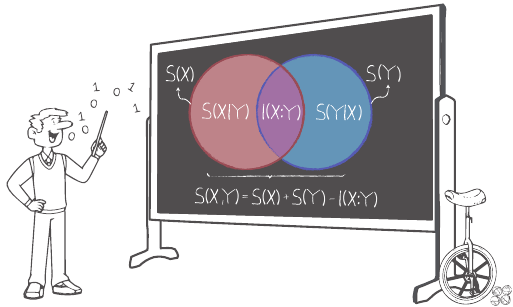}
    \caption{\emph{Information diagram}. Venn diagram illustrating the relationships among the joint entropy $S(X, Y)$, individual entropies $S(X)$ and $S(Y)$, conditional entropies $S(X | Y)$ and $S(Y | X)$, and mutual information $I(X: Y)$. Next to it, we see Shannon's unicycle and some of his juggling balls, as he was known for often doing both simultaneously.}
    \label{F-shannon}
\end{figure}

Reasonably, one can define a whole zoo of entropies based on the above formulation. For example, when we discuss information—particularly in the context of extracting it from a system—we are usually concerned with determining the state $X$ of the system by measuring a related quantity $Y$. The measurement of $Y$ yields information about $X$, thus reducing the uncertainty associated with $X$. The degree to which uncertainty is reduced represents the amount of information that $Y$ provides about $X$. Mathematically, this can be expressed in terms of the mutual information:
\begin{equation}\label{Eq:mutual-information}
I(X:Y) := S(X) - S(X|Y),
\end{equation}
where $S(X|Y)$ represents the uncertainty of the posterior probability distribution $p(x,y)$ averaged over the possible outcomes, defined as:
\begin{align}\label{Eq:posterior-prob}
    S(X \vert Y) :&= \sum_y p_Y(y)\qty[-\sum_x p_{X|Y}(x|y)\log p_{X|Y}(x|y)] \nonumber \\
    &= - \sum_{x,y} p_{XY}(x,y) \log p_{X|Y}(x\vert y).
\end{align}
Observe that we previously defined mutual information [Eq.~\eqref{Eq:mutual-information}] using the von Neumann entropy in Sec.~\ref{Sec:quantum-mechanics} when discussing correlations between two subsystems. Using Bayes’ formula,
\begin{equation}\label{Eq:Bayes}
    p_{XY}(x,y) = p_{X|Y}(x|y)p_Y(y),
\end{equation}
the mutual information can be rewritten as
\begin{equation}\label{Eq:mutual-information-2}
    I(X : Y) := - \sum_{x,y} p_{XY}(x,y) \log \qty[\frac{p_{XY}(x, y)}{p_X(x)p_Y(y)}].
\end{equation}
From this expression several important properties of mutual information can be derived. First, it is symmetric, meaning that the amount of information $X$ provides about $Y$ is identical to what $Y$ provides about $X$. Second, by applying the properties of logarithms, it can be shown from Eq.~\eqref{Eq:mutual-information-2} that $I(X )$ is always non-negative and becomes zero only when $X$ and $Y$ are statistically independent. This makes mutual information a reliable measure of the correlation between $X$ and $Y$. Third, if we measure a quantity $Y = f(X)$ perfectly, without errors, then $S(X|Y) = 0$ and $I(X ) = S(Y)$. In other words, the information gained from an error-free measurement equals the uncertainty of the measured outcome. Lastly, Eq.~\eqref{Eq:mutual-information-2} offers three different ways to represent mutual information:
\begin{align}
    I(X:Y) &= S(X) - S(X|Y) \\ &= S(Y) - S(Y|X) \\ &= S(X) + S(Y) - S(X,Y) \geq 0,
\end{align}
where $S(X,Y) := - \sum_{x,y} p_{XY}(x,y) \log p_{XY}(x,y)$ is the joint Shannon entropy. The last equality shows that correlations cause the entropy to be sub-additive, meaning $S(X,Y) = S(X) + S(Y) - I(X:Y)$. These relationships can be easily visualised using a Venn diagram, as shown in Shannon's blackboard in Fig.~\ref{F-shannon}. This expression will be useful in Sec.~\ref{Sec:exorcisim} when we explore the physical nature of Maxwell or Szilárd demons and interpret a measurement as the creation of correlations between the state of the demon and the state of the system.

As a final question in this section, one might ask whether there is a connection between Shannon, von Neumann, and thermodynamic entropy. The third example discussed in Sec.~\ref{Sec:thermodynamic-nutshell} shows that the von Neumann entropy (or Shannon entropy, when the density operator is expressed in the energy eigenbasis) coincides with the thermodynamic entropy when the probability distribution corresponds to the canonical ensemble. Note that this conclusion holds only if the system satisfies the equivalence of the ensembles~\cite{Touchette2015}. What about for general states? Attributing entropy to a nonequilibrium state has, in fact, been a prominent topic of research for
many years~\cite{Strasberg2021}. In general, identifying thermodynamic entropy with von Neumann entropy is incorrect—a point originally noted by von Neumann himself~\cite{Neumann1929}.

As shown in Sec.~\ref{Sec:quantum-mechanics}, a pure energy eigenstate of an isolated system has zero von Neumann entropy, whereas its equilibrium description—via a microcanonical or canonical ensemble—generally has nonzero thermodynamic entropy. A qualitatively different example arises in strongly coupled or non-Markovian setups, where the system and environment cannot be cleanly separated into distinct energy contributions. In such regimes, even if the global state has well-defined thermodynamic properties, the von Neumann entropy of the subsystem may not coincide with its thermodynamic entropy\cite{Strasberg2016,Perarnau2018}. Further discrepancies appear in bipartite entangled systems: although the global state may be pure (with zero von Neumann entropy), each subsystem can exhibit nonzero von Neumann entropy. In such cases, the thermodynamic interpretation is ambiguous without an explicit equilibrium framework. These diverse scenarios highlight that, outside of the weak-coupling, equilibrium regime, von Neumann entropy generally does not coincide with thermodynamic entropy.

However, in both stochastic~\cite{seifert2008stochastic,Sekimoto2010,VandenBroeck2015} and quantum thermodynamics~\cite{Kosloff2013,Skrzypczyk2014,Vinjanampathy2016} the von Neumann (and Shannon) entropy has shown a clear physical meaning in specific contexts: it governs the energetics of nonequilibrium processes for systems weakly coupled to one or more thermodynamic reservoirs. If we now wonder whether there is an entropic function capable of describing the thermodynamic entropy, it has been discussed that the observational entropy is the most appropriate candidate as it unifies the Gibbs-Shannon-von Neumann entropy and the Boltzmann one~\cite{Anthony2019,Dominik2019,Dominik2020,Faiez2020,Schindler2020}. Observational entropy is defined with respect to a coarse-graining of the system’s Hilbert space into a set of orthogonal projectors $\mathcal{X} = {\mathbbm{P}_x}$. Given a quantum state $\rho$, it is expressed as
\begin{equation}
    S^{\mathcal{X}}_{\text{obs}} = \sum_x p_x(-\log p_x+\mathcal{V}_x)
\end{equation}
where $p_x = \mathrm{tr}(\mathbbm{P}_x \rho)$ is the probability of obtaining outcome $x$, and $\mathcal{V}_x = \mathrm{tr}(\mathbbm{P}_x)$ quantifies the volume of the subspace associated with $\mathbbm{P}_x$. This expression captures both how the state is distributed across different macrostates and how large each macrostate is. By choosing the coarse-graining appropriately, observational entropy can reproduce the von Neumann entropy (when the coarse-graining is very fine) or the Boltzmann entropy (when the state is localised in a single macrostate). Unlike the von Neumann entropy, it can increase under unitary evolution when the coarse-graining is fixed, making it especially suitable for describing entropy production in isolated systems. For this reason, it has been proposed as a generalised definition of thermodynamic entropy, applicable even far from equilibrium. For a detailed discussion of observational entropy, we suggest the tutorial by Strasberg \& Winter~\cite{Strasberg2021}, which provides an accessible introduction and reviews how various entropy definitions behave in quantum thermodynamics. We also point to Ref.~\cite{schindler2025unificationobservationalentropymaximum}, which very recently introduced a general framework unifying observational entropy with maximum entropy principles, and showed how it captures entropy production across a broad range of scenarios.

\subsection{Thermodynamics of information \label{Sec:thermodynamics-information}}

Let us now combine elements of information theory, such as the notions of Shannon and von Neumann entropy, with thermodynamic concepts. Our main motivation for doing so comes from the fact that Maxwell's work revealed an interplay between thermodynamic entropy and information. However, traditional formulations of the second law, such as those of Clausius and Kelvin, do not address information. Here, we first provide elements for explicitly incorporating information into thermodynamics and then investigate the possible thermodynamic costs associated with manipulating information, including processes such as measurement, erasure, copying, and feedback. The following discussion is focused on our goal of resolving Maxwell's demon. For a broader discussion of the thermodynamics of information, we recommend the following readings~\cite{Sagawa2008, Sagawa2009,Sagawa2010,Parrondo2015,minagawa2023universalvaliditysecondlaw}.

In Sec.~\ref{Sec:quantum-mechanics}, we extended the concept of free energy beyond equilibrium states [Eq.~\eqref{Eq:noneq-F}]. Specifically, for a system described by a Hamiltonian $H$ and prepared in a state $\rho$ in the presence of a heat bath at an inverse temperature $\beta$, this quantity is defined as~\cite{Esposito2010}:
\begin{equation}\label{Eq:nonequilibrium-free}
    \mathcal{F}(\rho,H) := \langle H \rangle_{\rho}-\frac{1}{\beta}S(\rho),
\end{equation}
where the first term $\langle H \rangle_{\rho}$ denotes the average energy and $S(\rho)$ the Shannon or von Neumann entropy, depending on the context. For classical systems, where $x$ labels the microstate and $\rho(x, t)$ is the probability density over phase space, the average energy becomes $\langle H \rangle_{\rho} = \int \mathrm{d} x\, \rho(x, t) H(x)$ and $S$ corresponds to the Shannon entropy. In contrast, for quantum systems, $\rho$ is a density matrix and the expressions reduce to $\langle H \rangle_{\rho} = \tr(\rho H)$ and $S(\rho)$ becomes the von Neumann entropy.

As we previously discussed, the nonequilibrium free energy provides a bound for the minimum average work required to isothermally drive the system from one arbitrary state to another (or the maximum average work that can be extracted from the system when it is in an out-of-equilibrium state):
\begin{equation}\label{Eq:free-energy-bound-2}
    W \geq \Delta \mathcal{F}.
\end{equation}

\begin{figure}[t]
    \centering
    \includegraphics{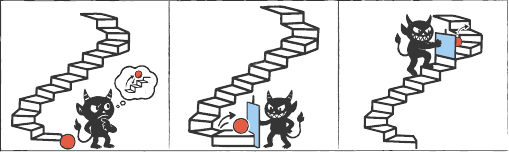}
    \caption{\emph{Information-driven motion}. A microscopic particle moves within a spiral-staircase potential, where thermal fluctuations cause it to randomly jump between steps. By measuring the particle’s position at regular intervals, we can determine when it jumps upwards. This allows us to implement a feedback mechanism by placing a block behind the particle, preventing subsequent downward jumps. Repeating this process allows the particle to move up the staircase. Ideally, the energy required to position the block is negligible, meaning the particle’s motion is driven purely by the information obtained from measuring its position. Figure inspired by Ref.~\cite{Toyabe2010}.}
    \label{F-staircase}
\end{figure}

To give a gist of what is to come, we consider a motivating example, which was experimentally realised in Ref~\cite{Toyabe2010}: a microscopic particle on a potential shaped like a spiral staircase (see Fig.~\ref{F-staircase}). The height of each step is comparable to $k_B T$. The particle, driven by thermal fluctuations, jumps between steps—sometimes moving up, sometimes down. Now, consider the following feedback control: the particle’s position is measured, and if an upward jump is observed, a thin partition is placed behind it to prevent a downward jump. This partition can be installed at negligible energy cost. By repeating this control at every jump, the particle is expected to keep climbing up the stairs.

Again, at first glance, it seems as though the particle is gaining free energy from nowhere, apparently violating the second law of thermodynamics, much like Maxwell’s demon or Smoluchowski/Feynman’s ratchet. However, the particle’s movement is actually driven by the information obtained from measuring its position. Here, the feedback control rectifies thermal fluctuations. The energy gained from the information is balanced by the energy cost to the demon for manipulating this information. When we consider the total system, including both the particle and the demon, the second law holds. In this scenario, the demon simply consists of macroscopic devices, like our computers.

We now introduce the main tools for modelling thermodynamic processes when information is involved.

\begin{coolexample}[Feedback process]

Consider a protocol in which a system, interacting with a heat bath at inverse temperature $\beta$, is measured, and the obtained information is then used to perform an isothermal process. To determine the thermodynamic cost of acquiring this information, we assume a classical system with a continuous random variable $x$ representing its microstate. Initially, the system is described by a Hamiltonian $H(x, t)$ and a probability density $\rho_X(x, t)$. At the very start, before any measurement or external interaction, the system's free energy is $\mathcal{F}_{\text{in}}$. At time $t = \tau$, an agent measures a quantity $M$ and gets a discrete outcome $m$. Just right before the measurement, its free energy, denoted by \mbox{$\mathcal{F}_{\text{pre}} := \mathcal{F}[\rho_X(x,\tau),H(x,\tau)]$}, is given by
\begin{equation}\label{Eq:free-pre}
    \mathcal{F}_{\text{pre}}= \langle H(x,\tau)\rangle_{\rho_X}-\frac{1}{\beta}S(X),
\end{equation}
where $\langle H(x,\tau)\rangle_{\rho_X} = \int \textrm{d}x \: H(X) \rho_{X}(x,\tau)$. Typically, we distinguish between the initial and pre-measurement free energies, as the system might not remain static; it could undergo controlled or natural dynamics leading up to the measurement point, thereby changing its state and, consequently, its free energy. The system's state after the measurement is then updated using Bayes' rule, resulting in:
\begin{equation}
    \rho_{X|M}(x|m) = \frac{p_{M|X}(m|x) \rho_X(x,\tau)}{p_M(m)}. 
\end{equation}
Since we are in a classical setting, the measurement does not disturb the system. This is captured by demanding that the system's state remains unchanged after the measurement 
\begin{equation}
    \sum_m p_M(x) \rho_{X|M}(x|m) = \rho_X(x,\tau).
\end{equation}
Using Eq.~\eqref{Eq:nonequilibrium-free}, we can write the free energy of the system after the measurement for a given outcome $m$ as
\begin{equation}
    \mathcal{F}[\rho_{X|M}(x|m),H(x,\tau)]= \langle H(x,\tau)\rangle_{\rho_{X|M}}-\frac{1}{\beta}S[\rho_{X|M}(x|m)],
\end{equation}
where $\langle H(x,\tau)\rangle_{\rho_{X|M}} = \int \textrm{d}x \: H(x,\tau) \rho_{X|M}(x|m)$. If we take the average over all possible outcomes, we obtain the nonequilibrium free energy of the system after the measurement
\begin{align}\label{Eq:free-pos}
    \mathcal{F}_{\text{pos}}:&=\sum_m p_M(m) \mathcal{F}[\rho_{X|M}(x|m), H(x,\tau)] \nonumber \\ &= \int \textrm{d}x \: H(x,\tau) \rho(x,\tau) - \sum_p p_M(m) S[\rho_{X|M}(x|m)] \nonumber \\
    &\overset{\eqref{Eq:posterior-prob}}{=}\langle H(x,\tau)\rangle_{\rho_X} -\frac{1}{\beta}S(X|M).
\end{align}
Thus, the free-energy difference between after and before the measurement is determined using Eqs.~\eqref{Eq:free-pos}-\eqref{Eq:free-pre}:
\begin{align}\label{Eq:W-meas}
    \Delta \mathcal{F}_{\text{meas}}:&= \mathcal{F}_{\text{pos}} -\mathcal{F}_{\text{pre}} = \frac{1}{\beta}[S(X)-S(X|M)]\nonumber\\& = TI(X:M).
\end{align}
Since mutual information is non-negative, measurement (or information acquisition) always increases the free energy, thus raising the amount of work that can be extracted isothermally. After the measurement, the agent gains information and uses it to perform the subsequent steps, a process we refer to as a \emph{feedback process}. Consequently, we define the system's final free energy, $\mathcal{F}_{\text{fin}}$, after the feedback control has been applied based on the measurement outcome. Similarly to before, $\mathcal{F}_{\text{fin}}$ may differ from $\mathcal{F}_{\text{post}}$ because the feedback operation can either perform work on the system or extract work from it, depending on the protocol. 

Finally, we can bound the work associated with the feedback process using the free energy bound given by Eq.~\eqref{Eq:free-energy-bound-2}. The total work done throughout the entire feedback process includes both the stages before and after the measurement. Consequently, we can break down the process into two subprocesses: from the initial to the pre-measurement state, $\mathcal{F}_{\text{init}} \to \mathcal{F}_{\text{pre}}$, and from the post-measurement to the final state, $\mathcal{F}_{\text{post}} \to \mathcal{F}_{\text{fin}}$. By applying Eq.~\eqref{Eq:free-energy-bound-2} to each subprocess, we obtain:
\begin{equation}\label{Eq:W-fb}
    W_{\text{fb}} \geq (\mathcal{F}_{\text{fin}}-\mathcal{F}_{\text{post}}) +(\mathcal{F}_{\text{pre}} - \mathcal{F}_{\text{init}}) = \Delta \mathcal{F}-\Delta \mathcal{F}_{\text{meas}}
\end{equation}
with $\Delta \mathcal{F} := \mathcal{F}_{\text{fin}} - \mathcal{F}_{\text{init}}$ being the difference between the final and initial free energy states immediately after and before the measurement, respectively. Combining Eqs.~\eqref{Eq:W-fb}~and~\eqref{Eq:W-meas}, the work performed in the feedback process can be written in terms of mutual information:
\begin{equation}\label{Eq:second-law-feedback}
    W_{\text{fb}} \geq \Delta \mathcal{F} - \frac{1}{\beta} I(X:M).
\end{equation}
\end{coolexample}

The above equation is known as the second law for feedback processes~\cite{Sagawa2008, Sagawa2012} and was experimentally verified in the motivating example discussed above~\cite{Toyabe2010}. It is fascinating to see how mutual information appears in the expression for feedback work, capturing the correlations established between the system and the measurement process, which lead to an increase in free energy. In particular, the examination of the thermodynamic costs inherent in the quantum acquisition of ``knowledge'' is a much studied field~\cite{Sagawa2008,Sagawa2009,Sagawa2012,Micadei2013,Guryanova2020idealprojective,Danageozian2022,minagawa2023universalvaliditysecondlaw,jake2024knowlegde}.

As a last observation, note that Szilard's engine can be cast in terms of the ideas discussed above. But Eq.~\eqref{Eq:second-law-feedback} does not resolve the paradox  because it constrains only the feedback stroke and therefore leaves out the energetic cost of creating the correlation $I(X:M)$ during the measurement and of resetting the demon’s memory; when those contributions are added [Sec. \ref{Subsec:TIA}], the total work over a full cycle obeys $W_{\text{tot}}\ge 0$, reinstating the second law. While this resolution is consistent with the widely adopted framework based on memory erasure, it is not the only possible perspective. Some researchers argue that erasure is not a fundamental requirement for restoring the second law. Instead, they advocate for an explicit accounting of the work and free energy involved in generating and using correlations. In this alternative approach, resetting the demon’s memory to a reference state is viewed as a conventional step, not a thermodynamic necessity. This viewpoint is articulated in Ref.~\cite{Ouldridge2019}, which presents a detailed analysis of a molecular Szilard engine and argues that the apparent paradox arises from implicit assumptions rather than any deep physical necessity.

\vspace{1cm}
\begin{mybox}{Photonic Maxwell's demon \& feedback control~\cite{Vidrighin2016}}{Photonic Maxwell's demon \& feedback control}

This experiment realises a photonic Maxwell's demon, demonstrating that the amount of work extracted is fundamentally bounded by the information acquired through measurement, namely $|W| \propto \sqrt{I}$, with $I$ being the single-measurement mutual information. This sublinear scaling reflects practical limitations in work extraction, while remaining consistent with the broader thermodynamic bound given by Eq.~\eqref{Eq:second-law-feedback}, which governs idealised, fluctuation-free systems.
\vspace{0.2cm}

The experiment starts with a light mode prepared in a thermal state. Measurement is performed using a high-transmittance beam splitter (BS) and avalanche photodiodes (APDs): thermal light passes through the BS, where a small fraction is reflected to a highly sensitive APD. The APD provides a binary outcome—either a click (photon detected) or no-click (no photon detected). Based on the detection outcomes, the energy in each mode is inferred, and the modes are ``labelled'' as more or less energetic. A conditional operation (feedback) is then applied, directing these modes to two photodiodes with opposite polarities to create an energy imbalance. Both photodiodes are connected to a capacitor, which stores the extracted work by charging through this controlled energy gradient.
\vspace{0.2cm}

The two modes are then directed to photodiodes connected to a capacitor. When the feedback is applied, the capacitor charges due to the controlled imbalance. The experiment derives the following bound:
\begin{equation*}
    \frac{|\langle W\rangle |}{\sigma(W)_0} \leq \sqrt{2\langle I \rangle},
\end{equation*}
where $\sigma(W)_0$ quantifies fluctuations in the absence of feedback. This shows that the signal-to-noise ratio of the extracted work scales with $\sqrt{I}$.
\end{mybox}

\subsection{Computation in the smallest nutshell \label{Subsection:computation-nutshel}}

Shannon's theory provided the mathematical and conceptual foundation for understanding and optimising how information is represented, transmitted, stored, and processed. The backbone of computing is the execution of logical operations for a given task~\cite{Turing2004}—something Shannon showed could be achieved using relays and switches. However, this architecture was slow, bulky, and prone to mechanical failure due to its moving parts. Relays consist of a coil that generates a magnetic field to open or close a mechanical contact, and this movement takes time—typically measured in milliseconds. It was soon realised that relays could be replaced by vacuum tubes, which control the flow of electrons through a vacuum using an electric field, without the need for moving mechanical components. This allowed vacuum tubes to switch states (\textbf{ON}/\textbf{OFF}) much faster than relays. In fact, they operated on the order of microseconds, as their switching time was limited only by how quickly the electric field could influence electron flow, far faster than the mechanical movement of relay parts. Beyond speed, vacuum tubes offered other advantages over relays. Relays were not suitable for signal amplification or processing analog signals. The inertia of their moving parts meant that repeatedly switching them on and off (as required for fast computations) was mechanically taxing and slow. Conversely, vacuum tubes could switch rapidly between states and also amplify signals and process analog data, making them ideal for both switching circuits and more complex tasks like signal amplification and processing. However, not all challenges were resolved. The vacuum tubes generated significant heat and required large amounts of power to operate. Because they worked by heating a filament to produce electrons, this process consumed a great deal of energy and produced excessive heat.

An important element in discussing early computation is the notion of manual programming and operation. Early computers, such as those using relays or vacuum tubes, required manual programming and configuration. Their programs were essentially hardwired into the machine until someone physically reconfigured the hardware. Switching tasks could take hours or even days, as programming involved more than just writing code—it meant altering the machine's setup itself. Once configured, the machine could automatically run tasks, but the manual setup process severely limited flexibility and made it difficult to change programs quickly. This paradigm changed with von Neumann’s 1945 report, ``First draft of a report on the EDVAC" (Electronic Discrete Variable Automatic Computer)~\cite{von1993first}. In this document, von Neumann introduced the groundbreaking concept of \emph{stored-programme architecture}.

The central feature of von Neumann's architecture is that both the instructions (the program) and the data are stored in the same memory. Instructions are executed sequentially, one after the other, unless the programme explicitly instructs the computer to jump to a different part of the code. This design allows computers to switch between programmes without the need for manual reconfiguration. Importantly, the memory in this system is a physical component, such as vacuum tubes, that stores both data and programme instructions, enabling the computer to dynamically access and modify the program as needed.

Computing was in its infancy, and the main problem at the time was finding ways to make computers process data and perform operations as quickly as possible. Naively speaking, this depended on improvements in physical systems, as these are the building blocks for switching between states, from 0 to 1 and vice versa. While early engineers and scientists understood that faster computation generated more heat and that physical systems had limitations, the deeper connection between information theory and thermodynamics had not yet been fully realised.

What is the real connection between information and computation, beyond the fact that both deal with the same 0’s and 1’s that Shannon’s information theory uses and that drive the operations inside a computer? Shannon’s theory was all about measuring and transmitting information, while computation is focused on processing data using logic and instructions. As we have seen, both processing and transmitting information happens through physical systems, and they can roughly be viewed as computation tasks. Whether a computation is fast or slow depends on the physical system being used. But is computation just a mathematical abstraction, or is there a deeper, intrinsic connection between computation and information itself?

\section{Thermodynamics of computation \label{Sec:thermodynamics-of-computation}}

What is a computer if not an engine that converts free energy into waste heat and mathematical work~\cite{bennett1982thermodynamics}? By the 1950s, this idea was taking shape, particularly with von Neumann's remark during a 1949 lecture~\cite{von1966theory}, where he stated that a computer operating at temperature $T$ must dissipate at least $k_B T \log 2$ of energy per elementary act of information. In this section, we discuss the fundamental thermodynamic cost of computation, specifically a lower bound of order $k_B T$ for certain data operations.

\subsection{Landauer's principle \label{Subsec:Landauer}}

\begin{wrapfigure}{l}{0.1\textwidth}
     \includegraphics[width=0.08\textwidth]{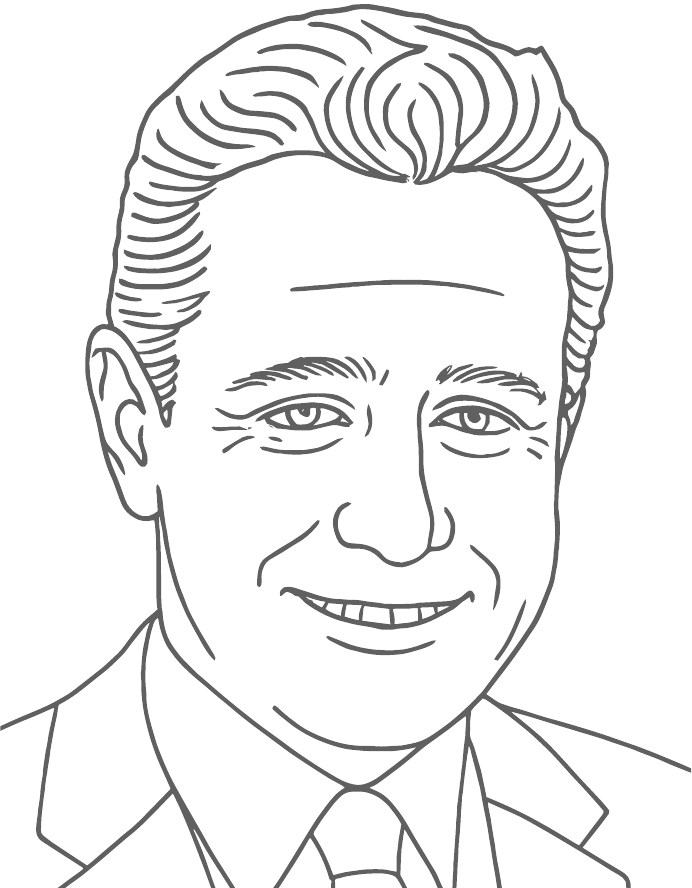}
\end{wrapfigure} 
\emph{Information is physical!~\cite{landauer1991information}}

\noindent Despite being a short and self-explanatory statement, the above quote highlights the fact that information is stored in physical systems, such as books, hard disks, or even colourful stickers. The same systems are also used to transmit and process information. Whether through electrical or optical signals, they are inevitably bound by the laws of physics. Formally, one might ask an equivalent question: what is the interplay between performing a logical operation and its associated thermodynamic cost? This question is reinforced by the fact that information processing occurs in a physical system, and thermodynamic processes—especially beyond the thermodynamic limit (whether due to finite-time processes or finite systems)—involve energy dissipation. In his famous 1961 paper~\cite{landauer1961irreversibility}, Rolf Landauer relates entropy decrease and heat dissipation during logically irreversible processes, i.e., a minimum cost that must be paid to erase information.

\begin{figure}[t]
    \centering
    \includegraphics{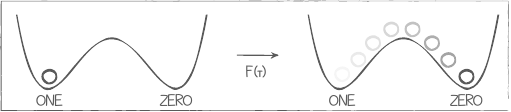}
    \caption{\emph{Restoring to \texttt{ZERO}}. A bistable potential with a particle can be recast as an information-restoring process by assuming the particle represents a bit, with its two states corresponding to the left (\texttt{ONE}) and right (\texttt{ZERO}) positions. The task consists of moving the particle from left to right using a conservative force $F(t)$ regardless its initial position. However, such a task is not possible as this  process cannot be reversed.}
    \label{F-bistable}
\end{figure}

As a warm-up, one can revisit the motivating example discussed by Landauer, where the task consists of restoring a bit to a given state. We can envision a binary device as a particle in a bistable potential (see Fig.~\ref{F-bistable}) and define the operation of restoring as the process that moves the particle to one of the two sites of the potential. For convenience, we assume that the left site corresponds to the state \texttt{ZERO}, while the right site represents the state \texttt{ONE}. Thus, the task is to move the particle to the state \texttt{ZERO}. Observe that if we know the particle's position, the task becomes simple. If the particle is already in the desired state, we do not need to do anything, and no energy is expended. However, if we know that the particle is in the \texttt{ONE} state, we can apply a force to push it over the barrier, and once it passes the maximum, we can apply a retarding force to slow it down. This ensures that when the particle reaches the \texttt{ZERO} state, it has no excess kinetic energy and no energy is expended throughout the process. Although it might be seen that restoring the particle's position to \texttt{ONE} can be done without expending any energy, we have actually used two different protocols depending on the particle's initial state. This leads us to ask whether a general procedure exists that can always perform the restoring action, regardless of the particle's state, without any energy cost. 

In Landauer's motivating example, this question is translated into asking whether it is possible to construct a time-varying force capable of moving the particle to the \texttt{ONE} state, regardless of its initial position. However, since we are dealing with a conservative system, we can easily conclude that this is impossible. This becomes clear by imagining the reversed scenario. When time is reversed, the same force should work backward. This means that starting from the \texttt{ONE} state, the particle would retrace its steps. Since it could originally have come from either the \texttt{ZERO} or the \texttt{ONE} state, in the reversed situation, it could potentially end up in either state. This creates a paradox: starting from the \texttt{ONE} state, the particle would need to end up in two different places. However, according to the laws of classical mechanics, which are deterministic, a particle starting from a specific position and velocity can have only one future outcome, not two. In conclusion, the seemingly simple task of restoring a particle's position inevitably results in energy dissipation. 

The motivating example can be generalised to a broader range of logical operations that have similar consequences—energy dissipation. This leads us to categorise two types of computation. A task is called \emph{logically irreversible} if its output does not uniquely determine its inputs. In Sec.~\hyperref[subsec:reversible-computation]{\ref{subsec:reversible-computation}}, we will discuss that any logical operation that does not preserve enough information about its inputs (i.e., it is logically irreversible) implies a thermodynamic cost, such as energy dissipation. Conversely, when we can reverse the computation—such that every output can be traced back to a unique input—we call the task (or computation) reversible. The first two examples shown in Fig.~\ref{F-shannon-scheme}, AND and OR, are cases where it is impossible to uniquely determine the inputs from the outputs, making them irreversible functions. In contrast, in the last example, XOR outputs 1 if exactly one of the inputs is 1. If we know one of the inputs, we can reverse the process and determine the other input from the output, making this function reversible.

To clarify the connection between logical irreversibility and entropy, Landauer revisits the example of the reset operation by imagining a collection of bits to be reset. This can be likened to a physical system where each bit is analogous to a spin in an ensemble, and the task is to align all the spins in the same direction. If the spins begin in thermal equilibrium (randomly distributed between 0 and 1) relative to a thermal reservoir at temperature $T$, resetting all the bits to \texttt{ZERO} effectively reduces the number of possible configurations the system can occupy. According to Shannon entropy~[Eq.~\eqref{Eq:shannon-entropy}], a two-state system containing a bit of information has entropy $\log 2$. After reset, the entropy is zero. Thus, this reduction corresponds to a decrease in entropy of order $k_B \log 2$ per bit. Since the total entropy of a closed system cannot decrease, this loss of entropy must be compensated elsewhere, appearing as heat dissipated into the environment. Using the Clausius inequality to express the heat flow into the reservoir, we find that the minimum heat generated during the reset process is at least 
\begin{equation}\label{Eq:Landauers-principle}
    \beta\Delta Q_{\ms E} \geq -\Delta S_{\ms S}
\end{equation}
where in the present example $\Delta S_{\ms S} = \log 2$. This value represents the lower bound for energy dissipation required for the reset operation under these conditions. It is important to note that the inequality in Eq.~\eqref{Eq:Landauers-principle} is saturated in the ideal isothermal case, which assumes an infinite time span for the process to be realised. While isothermality is an assumption needed to reach the bound, it is not what makes it unattainable. The bound becomes practically unattainable due to the requirement for an infinite duration. However, we can get remarkably close to this bound, as demonstrated in the first experimental realisation in Sec.~\ref{Subsec:Landauer-experimental}. Later, in Sec.~\ref{Landauer-principle-finite-time}, we analyse the finite-time regime and explore possible corrections to this bound.
\vspace{1cm}
\begin{mybox}{Landauer erasure with a molecular nanomagnet~\cite{gaudenzi2018quantum}}{Landauer erasure with a molecular nanomagnet}

This experiment realises a Landauer erasure process in the quantum realm.

\vspace{0.2cm}

A crystal of molecular magnets ($\text{Fe}_8$) is used as a quantum memory, where the collective spin of each $\text{Fe}_8$ molecule is $S = 10$. The spin can align in one of two main orientations, $S_z = \pm 10$, corresponding to the classical bit states \texttt{ZERO} and \texttt{ONE}, making each $\text{Fe}_8$ molecule a qubit. The system is described by the Hamiltonian:
\begin{equation}
    H = -D S^2_z + E(S^2_x-S^2_y)-g\mu_B \, \textbf{S}.\textbf{B}
\end{equation}
where $D$ is the anisotropy, $g$ is the Landé g-factor, and $\mu_B$ is the Bohr magneton. The first term, $D S_z^2 + E(S_x^2 + S_y^2)$, defines the energy landscape, creating a barrier that stabilises the spin in the $S_z = \pm 10$ orientations. The third term provides external control over this landscape through the magnetic field $\mathbf{B}$.

\vspace{0.2cm}

The protocol involves three steps: \emph{(i)} A magnetic field along the $y$-axis lowers the energy barrier, allowing quantum tunnelling between $S_z = +10$ and $S_z = -10$, erasing the bit by making the spin delocalised between \texttt{ZERO} and \texttt{ONE}. \emph{(ii)} A magnetic field along the $z$-axis biases the system, resetting all spins to the $S_z = +10$ (\texttt{ZERO}) state. \emph{(iii)} Finally, the fields are removed, locking the spins in the $S_z = +10$ state, completing bit storage.

\end{mybox}

Therefore, logically irreversible operations, such as resetting a bit, are fundamentally tied to physical processes that result in energy dissipation. Specifically, these operations must follow thermodynamic laws, meaning that every time information is erased, a minimum amount of heat must be generated and released into the environment. This notion is now known as Landauer's principle.

\subsection{How general is Landauer's principle? \label{Subsec:Landauer-proof}}

To explain and discuss Landauer's principle, we used two specific models: the bistable potential and the spin system, each under certain assumptions that are subject to debate. A natural question arises: How general is this principle, and what are the minimal assumptions necessary to satisfy Eq.~\eqref{Eq:Landauers-principle}? This is particularly important because there have been both theoretical~\cite{Allahverdyan2001, norton2011waiting, alicki2012quantum} and experimental~\cite{orlov2012experimental} reports claiming situations where Landauer's principle is apparently violated. These claims of potential violations seem to stem from the lack of a general statement formally written or rigorously proven. To address this, we revisit the work of Reeb and Wolf~\cite{Reeb2014}, where Landauer's principle is derived and refined using a general and minimal framework. Rather than repeating all the detailed arguments and reasoning from Ref.~\cite{Reeb2014}, we will present a simplified but consistent version of their proof.

We consider a minimal setup consisting of a system and a heat bath. The three main assumptions required to derive Landauer's principle are as follows. First, while the state of the system can be arbitrary, the heat bath initially starts in a thermal (Gibbs) state at inverse temperature $\beta$:
\begin{equation}\label{Eq:thermal-state}
    \gamma_{\ms{B}} = \frac{e^{-\beta H_{\ms B}}}{\tr(e^{-\beta H_{\ms B}})},
\end{equation}
where $\gamma_{\ms{B}}$ denotes the state of the heat bath, and $H_{\ms B}$ represents its Hamiltonian, about which we make no specific assumptions. Second, we assume that the system and the bath are initially uncorrelated
\begin{equation}\label{Eq:product-state}
    \rho_{\ms{SB}} = \rho_{\ms S}\otimes \gamma_{\ms B}.
\end{equation}
Although the product state assumption from Eq.~\eqref{Eq:product-state} is standard in both thermodynamics and quantum mechanics, we emphasise its crucial role in ensuring the validity of Landauer's principle. Reported violations of Landauer's bound~\cite{Allahverdyan2001,orlov2012experimental} can be explained by recognising that the product state assumption is not taken into account. Finally, the last assumption is that the process undergone by the system and heat bath is governed by a unitary evolution:
\begin{equation}
    \sigma_{\ms{SB}}:=U(\rho_{\ms S}\otimes \gamma_{\ms B})U^{\dagger}.
\end{equation}
Importantly, the unitary assumption implies that all elements and resources are accounted for, meaning that no unspecified environment $\ms{B}$ can participate in the process or contribute to entropy.

These three assumptions are all we need to arrive at Eq.~\eqref{Eq:Landauers-principle}. Notice that while we state Landauer's principle as a consequence of the information erasure process, we have not imposed any specific constraints on our protocol (the unitary evolution) to enforce this process. This is because our goal is to formulate the problem in a way that applies to any process governed by unitary evolution, not just erasure. Erasure is simply a special case that can be related to the result we will present.

Let us begin by deriving an important relation between the initial and final entropies after the process, while introducing an information-theoretic quantity. This relation can be obtained by leveraging the properties of von Neumann entropy: additivity under the tensor product, invariance under unitary evolution, and finally, subadditivity. More precisely:
\begin{align}\label{Eq:entropic-relation}
    S(\rho_{\ms S}) + S(\gamma_{\ms B}) \overset{\eqref{Eq:entropy-additive}}{=} S(\rho_{\ms S}\otimes \gamma_{\ms B}) &\overset{\eqref{Eq:entropy-invariance}}{=} S[U(\rho_{\ms S}\otimes \gamma_{\ms B})U^{\dagger}] \nonumber\\&\overset{\eqref{Eq:entropy-subadditive}}{\leq} S(\sigma_{\ms S})+S(\sigma_{\ms B}).
\end{align}
Now, by manipulating Eq.~\eqref{Eq:entropic-relation}, we can introduce and express the following quantity:
\begin{align}
    S(\sigma_{\ms S})+S(\sigma_{\ms B})-S(\rho_{\ms S}\otimes \gamma_{\ms B}) &\overset{\eqref{Eq:entropy-invariance}}{=}  S(\sigma_{\ms S})+S(\rho_{\ms S}) - S(\sigma_{\ms{SB}})\nonumber \\&\overset{\eqref{Eq:mutual-information-quantum}}{=} I(\ms{S}:\ms{B})_{\sigma_{\ms{SB}}} \geq 0.
\end{align}
Finally, the above equation can be further expressed as 
\begin{equation}\label{Eq:entropy-relation}
   -\Delta S_{\ms{S}}  +I(\ms{S}:\ms{B})_{\sigma_{\ms{SB}}} \geq 0 = \Delta S_{\ms B},
\end{equation}
where $\Delta S_{\ms X}:= S(\sigma_{\ms X})-S(\rho_{\ms X})$ for $\ms{X} \in \{\ms{S}, \ms{B}\}$.
Up to this point, we have merely applied the properties of von Neumann entropy to derive a relation connecting the initial and final entropies of the system and bath with the mutual information [Eq.~\eqref{Eq:entropy-relation}]. We will now use this relation to prove the Landauer's principle in its equality form, which reduces to the well-known Landauer bound when two negative terms are discarded. 

Let us focus on the right-hand side of Eq.~\eqref{Eq:entropy-relation} and use the fact that one can explicitly write the term $\Delta S_{\ms B}$ by using the assumption that the bath is initially prepared in a thermal state as given in Eq.~\eqref{Eq:thermal-state}. This means that one can write the entropy of the initial state as in Eq.~\eqref{Eq:entropy-thermal-state}: $S(\gamma_{\ms B}) = \beta \tr(\gamma_{\ms B} H_{\ms B}) +\tr(e^{-\beta H_{\ms B}})$. Note that if we add and subtract the average energy of the final state, we can write the entropy difference in terms of thermodynamic and informational quantities:
\begin{align}\label{Eq:entropy-bath}
    \Delta S_{\ms B} &= S(\sigma_{\ms B}) - \beta \tr(\gamma_{\ms B} H_{\ms B}) \!-\!\tr(e^{-\beta H_{\ms B}})+\tr(\sigma_{\ms B}H_{\ms B})\!-\!\tr(\sigma_{\ms B}H_{\ms B}) \nonumber \\
    &=S(\sigma_{\ms B})- \tr{\sigma_{\ms B}\log\qty[\frac{e^{-\beta H_{\ms B}}}{\tr(e^{-\beta H_{\ms B}})}]}+ \beta\tr[H_{\ms B}(\sigma_{\ms B}-\gamma_{\ms B})]\nonumber\\
    &= S(\sigma_{\ms B})-\tr(\sigma_{\ms B}\log\gamma_{\ms B})+\beta \Delta Q \nonumber \\
    &\overset{\eqref{Eq:relative-entropy}}{=} -S(\sigma_{\ms B}\|\gamma_{\ms B})+\beta \Delta Q_{\ms B}.
    \end{align}
Finally, substituting Eq.~\eqref{Eq:entropy-bath} into Eq.~\eqref{Eq:entropy-relation} allows us to derive a general expression for the heat exchange in a unitary process, namely:
\begin{equation}\label{Eq:sharpened-Landauder}
    \beta \Delta Q_{\ms B} = -\Delta S_{\ms S}+I(\ms{S}:\ms{B})_{\sigma_{\ms{SB}}} + S(\sigma_{\ms B}\|\gamma_{\ms B}).
\end{equation}
Since both mutual information and relative entropy are non-negative, discarding these two terms immediately leads to Landauer's bound.
\begin{equation}\label{Eq:Landauer-bound}
    \beta \Delta Q_{\ms B} \geq -\Delta S_{\ms S}.
\end{equation}
The equality in Eq.~\eqref{Eq:sharpened-Landauder} was previously derived in a different context~\cite{Esposito2010}, focussing on distinguishing between reversible and irreversible contributions to entropy change, but no connection to Landauer's principle was established.

As a final remark, note that Landauer’s principle can naturally be written in terms of the entropy production framework since information erasure is fundamentally an irreversible process. Specifically, Eq.~\eqref{Eq:Landauer-bound} can be cast as
\begin{equation}
   \Sigma:= \beta \Delta Q_{\ms B} + \Delta S_{\ms S} \geq 0,
\end{equation}
where, reversible processes satisfy $\Sigma =0.$

Finally, it is worth noting that Landauer erasure can be understood as the task of cooling a thermal state down to its ground state. According to the third law of thermodynamics, cooling a system to its ground state within a finite amount of time using finite resources is impossible. However, Landauer's bound is ideally achieved only in infinite time, thereby identifying time itself as a resource in the context of the third law~\cite{Masanes2017,Wilming2017}. Recently, a generalised and unified Landauer’s bound was put forward in Ref.~\cite{Taranto2023}. Surprisingly, finite energy and time suffice to perfectly erase information (or cool a quantum system). However, a hidden resource, termed control complexity, must diverge. This notion refers to the complexity of operations required to control the interactions between a quantum system and auxiliary systems (referred to as machines) that are specifically designed to achieve the target transformation. 

\begin{coolexample}[Erasing with a single-swap]
    Consider the problem of mapping an initially maximally mixed state $\rho = \frac{1}{2}(\ketbra{0}{0}+\ketbra{1}{1})$, described by a trivial Hamiltonian $H=0$, to a final pure state $\sigma = \ketbra{0}{0}$. Assume that we are allowed to use a two-dimensional thermal ancilla, prepared at some inverse temperature $\beta$ with energy gap $E$, $\gamma = (\ketbra{0}{0} + e^{-\beta E}\ketbra{1}{1})/Z$, with $Z = 1+e^{-\beta E}$ being its partition function. A protocol for achieving this transformation consists of applying a unitary SWAP $U_{\rm swap} = \ketbra{01}{10}+\ketbra{10}{01}+\ketbra{00}+\ketbra{11}$, which flips both states and increase the energy gap $E$. However, one can easily see that the work cost for performing this protocol, i.e., $W = \Delta E_{\ms B} = (\frac{1}{2} - \frac{e^{-\beta E}}{Z}) E$, diverges as $E$ tends to infinity. This shows that in this specific setup, perfect erasure comes at the cost of diverging energy.
\end{coolexample}

\subsection{Landauer's principle in the lab \label{Subsec:Landauer-experimental}}
\begin{figure*}
    \centering
    \includegraphics{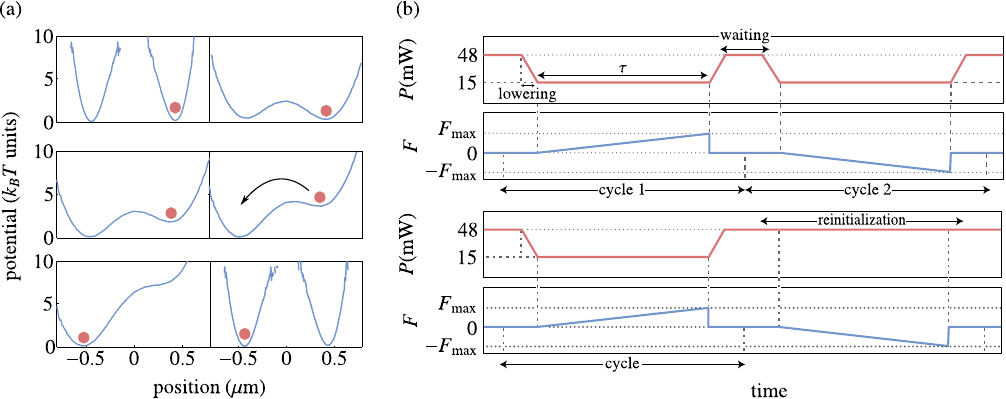}
    \caption{\emph{Experimental verification of Landauer's principle}. Panel (a) illustrates the erasure of one bit of information stored in a bistable potential. The process involves first lowering the central barrier, followed by applying a tilting force, which drives the particle from the left to the right. The protocol works regardless of the particle's initial position. Panel (b) (top) details the erasure protocol for when the particle transitions from 0 to 1 (or vice versa). The protocol for measuring the heat when the particle remains in the same well is depicted in the bottom of panel (b). These figures have been adapted from Ref~\cite{berut2012experimental}, with minor edits for consistency with this tutorial, while retaining their original meaning.}
    \label{F-experimental-landauer-1}
\end{figure*}
Let us roll up our sleeves and discuss the first experiment that verified Landauer’s principle using a generic model of a one-bit memory~\cite{berut2012experimental}. In this experiment, a single colloidal particle is trapped in a double-well potential [similar to the warm-up example discussed by Landauer and depicted in Fig.~\eqref{F-bistable}]. Unlike the previous section, this setup is entirely described by classical mechanics and stochastic thermodynamics~\cite{seifert2008stochastic}. Therefore, assumptions such as uncorrelated states or unitary evolution, are not relevant in this context. Additionally, the notion of entropy used here is the Shannon one. Most importantly, the conclusion of this section is that, in the limit of slow erasure processes, the mean dissipated heat saturates Landauer’s bound.

The setup comprises a silica bead, approximately $2 \, \mu$m in diameter, trapped in an optical tweezer~\cite{moffitt2008recent}. The double-well potential is created by rapidly alternating the focus of the laser between two distinct positions, ensuring that the particle experiences an effectively stable potential over time, despite the alternating positions. The barrier separating the two wells is high compared to the thermal energy $k_B T$, which keeps the particle trapped at one site of the well. Conversely, when the barrier is low, the particle can reach the other site. As a result, the state of the memory can be assigned a value of \texttt{ZERO} if the particle is in the left-hand well, or \texttt{ONE} if it is in the right-hand one. The memory is said to be erased when its state is reset to \texttt{ZERO} (or alternatively \texttt{ONE}), regardless of its initial state.

The experiment starts with the double-well occupied with equal probability, resulting in an initial entropy of $S = k_B \log 2$. To reset the memory, the barrier height (initially larger than $8k_B T$) is lowered to $2.2 k_B T$ over a period of 1 second and kept low for a time $\tau$. Since the particle’s location is uncertain, a tilting force is applied to remove the ambiguity. More precisely,

the force is linearly increased up to a maximal amplitude, effectively ``tilting'' the landscape of the potential~(see Fig.~\hyperref[F-experimental-landauer-1]{\ref{F-experimental-landauer-1}a}). The process ends by turning off the tilt and restoring the barrier to its original height, again over 1 second. The total duration of the erasure protocol is $\tau_{\text{cycle}} = \tau + 2 s$. 

Before turning to the quantitative results, it is useful to compare these durations with the bead’s own response times. A $1\,\text{s}$ change in the barrier is already longer than the bead needs to relax inside a single well, a fact noted by the original authors as ``long compared with the relaxation time of the bead''. When the barrier is low ($2.2\:k_{\mathrm{B}}T$), the mean thermal-hopping time between wells is roughly $10,\text{s}$; the erasure runs that probe Landauer’s limit keep the tilt on for $\tau \simeq 25$–$40\:\text{s}$, i.e., several hopping times, so the system remains close to equilibrium. Only the deliberately shortest protocols (with $\tau$ of a few seconds) explore a faster, more dissipative regime. It is important to note that time plays a crucial role in this process: entropy is produced, and the minimum entropy production occurs when the procedure is performed very slowly. Therefore, approaching the Landauer limit requires minimising dissipation (and entropy production) as much as possible.

Having sketched the main ideas of the experimental protocol, let us now focus on some specific details. The tilting force is created by moving the small chamber (or `cell') that holds the single bead relative to the laser, using a piezoelectric motor. Shifting the position of the chamber changes the bead's location within the laser's trapping region, which effectively tilts the double-well and guides the bead from one site to the other. The particle’s trajectory, from \texttt{ZERO} to \texttt{ONE} or vice versa, is captured using a fast camera that tracks the transition during the cycle. When the state of the memory is changed, a series of double cycles is used, which moves the bead from one well to the other and back (see Fig.~\hyperref[F-experimental-landauer-1]{\ref{F-experimental-landauer-1}b} for a scheme of the experimental procedure and a more detailed discussion). In the opposite case, where the state of the memory remains unmodified, the system undergoes a reinitialisation phase. This step consists of a single cycle used to ensure that the bead remains in the same well, resetting the system for the next cycle (see Fig.~\hyperref[F-experimental-landauer-1]{\ref{F-experimental-landauer-1}b}). The next question is how can we write down the explicit quantities appearing in Landauer's bound?

Since we are dealing with a microscopic system, fluctuations cannot be neglected, and as a result, thermodynamic quantities become stochastic variables. In this context, we use lowercase letters for quantities representing individual realisations along a trajectory, and uppercase letters for their averages. The dissipated heat $q$, along a given trajectory $x(t)$, can be derived from the first law of thermodynamics $dU = \delta w + \delta q$. For a colloidal particle trapped in a potential $U(x,t)$, the work done by the system is associated with the particle’s motion, given by $\dot{x}(t)$, where $\dot{x}(t)$ is the velocity of the particle along the trajectory. The heat dissipated during this motion is related to the change in potential energy, expressed as $\delta q = - \dot{x}(t) [\partial U(x,t)/\partial x]\textrm{d}t$. The negative sign arises because heat dissipation occurs when the system does work on the surroundings. To obtain the total dissipated heat over a full cycle, we integrate this expression over time, resulting in the following equation for the dissipated heat: 
\begin{equation}
    q = - \int_{0}^{\tau_{\text{cycle}}} \textrm{d}t \dot{x}(t) \frac{\partial U(x,t)}{\partial x}.
\end{equation}
The average dissipated heat, obtained by averaging over all trajectories (over 600 cycles in the experiment), is always greater than the entropy difference: $Q:=\langle q \rangle \geq T \Delta S = k_B T \log 2$.
\begin{figure}[t]
    \centering
    \includegraphics{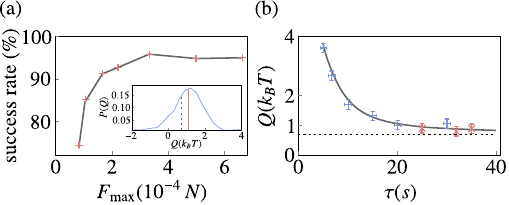}
    \caption{\emph{Success rate and dissipated heat}. Panel (a) shows the success rate of the erasure cycle as a function of the maximum tilt. The inset presents the heat distribution, with the solid red line indicating the average dissipated heat, and the dashed vertical line marking the Landauer bound. Panel (b) displays the average dissipated heat as a function of the protocol duration, measured for three success rates: blue for $r \geq 0.9$, red for $r \geq 0.85$, and green for $r \geq 0.75$. The horizontal dashed line represents the Landauer bound. These figures have been adapted from Ref~\cite{berut2012experimental}, with minor edits for consistency with this tutorial, while retaining their original meaning.}
    \label{F-experimental-landauer-2}
\end{figure}
Reaching Landauer's bound depends primarily on two factors, the duration of the tilt $\tau$ and its maximal amplitude $F_{\text{max}}$. If the tilt force is too weak to push the bead over the barrier, erasure will not occur (see Fig.~\hyperref[F-experimental-landauer-2]{\ref{F-experimental-landauer-2}a} for the trade-off between the success rate and the maximal amplitude). Conversely, for long durations, the mean dissipated heat does not saturate at Landauer's limit, and incomplete erasure will result in less dissipated heat. For a success rate $r$, the Landauer bound can be generalised as:
\begin{equation}
    Q(r) = k_B T[\log 2 + r\log r + (1-r) \log (1-r)].
\end{equation}
This equation shows that no heat is dissipated when $r=1/2$, meaning the memory is left unchanged by the protocol and the transformation is quasi-reversible. In an ideal quasi-static erasure process ($\tau \to \infty$), the dissipated heat equals the Landauer bound. For large but finite $\tau$, the asymptotic approach to the Landauer bound is described by $Q = Q_{\text{Landauer}} + \alpha/\tau$, where $\alpha$ is a positive constant~\cite{sekimoto1997complementarity}. For shorter times, the dissipated heat follows an exponential relaxation: $Q = Q_{\text{Landauer}} + A e^{t/\tau_K} + \alpha/\tau$, where $\tau_K$ is the Kramers time (a characteristic time at which a Brownian particle escapes from a
potential well above a potential barrier). The erasure rate and its approach to the Landauer bound are shown in Fig.~\hyperref[F-experimental-landauer-2]{\ref{F-experimental-landauer-2}b}.

Importantly, the attainability of the Landauer bound has been experimentally verified across various platforms. For classical systems, these include colloidal particles~\cite{berut2012experimental,Jun2014,Brut2015,Gavrilov2016}, optomechanical systems~\cite{Ciampini2021}, and micromechanical oscillators~\cite{Dago2021,Dago22}. For quantum systems, the platforms range from nanomagnets~\cite{Hong2016,Martini2016,gaudenzi2018quantum} and superconducting devices~\cite{Saira2020} to nuclear magnetic resonance~\cite{Peterson2016}, ion traps~\cite{Yan2018,roldan2014universal} and semiconductor
quantum dots~\cite{Barker2022,Scandi2022}.

\subsection{Reversible computation \label{subsec:reversible-computation}}

So far, we have hinted that typical computation is logically irreversible. But is that an unavoidable feature of computers? According to Landauer, whenever a computational task discards information about its previous state, it generates the corresponding amount of entropy. On the other hand, we might imagine that if we could somehow save all the information and steps of the computation, we could avoid this loss. For example, if we had an extra tape, initially blank, where we recorded each operation as it was performed, we could, in theory, make the computation reversible. However, as Landauer himself pointed out, this approach only postpones the problem of discarding information, especially if we plan to reuse that tape again. Consequently, a useful reversible computer would be one that, instead of storing all intermediate steps permanently, is designed to erase any unnecessary data once it is no longer needed—leaving behind only the initial input and the final output. Of course, this process still generates entropy while clearing out unnecessary intermediate information. However, the computation can still be reversed since the input and output remain intact. Surprisingly, it was shown by Bennett~\cite{bennett1973logical} that reversible computers that meet these requirements indeed exist.
\begin{figure}[t]
    \centering
    \includegraphics{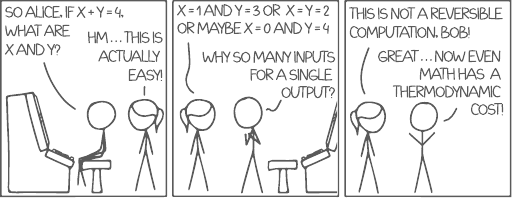}
    \caption{\emph{Without a history tape, no such a thing as a free lunch}. Imagine a machine that does nothing but adding numbers. If the machine outputs the number four, and Bob asks Alice to figure out the inputs that resulted in this number by running the process backward, she immediately encounters a problem. The inputs could have been one and three, or zero and four, among many other combinations. This fundamentally illustrates the concept of irreversibility: there is no single inverse solution, but rather a range of possibilities that could lead to the same outcome of four. Illustration inspired by XKCD~\cite{XKCD}.}
    \label{F-reversibility}
\end{figure}
We will not reproduce Bennett's formal demonstration but instead offer a heuristic approach to support his claims. Let us begin with some basic facts: typically, computers perform operations and discard information about their history, leaving the machine in a state where it is unclear what the previous step was. When this happens, the computation is said to be logically irreversible, meaning it lacks a single-valued inverse (see Fig.~\ref{F-reversibility}, where Alice and Bob realise that they cannot trace back the initial steps of the computation). Notice that if we were to record each operation on a blank tape, this issue could be avoided. However, not every step needs to be saved, only those that would allow the process to be reversed, ensuring that a given output can be traced back to a unique input.

Suppose that we have a reversible computer and an initially blank tape, where every step of the computation is recorded. After the computer runs a long computation, resulting in a lengthy tape history, the question arises: can we erase the tape while preserving the final output and avoiding entropy generation? 

First, by running the computation in reverse, the machine would undo each step one by one. This would eventually return the tape to its original blank state, as if the computation had never occurred. Since the forward computation was reversible, the reverse process would also be reversible. However, running the computation backward would also undo the final result, turning it back into the original input, which interferes with the purpose of performing the computation. This issue can be easily solved by making a copy of the final output on a separate tape before starting the reverse process. In this way, we can keep the output safe while reversing the rest of the computation.

During the copying process, we stop recording to the history tape to avoid generating unnecessary data. Once the output is securely copied, we can proceed with reversing the computation. While the original output will be erased during this reversal, the history tape will also be erased, restoring it to its blank state. In the end, we have the original input restored (as if nothing had been done), the final output copied preserved, and no remaining history on the tape. Even though the history tape is erased, the computation remains reversible and deterministic, as each step of the process can still be traced and reversed. This argument was formally addressed and rigorously proven by demonstrating that, given an ordinary Turing machine, a reversible three-tape Turing machine can be constructed to emulate the original on any input, while leaving behind only the input and the desired output at the end of its computation. Importantly, this argument is not limited to three-tape Turing machines but can be applied to any form of deterministic computation, whether finite or infinite, as long as it has memory to record the history.
\begin{table}[t]
\label{T:scheme-computation}
\centering
\begin{tblr}{
  row{even} = {c},
  row{3} = {c},
  row{5} = {c},
  row{7} = {c},
  row{9} = {c},
  row{11} = {c},
  cell{1}{1} = {c},
  cell{13}{2} = {c},
  cell{13}{3} = {c},
  cell{13}{4} = {c},
  vline{2,5} = {1-2}{},
  vline{1-2,5} = {3-5,7-9,11-13}{},
  hline{1-2} = {2-4}{},
  hline{3,6-7,10-11,14} = {-}{},
}
\textbf{Stage}   & \tikzmark{startV}            & \textbf{Tapes} &             \\
        & \emph{work tape}  & \emph{history tape} & \emph{output tape} \\
\tikzmark{startH} & \texttt{\_INPUT}   & \_           & --          \\
\textbf{Forward} & \texttt{WO\underline{R}K}   & \texttt{HIST}\_       & --          \\
        & \texttt{\_OUTPUT}   & \texttt{HISTORY}\_    & --          \\
        &            &              &             \\
        & \texttt{OUT\underline{P}UT} & \texttt{HISTORY\_}    & --          \\
\textbf{Copy}    & \texttt{\_OUTPUT}   & \texttt{HISTORY\_}    & \texttt{\_OUTPUT }  \\
        & \texttt{\_OUTPUT}  & \texttt{HISTORY\_}    & \texttt{\_OUTPUT}   \\
        &            &              &             \\
        & \texttt{\_OUTPUT}   & HISTORY\_    & \texttt{\_OUTPUT}    \\
\textbf{Reverse} & \texttt{WO\underline{R}K}   & \texttt{HIST\_ }      & \_OUTPUT    \\
        & \texttt{\_INPUT}    & --           & \texttt{\_OUTPUT}    
\end{tblr}
\tikz[remember picture, overlay] {
    \draw[->, thick] ([yshift=-5.5cm,xshift=-0.8cm] {{pic cs:startH}}) --++ (7.15cm, 0) node[midway, above, yshift=-0.5cm] {Forward, Copy, Reverse};

    \draw[->, thick] ([yshift=0.35cm,xshift=5.5cm] {{pic cs:startV}}) --++ (0, -7cm) 
        node[midway, anchor=center, rotate=-90,yshift=0.2cm] {progression of computation};
}

\end{table}

A crucial nuance in this discussion lies in the role of the input itself. While reversible computation ensures that no entropy is generated during the processing of a given input, the preparation of that input—whether acquired from a user, a sensor, or a quantum random number generator (QRNG)—comes with its own thermodynamic costs. For instance, starting a tape with a specific input (e.g., erasing prior data to write new bits) invokes Landauer’s principle, as does storing useful information for later computation. Similarly, the utility of the output matters: computations are rarely performed on random strings but on inputs meaningful to a task, and the act of selecting or generating these inputs often involves irreversible steps. Thus, while reversible algorithms avoid entropy generation during computation, the broader thermodynamic footprint of a computational task must account for the acquisition, preparation, and eventual use of inputs and outputs. This is consistent with Bennett’s observation that reversible computing preserves the logical reversibility of the computation itself, but practical implementations remain subject to thermodynamic constraints at the boundaries of the system~\cite{Bennett1989, Wolpert2020}.

Bennett's technique for performing an arbitrary computation reversibly is illustrated in Table~\hyperref[T:scheme-computation]{1}. The process involves three tapes and three stages. The first tape, known as the \emph{work tape}, stores the \texttt{INPUT}, intermediate steps, and \texttt{OUTPUT}. The second tape, \emph{history tape}, logs the entire computation, recording each step taken by the machine. The third tape, \emph{output tape}, holds the final result. The stages consist of performing the forward computation, copying the output, and reversing the computation. The detailed procedure is explained below.

Everything starts with an \texttt{INPUT} on \emph{work tape}, which is processed through a series of computational steps and operations, eventually leading to an output. We refer to this \texttt{INPUT} processing as \texttt{WORK}. For instance, if the \texttt{INPUT} consists of numbers, the \texttt{WORK} could involve operations such as addition, subtraction, multiplication, or more complex logical functions. As the machine performs its computations, it records each step on the \emph{history tape}. The machine’s time-steps are illustrated in Table~\hyperref[T:scheme-computation]{1}, where the underbar symbol indicates the position of the tape head. At the end of the forward stage, the \texttt{INPUT} has been fully processed into the \texttt{OUTPUT} on the \emph{work tape}, while the \emph{history tape} has captured all the steps taken during computation. In the second stage, the \texttt{OUTPUT} from the forward computation remains on the \emph{work tape}, and the \emph{history tape} still holds the entire computation record. The machine then copies the \texttt{OUTPUT} from the \emph{work tape} onto the \texttt{OUTPUT} tape, ensuring the final result is stored on both tapes. Finally, in the third stage, the machine reverses the computation on the \emph{work tape}, using the \emph{history tape} to ``uncompute'' the steps. Starting at \texttt{OUTPUT}, the tape head moves backwards, undoing each step until \texttt{INPUT} is fully restored. The \emph{history tape} is read in reverse, eventually becoming blank again. Throughout this reversal, the final result remains preserved in \emph{output tape}~(see Fig.~\ref{F-reversibility-2}).

So far, the main message of this section has been that any computation can be performed reversibly. But can we find examples of reversible or irreversible computation in nature? Quite surprisingly, the biosynthesis and breakdown of messenger RNA provide such examples. RNA synthesis, where the molecule is built step by step from DNA, is a logically reversible process—each step can, in theory, be undone without losing information (see Ref.~\cite{bennett1973logical} for more details). However, when cells break down RNA, they do so irreversibly, destroying the RNA without preserving any record of how it was created, resulting in the loss of information. Following Bennett's work, several studies at the intersection of computation and thermodynamics demonstrated that the second law is safe, even when considering ``intelligent beings,'' as long as their information processing is governed by the same laws as universal Turing machines~\cite{Zurek1984,Zurek1989,Bennett1989}.
\begin{figure}[t]
    \centering
    \includegraphics{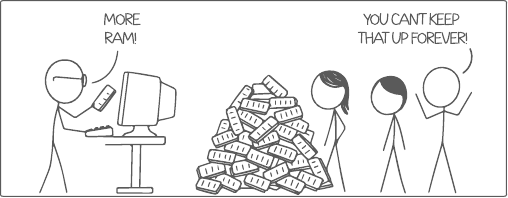}
    \caption{\emph{Without erasure, there is no end to lunch.}. Imagine a computer scientist determined to avoid the thermodynamic cost of erasure. Instead of deleting information, they endlessly accumulate RAM, offloading used memory into an ever-growing pile. While reversible computation avoids heat dissipation, storing every intermediate state demands increasingly more resources. Eventually, the physical burden becomes unsustainable. Illustration inspired by XKCD~\cite{XKCD}.}
    \label{F-ram-2}
\end{figure}

However, a critical distinction arises between reversible algorithms and physical implementations of measurement. Bennett argued that measurement itself could, in principle, be performed reversibly if the observer retains a full record of the measurement outcomes~\cite{bennett1982thermodynamics}. Yet, for systems like Maxwell’s demon, practical constraints intervene: while the demon could theoretically measure a particle’s state reversibly, it cannot indefinitely store the acquired information without eventually erasing it to reset its memory. This erasure—required to reuse the demon’s finite memory—invokes Landauer’s principle, which generates entropy and restores the second law. Thus, even though reversible computation avoids entropy generation by preserving intermediate steps, physical agents like the demon face unavoidable thermodynamic costs when interacting with the environment, as they cannot ``unmeasure" without eventually discarding information.

Inspired by the concept of reversible computing—where computations are, in principle, reversible and all intermediate states can be recovered—one might ask whether the entropy increase from erasure could be avoided by simply accumulating memory. As illustrated in Fig. \ref{F-ram-2}, this could involve endlessly adding RAM or memory modules to store every computational state. Although reversible computing may prevent heat dissipation, it demands infinite memory, which is physically unfeasible. Thus, the entropic cost, while locally or temporarily deferred, inevitably resurfaces in another form. While thermodynamic entropy in the environment might be avoided by preventing heat release, informational entropy still increases within the system—not as heat, but as mounting complexity and resource consumption embedded in the memory device’s architecture.

We conclude this section by noting that similar conclusions about reversible computation were independently reached by Fredkin~\cite{fredkin1982conservative}. This became known as the billiard-ball model of computation, a prime example of a ballistic reversible computer. Conversely, the model discussed in this section is categorised as a Brownian computer, where the computation is driven by random thermal motion and energy dissipation is minimised by operating near thermal equilibrium. Ballistic computation, on the other hand, relies on the deterministic motion and collisions of particles, ideally assuming that energy is conserved through perfectly elastic collisions.

The question of how thermodynamics constrains our ability to process and manipulate information has long been a topic of exploration, approached from various perspectives~\cite{Benioff1980, Benioff1982, fredkin1982conservative, Fredkin2002, Deffner2013c, Boyd2018, Faist2018, Wolpert2020, Korzekwa2021, conte2019thermodynamic}. Today, this remains a vibrant area of research (see Ref.~\cite{conte2019thermodynamic} for a recent review). For instance, stochastic computation~\cite{Wolpert2019} leverages the full toolkit of stochastic thermodynamics to investigate the energetic costs of implementing computational tasks that are generally more complex than simple bit erasure~\cite{Wolpert2020,Ouldridge2023,Manzano2024}. Another prominent example is computation based on autonomous quantum thermal machines (see Ref.~\cite{Mitchison2019} for a review of quantum thermal machines). In this approach, computational tasks are encoded in the dynamics of open quantum systems, such as a few qubits interacting with multiple thermal baths~\cite{LipkaBartosik2024,Xuereb2023, Xuereb2024b,meier2024autonomous}.
\begin{figure}[t]
    \centering
    \includegraphics{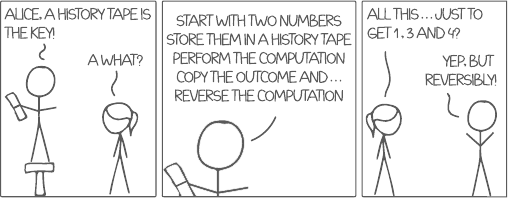}
    \caption{\emph{With a history tape, there is such a thing as a free lunch}. Imagine that the machine now adds and subtracts numbers. If the machine outputs the numbers four and two, Alice and Bob would immediately realise that the machine uniquely defines the inputs, making the computation reversible. Illustration inspired by XKCD~\cite{XKCD}.}
    \label{F-reversibility-2}
\end{figure}

\section{Exorcising the demon} 
\label{Sec:exorcisim}

After our grand information-computation interlude (and, of course, a historical one too), we now have all the essential tools and theoretical framework needed to resolve the paradox of Maxwell's demon. For simplicity, we will focus on Szilard's version. We begin with a less technical, heuristic (but historical) argument, then address the same problem more systematically, and conclude by presenting modern arguments for resolving this apparent paradox. The resolution is attributed to Charles Bennett~\cite{bennett1982thermodynamics}. However, it is important to mention that Penrose, independently and a decade before Bennett, had already pointed out that the crucial aspect in solving Maxwell's demon lay in erasing the information acquired and stored in a memory system; this would then entail an entropy cost. This is discussed in Chapter six of Ref.~\cite{PENROSE1970}.

\subsection{Heuristic approach}

To start with, each cycle step is thermodynamically reversible if carried out quasi-statically. This also means that the demon’s measurement of the particle’s position can, in principle, be performed reversibly, adding no net entropy to the universe. Although we could assume that the measurement is irreversible (which would lead to an entropy increase), our goal is to show that even if the measurement is reversible, the entropy cost will match Landauer's prediction. This reasoning echoes a crucial point stressed in Bennett's foundational review~\cite{bennett1973logical}, where he argues that the entropy cost associated with Maxwell's demon arises not from the measurement step itself, but from the erasure or overwriting of information. Specifically,

\begin{wrapfigure}{l}{0.11\textwidth}
     \includegraphics[width=0.1\textwidth]{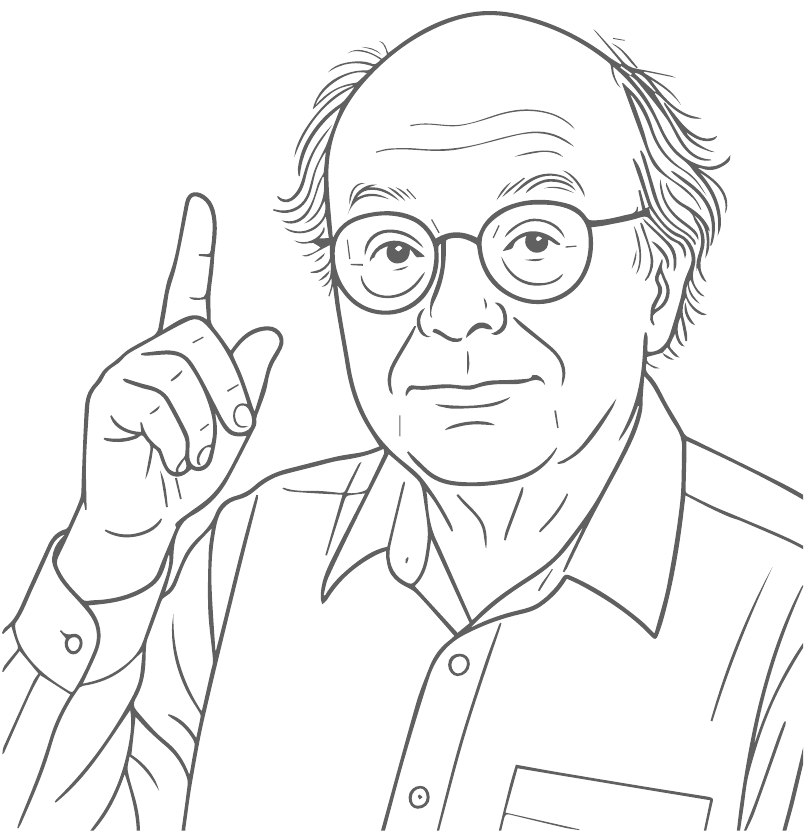}
 \end{wrapfigure}
\noindent\emph{This forgetting of a previous logical state, like the erasure or overwriting of a bit of intermediate data generated in the course of computation, entails a many-to-one mapping of the demon’s physical state, which cannot be accomplished without a corresponding entropy increase elsewhere}~\cite{bennett1982thermodynamics}
\newline

This insight is directly connected to the idea of logical irreversibility discussed in Section~\ref{Subsec:Landauer}, where we observed that many logical operations, such as resetting a bit, correspond to many-to-one mappings, and thus, by Landauer’s principle, are necessarily accompanied by entropy production. In contrast, reversible computations—where each output maps uniquely back to an input—can, in principle, be carried out without dissipation. Let us now proceed by breaking the cycle into small pieces.

Before the demon measures the particle’s position, the demon’s memory and the molecule’s position are uncorrelated: the demon has no information about which side of the partition the molecule occupies. Once the (reversible) measurement occurs, the demon’s memory and the molecule become correlated. Under perfect reversibility, this correlation does not reduce the total entropy of the demon–molecule system; rather, the measurement merely transforms their joint state from a product state to a correlated state. The crucial change is that the demon acquires knowledge of the molecule’s location without any net entropy cost to (demon + molecule) at this stage.

This subtle point—that measurement can, in principle, be performed reversibly—is important. As Bennett also noted, blaming measurement for entropy production mixes up two different things: the physical act of measuring and the logical act of erasing or forgetting information:

\begin{wrapfigure}{l}{0.11\textwidth}
     \includegraphics[width=0.1\textwidth]{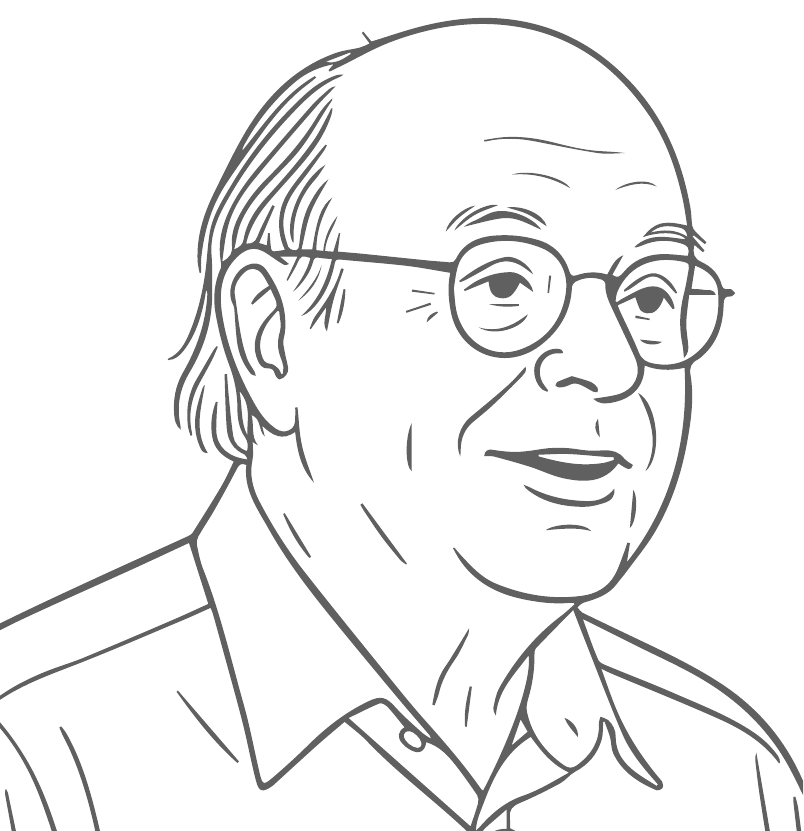}
 \end{wrapfigure}
\noindent\emph{...it is important... to attribute the entropy cost to logical irreversibility, rather than to measurement, because in doing the latter one is apt to jump to the erroneous conclusion that all transfers of information... have an irreducible entropy cost of order $k_B T \log 2$ per bit..~\cite{bennett1982thermodynamics}}
\newline

In our setup, the measurement step merely establishes a correlation between the system and the memory and does so in a thermodynamically reversible way—provided it is implemented via a reversible process, such as a unitary operation on a larger system. The key is that the demon does not discard or overwrite the acquired information at this stage.

Next comes the isothermal expansion, during which the molecule is allowed to move freely in the available space. By exploiting its knowledge of the molecule’s initial position, the demon can extract work during this expansion. As the molecule spreads out, its position becomes effectively randomised again, destroying the earlier correlation. In this process, the environment’s entropy is reduced by one bit (since the extracted work implies less heat is dissipated), while the entropy of the demon + molecule system increases from one bit to two bits. By the end of this step, the demon’s memory and the molecule appear as two independent systems, each carrying one bit of entropy: the demon retains its recorded bit, and the molecule is again equally likely to be on either side.

Finally, in order to repeat the cycle, the demon’s memory must be reset to its initial blank state. According to Landauer’s principle, erasing one bit of information necessarily generates at least one bit of entropy ($k_B T \log 2$) in the environment. This reset step increases the environment’s entropy by one bit, cancelling the reduction that occurred during the expansion. Consequently, while the demon and the molecule can be returned to their original states (with no net entropy change for demon + molecule), the environment also returns to its original entropy level. The full cycle therefore results in no overall entropy change in the demon, the molecule, or the environment, in accordance with the second law of thermodynamics.

In this way, the resolution of the paradox is tied directly to the thermodynamic cost of logically irreversible operations, as discussed in Section~\ref{subsec:reversible-computation}. It is not the act of knowing that incurs a cost, but the act of forgetting.

\subsection{Thermodynamics of information approach} \label{Subsec:TIA}

To restore the second law in the Szilard engine, we consider the physical nature of the demon by assigning it a memory to record the measurement outcome. Operating within the classical framework (see Sec.~\ref{Sec:thermodynamics-information}), we make certain assumptions about the measurement process. First, the system and the measurement apparatus are initially uncorrelated: \begin{equation} 
    \rho_{XY}(x,y) = \rho_X(x) \rho_Y(y). 
\end{equation}
Second, they do not interact before or after the measurement. Additionally, we assume that the measurement does not affect the system; although this is not strictly true in a quantum description, this approximation is useful in our context.

The measurement process creates a correlation between the system $X$ and the state of the apparatus after the measurement $Y'$. To account for this, we update the joint state of the system and apparatus after measurement to capture these correlations. The state of the system is no longer independent because the measurement introduces new information. This leads us to the correlated state: 
\begin{align}
\rho_{XY'}(x,y)
    &=\! \sum_{m\in\{\mathrm L,\mathrm R\}}
       \rho_X(x)\,p_{M|X}(m|x)\,p_{Y'|M}(y|m) \nonumber \\
    &= \rho_X(x)\!\left[
          p_{M|X}(\mathrm L|x)\,p_{Y'|M}(y|\mathrm L)
        + p_{M|X}(\mathrm R|x)\,p_{Y'|M}(y|\mathrm R)
       \right].
\end{align}
where $p_{Y'|M}(y|m)=\dfrac{p_Y(y)}{p_M(m)}$ when $m=m(y)$ and is zero otherwise.  Hence, for any fixed apparatus read-out $y$ only the term with $m=m(y)$ is non-zero.  
We label the two possible measurement outcomes by $\mathrm{L}$ (particle in the left half of the box) and $\mathrm{R}$ (particle in the right half). Note that the marginal density $\rho_X(x)$ is unchanged by the measurement, while the apparatus state changes from $Y$ to $Y'$.

Since the energy of the system does not change during the measurement, the non-equilibrium free energy of the global system after the measurement can be written as 
\begin{equation} \mathcal{F}(X,Y') = \mathcal{F}(X) + \mathcal{F}(Y') + \frac{1}{\beta} I(X:M).
\end{equation} 
Consequently, the work needed to perform the measurement satisfies
\begin{equation} W_{\text{meas}} \geq \Delta \mathcal{F}_{\text{tot}} = \Delta \mathcal{F}_Y + \frac{1}{\beta} I(X:M), 
\end{equation} 
where $\Delta \mathcal{F}_Y = \mathcal{F}(Y') - \mathcal{F}(Y)$. Since $I(X:M) \geq 0$, creating correlations between the two subsystems increases the free energy. If this increase is not balanced by a decrease in $\Delta \mathcal{F}_Y$, work must be supplied, and heat dissipation occurs.

The demon can extract work $W_{\text{ext}} \geq \frac{1}{\beta} I(X:M)$ using the information acquired during measurement in a cyclic process where the system is returned to its initial state $X$. However, to complete the cycle, the apparatus must also be returned to its initial state $Y$. Therefore, the demon must perform work to reset the apparatus, given by $W_{\text{reset}} \geq \mathcal{F}(Y) - \mathcal{F}(Y') = -\Delta \mathcal{F}_Y$. As a result, the total work involved in the process is:
\begin{equation} W_{\text{tot}} = W_{\text{meas}} + W_{\text{fb}} + W_{\text{reset}}, 
\end{equation} 
where $W_{\text{fb}} = -W_{\text{ext}}$ is the work extracted during the feedback process. Therefore, the validity of the second law for feedback processes is restored when the entropy costs of measurement and resetting the demon's memory are taken into account. This generalises Bennett's analysis of the Szilard engine. Bennett discussed the case where $W_{\text{meas}} = 0$, and the demon must overwrite the outcome of the measurement, performing work $k_B T \ln 2$. This situation corresponds to $\Delta \mathcal{F}_Y = -T I(X:M)$.

From the previous discussion, we learn that the Szilard engine can be framed in terms of a simple exchange between work and the free energy stored in the correlations between the system and the demon. For example, if $\Delta \mathcal{F}_Y = 0$, the engine operates by creating correlations during measurement, an operation that requires work $T I(X:M)$ and increases the free energy by the same amount, and then destroying these correlations during feedback, where the same amount of work is extracted.

\subsection{General approach}

We now turn to a more abstract and modern approach. We aim to capture a scenario in which the system begins in a given (unknown) state, and the demon measures the system, using the acquired information to extract work, all while ensuring that the system is returned to its original state.

The setup consists of a system $\ms S$, a thermal environment $\ms E$, a memory system $\ms M$, and a work reservoir $\ms W$. While we do not make any assumptions about the initial state of the system or the memory, we assume that the environment acts as an ideal reservoir (in the classical sense). In addition, the work reservoir can be associated with a specific system in a particular state. This follows from the physical nature of work: when work is performed, it results in changes to the state of a physical system. Work must be stored in some form, such as the potential energy of a weight, the charge state of a battery, or other systems capable of storing energy. Hence, the work reservoir represents a physical system whose state encodes the amount of work performed, and this state is described by a density operator. This formalism also allows us to distinguish between two contributions: the work that stores energy without increasing entropy (zero entropic cost) and the part associated with maximum entropic cost (the thermal reservoir).

Each subsystem is described by a density operator $\rho_{\ms X}$ with $\ms X \in \{\ms S,\ms E,\ms M, \ms W \}$. We assume that the composite system is closed and evolves via an energy-preserving unitary $U$ as
\begin{equation}
   \sigma_{\ms{SEMW}}:= U(\rho_{\ms S} \otimes\rho_{\ms E}\otimes\rho_{\ms M} \otimes \rho_{\ms W})U^{\dagger}.
\end{equation}
Using the subadditivity of entropy, one can write the following inequality:
\begin{equation}
    S(\sigma_{\ms S})+ S(\sigma_{\ms E})+ S(\sigma_{\ms W}) +  S(\sigma_{\ms M}) \geq S(  \sigma_{\ms{SEMW}}).
\end{equation}
Since the von Neumann entropy is invariant under unitary transformations, we can replace $S(  \sigma_{\ms{SEMW}})$ by $S(\rho_{\ms{SEMW}})$. Then, using the fact that the initial state is uncorrelated, we have $S(\rho_{\ms{SEMW}}) = S(\rho_{\ms S})+ S(\rho_{\ms E})+ S(\rho_{\ms W}) +  S(\rho_{\ms M})$. This leads to the following inequality:
\begin{equation}\label{Eq:entropy-chain}
    \Delta S_{\ms S}+ \Delta S_{\ms E}+ \Delta S_{\ms W}+ \Delta S_{\ms M} \geq 0.
\end{equation}
Because the protocol brings the system back to its initial state, its entropy change vanishes, $\Delta S_{\ms S} = 0$. The work reservoir is modelled as an ideal deterministic battery: at every moment, it occupies a definite energy eigenstate—e.g. a weight at a well-defined height—so its state is given by $\rho_{\ms W}(t) = \ketbra{E(t)}$, and its entropy satisfies $S[\rho_{\ms W}(t)] = 0$ for all $t$. This implies $\Delta S_{\ms W} = 0$.
Given that we are dealing with an ideal thermal reservoir, the change in its entropy is related to the exchanged heat as $\Delta S_{\ms E} = \beta Q_{\ms E}$.
Combining this with the entropy balance in Eq.~\eqref{Eq:entropy-chain}, and using $\Delta S_{\ms S} = \Delta S_{\ms W} = 0$ as argued above, we obtain:
\begin{equation}\label{Eq:Landauers-memory}
\Delta S_{\ms M} \geq -\beta Q_{\ms E}.
\end{equation}
This inequality shows that the amount of heat extracted from the bath and converted into work must be compensated by a corresponding increase in the entropy of the memory system.

\section{What's next? \label{Sec:whatsnext}}

Maxwell's demon was resolved back in the 80s, but its implications continue to inspire novel theoretical frameworks that incorporate information into thermodynamics~\cite{Sagawa2012St,Parrondo2015}. The formalism for studying thermodynamics in both classical and quantum systems--especially in situations beyond equilibrium, where fluctuations are significant--is known as (quantum) stochastic thermodynamics~\cite{seifert2008stochastic,Strasberg2013} and quantum thermodynamics~\cite{Gemmer2004,Kosloff2013,Goold2016,Vinjanampathy2016,Alicki2018,Binder2018,Deffner2019}. In what follows, we bring a list (though not exhaustive) timeline of active areas influenced by Maxwell's demon. At the beginning of each section, we also mention the essential theoretical tools for readers who want to explore these topics further.

\subsection{Physical models \& feedback control}

Could an ``authentic'' Maxwell's demon be more than just an agent sitting in a gas chamber, selectively sorting particles? Although Maxwell's demon was originally conceived within a paradigm where it selectively sorts gas molecules based on their velocity, we discussed earlier how it can also be viewed as an information-gathering and processing device that seemingly extracts work at no cost. Consequently, various physical implementations of Maxwell's demon, both classical and quantum, have emerged in recent years~\cite{Milburn1998,Schaller2011,Esposito2011,Averin2011,mandal2012work,barato2013autonomous,Deffner2013Max,Strasberg2013,Bergli2013,Chapman2015,Shiraishi2015,Kutvonen2016,Krzysztof2018,Pekola2016,Najera2020,Andersson2020,Ciliberto2020,Seah2020,Freitas2021,Freitas2022,Campsi2022,Freitas2023,Bao2023,erdman2024artificiallyintelligentmaxwellsdemon,Annby2024}. They used different implementations to provide a complete thermodynamic description of a Maxwell demon model and the system on which it acts. However, as extensively discussed here, it is not possible to violate the second law of thermodynamics using a Maxwell's demon. Nonetheless, they can still be interpreted as a feedback system. That is, the demon observes the system, gathers information about its microscopic states, and uses this information to perform a thermodynamic process~\cite{Kim2007,Sagawa2008,Jacobs2009,Cao2009,Horowitz2010,Sagawa2010,Horowitz2011,Sagawa2012,Esposito2012,Abreu2012,Strasberg2013,Funo2013}.

As an example of both the physical implementation of a Maxwell demon and the thermodynamic behaviour of feedback-controlled systems, we briefly mention the results from Ref.~\cite{Strasberg2013}. In this work, a demon reduces the apparent entropy of a subsystem by processing information and influencing the electron flow. However, this reduction in entropy comes at a cost—the demon must dissipate energy and produce entropy to exert its influence, ensuring that the total entropy production of the entire system remains non-negative. This model demonstrates how feedback control and information flow can be incorporated into non-equilibrium thermodynamics, providing insights into the role of information in modifying thermodynamic processes.

The model consists of two single-level quantum dots, one of them being the main system $\ms{S}$ and the other acting as the demon $\ms D$. Both are described by fermionic annihilation/creation operators $c_{\ms X}/c^{\dagger}_{\ms X}$ and respective energy $\epsilon_{\ms X}$, where $\ms{X} \in \{\ms S, \ms D \}$. Both quantum dots interact via Coulomb repulsion $U$. The Hamiltonian describing this interaction is given by:
\begin{equation}
    H=  \epsilon_{\ms{S}}c^{\dagger}_{\ms S}c_{\ms S} + \epsilon_D c^{\dagger}_{\ms D} c_{\ms D} + Uc^{\dagger}_{\ms D} c_{\ms D} c^{\dagger}_{\ms S} c_{\ms S}. 
\end{equation}
The system dot $\ms S$ is connected to two heat baths $L$ and $R$, which have the same temperature but different chemical potentials. This creates a flow of electrons between the reservoirs through the dot, generating a particle current. This setup constitutes a single-electron transistor (SET). The demon dot $\ms D$ is coupled to a separate reservoir and is also capacitively coupled to the system dot. This means that dot $\ms D$ can sense and react to whether dot $\ms S$ is filled or empty—much like the original idea of Maxwell’s demon being able to ``see'' and act on individual particles.

The dynamics of the system are described using a Markovian master equation, which calculates the probabilities of electrons moving between different states in the dots over time. The figure of merit here is entropy production, which remains non-negative throughout the process. However, an interesting effect emerges: the demon alters the amount of entropy produced by manipulating the flow of electrons through the dot $\ms S$. More precisely, the demon’s ability to monitor and respond to the state of the system dot effectively lowers the apparent entropy of dot $\ms S$. This entropy reduction comes at a cost—dissipation in the demon dot and its reservoir compensates for the reduced entropy in dot $\ms S$, ensuring the total entropy production remains non-negative.

Several proposals for nanoelectronic circuits have been put forward since Ref.~\cite{Strasberg2013}—see Review~\cite{Pekola2015review} for recent experiments on quantum heat transport, fluctuation relations and implementations of Maxwell's demon) and have been implemented experimentally in Refs.~\cite{koski2014experimental, Koski2014b}. Beyond the single-electron transistor, there are many other ways to implement a Maxwell demon and explore feedback mechanisms.

Importantly, the model discussed above is closely related to an earlier model by Sánchez and Büttiker~\cite{Rafael2011}, where a similar mechanism was studied in the context of power generation. While the fundamental equations governing the system are the same, the explicit interpretation in terms of information exchange and feedback was developed in later works~\cite{Horowitz2014,Shiraishi2015b,Kutvonen2016b,Monsel2025}. Experimental realisations of similar setups have also been reported~\cite{Thierschmann2015,Hartmann2015}, with connections to the Feynman ratchet problem~\cite{Roche2015}. However, a crucial limitation of these models is that, while they exhibit apparent second-law violations in the conductor, they also violate the first law, making them not `strict' Maxwell demons. A more refined version was later proposed~\cite{Rafael2019}, where all demonic actions—including opening and closing gates—are explicitly incorporated. The criteria for identifying true Maxwellian demons have been further developed in Ref~\cite{Rafael2019b}, including considerations of internal current reversal and cross-correlations, recently explored in coupled qutrit systems~\cite{Picatoste2024}.

\vspace{1cm}
\begin{mybox}{Rectifying entropy production on Maxwell's demon~\cite{Camati2016}}{Rectifying entropy production on Maxwell's demon}

This experiment realises a Maxwell’s demon in the form of a feedback control mechanism.

\vspace{0.2cm}

The system, a spin-$\frac{1}{2}$ represented by a carbon nucleus ($^{13}\mathrm{C}$), is initially described by the Hamiltonian \mbox{$H_0 = \frac{1}{2}\hbar \omega \sigma_z$} and is prepared in a thermal state. The memory, modelled by a hydrogen nucleus ($^1\mathrm{H}$), starts in its ground state. The protocol involves three main steps. \emph{(i)} The Hamiltonian of the carbon nucleus changes rapidly from $H_0$ to $H_{\tau} = \frac{\hbar \omega}{2}\sigma_x$ by applying an external magnetic field. This quick change drives the system out of equilibrium. \emph{(ii)} To establish a correlation between the system and the memory, a CNOT operation is applied, which effectively ``links'' the states of the carbon and hydrogen nuclei. Next, a projective measurement is performed on the carbon nucleus, which captures information about its state. (\emph{iii}) Based on the measurement result, a controlled operation is applied to the system. This operation is chosen based on the outcome of the measurement and allows the protocol to control the production of entropy by directing the system evolution in a targeted way.
\end{mybox}

\begin{center}
    \small{\emph{Continuous feedback}}
\end{center}

Recently, variants of the original Maxwell demon that operate continuously in time have emerged, based on the concept of continuous quantum feedback~\cite{Krzysztof2019,RibezziCrivellari2019}. Unlike conventional feedback control, these demons continuously monitor the system and use the acquired information to perform a thermodynamic process.

A beautiful illustration of this idea is the concept of gambling demons introduced in~\cite{Manzano2021}. Roughly speaking, the demon invests work into a system to perform a thermodynamic process, guided by a gambling strategy. At various points during the process, it must decide whether to continue investing more work or stop and ``cash out'' the remaining work. Since the system's evolution is stochastic, its future behaviour remains inherently unpredictable. Remarkably, in specific gambling schemes, the demon can, on average, extract more free energy than the work invested over many iterations—a scenario forbidden by the standard second-law inequality~[Eq.~\eqref{Eq:free-energy-bound-2}]. However, by accounting for the information acquired during the process and the nature of the protocol, a generalised second-law-like inequality is derived.

The setup consists of a thermodynamic system, which may be in equilibrium or out of equilibrium, interacting with a inverse temperature heat bath $\beta$. The system is characterised by a Hamiltonian $H[\lambda(t)]$, which depends on an external parameter $\lambda(t)$, and its state is represented by a probability density $\rho(x,t)$, with $x$ denoting a microstate and $x_{[0,t]} = \{x(t) \}_{t=0}^{\tau}$ a given trajectory. The protocol involves varying deterministically the external parameter $\lambda(t)$ with a total duration of $\tau$. The system’s evolution is described within the framework of stochastic thermodynamics, where thermodynamic quantities are expressed as functionals of the stochastic trajectory $x_{[0,\tau]}$. The gambling strategy is defined via a generic stopping condition that depends on the information collected about the system up to the current time. In each run, the demon gambles by applying this prescribed stopping condition. Since the demon must decide to stop before or at the end of the non-equilibrium drive, stopping times satisfy $\mathcal{T}(x_{[0,\tau]}) \leq \tau$ for any trajectory $x_{[0,\tau]}$. For these systems, the inequality~\eqref{Eq:free-energy-bound}] is generalised to
\begin{equation}\label{Eq:general-gambling}
 \langle W \rangle_\mathcal{T} -  \langle \Delta F\rangle_{\mathcal T} \geq - k_B T\, \langle \delta \rangle_\mathcal{T},
\end{equation}
where the average is over taken over many trajectories $x_{[0, \mathcal{T}]}$, each stopped at a stochastic time~$\mathcal{T}$. The term on the right-hand side of Eq.~\eqref{Eq:general-gambling} is the stochastic distinguishability, defined as
\begin{equation}\label{Eq:distinguishibility-delta}
\delta(\mathcal{T}) := \ln \left[ \frac{\varrho(x(\mathcal{T}),\mathcal{T})}{\tilde{\varrho}(x(\mathcal{T}),\tau-\mathcal{T})} \right],
\end{equation}
where $\varrho(x(\mathcal{T}), \mathcal{T})$ is the probability density of the system’s state at the stopping time $\mathcal{T}$ in the forward process, and $\tilde{\varrho}(x(\mathcal{T}), \tau-\mathcal{T})$ corresponds to the probability density at the same stopping time in the reference time-reversed process. This term captures how distinguishable the forward and reverse trajectories are at stopping time. This leads to the following generalised fluctuation theorem, $\langle e^{-\beta (W- \Delta F)-\delta} \rangle_\mathcal{T} = 1$ [compare it with Eq.~\eqref{Eq:Jarzynski-relation}]. Importantly, the idea of a gambling demon can be extended to the quantum realm by considering quantum jump trajectories. The key difference in this case is that the fluctuation theorem derived earlier now includes an additional entropic term associated with the quantum measurement process.

\vspace{1cm}
\begin{mybox}{Gambling single-electron box~\cite{Manzano2021}}{Gambling single-electron box}

This experiment realises a gambling demon and verifies the modified second-law inequality with a 99.5\% of fidelity.

\vspace{0.2cm}

The main system consists of two metallic islands connected by a tunnel junction, forming a single-electron box (SEB). At low temperatures, the SEB can be approximated as a two-level system with charge number states $n = 0$ and $n = 1$. The system is driven by an external gate voltage $V_g$, which controls the offset charge $n_g$. The tunnelling of an electron between the islands corresponds to transitions between the states $n = 0$ and $n = 1$, determined by $n_g$ and associated with an energy cost. The system is continuously monitored by a SET, which is capable of distinguishing small changes in the electrostatic potential of the box caused by the presence or absence of an extra electron. This real-time monitoring allows for precise tracking of the system's state.
\vspace{0.2cm}

The protocol begins with the system in thermal equilibrium, where the initial energies are uniformly distributed. The energy split between the states is then tuned by controlling $n_g$. By continuously measuring the system’s charge state, the SET provides the information needed to reconstruct the stochastic trajectory of the system. The protocol is repeated multiple times to gather sufficient statistics. The gambling strategy involves stopping the dynamics at stochastic times when the work exceeds a predefined threshold. This strategy leverages the continuous monitoring and feedback provided by the SET to apply a stopping rule dynamically.
\end{mybox}

Maxwell's demon with continuous quantum feedback control has also been explored in the context of generating many-body entanglement in a main system. For instance, the demon operates by randomly selecting two qubits from the system using a roulette mechanism. It then implements quantum feedback control on the selected qubits, simultaneously reducing entropy and enhancing correlations between them. By continuously repeating this process of selection and feedback control, entanglement of many bodies can be generated~\cite{Ryu2022}. Beyond the two examples discussed here, the incorporation of continuous feedback has become a very active area of research~\cite{Horowitz2014,Andersson2020,Garrahan2023}.

\subsection{Maxwell's demon \& quantum properties}

Quantum mechanics introduces a range of features that are absent in classical systems. We have already explored some of these, such as the role of particle statistics in the Szilárd engine, and briefly mentioned that Maxwell's demon and continuous feedback control can generate entanglement. But what thermodynamically happens when Maxwell's demon interacts with a system that exhibits uniquely quantum properties, such as entanglement and coherence?

In the early 2000s, results exploring the interplay between work extraction and quantum correlations within a Maxwell's demon scenario began to emerge~\cite{Oppenheim2002,Zurek2003,Maruyama2005}. First, thermodynamic inequalities were derived to distinguish entangled states from classically correlated ones based on the amount of extractable work, leading to a work extraction protocol that acts as a separability criterion~\cite{Maruyama2005}. Second, the concept of efficiency between quantum and classical demons was introduced, with the difference in the extraction capabilities of the work quantified by quantum discord, a measure of ``quantumness'' in correlations~\cite{Zurek2003}. This efficiency difference demonstrated that quantum demons, by exploiting quantum correlations, could extract more work than classical demons, thus providing a thermodynamic advantage and complementing previous findings on the unique role of entanglement in work extraction. These later led to a number of works underlying the thermodynamics of correlations~\cite{Funo2013b,Karen2013,Braga2014,Huber2015,Bruschi2015,Perarnau2015,Friss2016,Lebedev2016,Brunelli2017,Ciampini2017,Zanin2022enhancedphotonic,AdeoliveiraJunior2024}

Extending Maxwell's demon to the quantum domain, or linking thermodynamic quantities and quantum features, often relies on assuming specific models for both the system and the demon's memory. However, a general and minimal setup that addresses both of these issues was proposed in~\cite{Adeoliveiraheat2024}. The authors considered a minimal configuration comprising a quantum system, a quantum memory, and a thermal environment. The main system is unknown and arbitrary, and while the initial state of the memory is also arbitrary, it is required that it begins and ends in the same state. This ensures that the memory affects the system’s dynamics without changing its energetics, so heat exchange occurs only between the system and its environment. As for the environment, no restriction on its dimension is imposed, other than that it was initially prepared in a thermal Gibbs state at temperature $T$. The composite system is assumed to be closed and evolves under an energy-preserving unitary~\cite{Janzing2000}. Without additional constraints, no assumptions are made about the strength of the interaction (whether weak or strong), its complexity (whether local or collective), or its duration (short or long) relative to natural time scales. As there is no additional source of energy or battery system, energy exchange between the main system and the environment is accounted for as heat.

Within this minimal setup, fundamental bounds on the exchange between the quantum system and the thermal environment led to a direct correspondence between thermodynamic and quantum features. Surprisingly, these results also revealed that quantum properties can be detected by monitoring heat exchange in a quantum process.

\subsection{Maxwell's demon \& quantum heat engines}

Quantum thermodynamics, an emerging field attempting to translate thermodynamics laws and understand how thermodynamic process are carried out in the quantum realm, embraced Maxwell's demon to itself. One of the first papers in the field suggested that a three-level maser could be regarded a heat engine~\cite{Scovil1959}. It did not take long after the study of quantum heat engines was formalised~\cite{Alicki1979,kosloff1984quantum} for Maxwell's demon to stir the pot.

\subsubsection{Maxwell's demon as an engine}

The analogy between Maxwell's demon and a quantum heat engine was first drawn in~\cite{Lloyd1997}, where it was emphasised that a quantum demon is nothing more than an interaction between two quantum systems that enables the controlled transfer of information from one to the other—essentially making it an information-processing quantum heat engine. 

In fact, a quantum heat engine can effectively be seen as the demon itself, with their operations emulating a Maxwell's demon. This was discussed by Kieu~\cite{Kieu2004,Kieu2006}, using a two-level system undergoing quantum adiabatic processes and energy exchanges with heat baths. By using quantum measurement and control processes to selectively transfer energy, the machine's functionality resembles the behaviour of Maxwell's demon, sorting molecules based on their temperature~\cite{Mandal2013}. Different models of heat engines implementing Maxwell's demon or the Szilárd engine also emerged~\cite{Boyd2917,Fu2021,Sun2024}.

\subsubsection{Maxwell's demon assisting an engine}

What if, instead of assuming that the heat engine is a demon per se, we allow a demon to assist a quantum heat engine? This problem was introduced by Quan et al.~\cite{Quan2006}, who proposed a new quantum heat engine model with a built-in quantum Maxwell demon that performs both the quantum measurement on the working substance and a feedback control for the system according to the measurement. This work led to two important concepts that continue to be actively explored today: quantum heat engines assisted by Maxwell’s demons~\cite{Quan2007,Cai2012,Elouard2017,Poulsen2022,Chan2022} and the use of the demon as feedback control.

\subsection{Landauer's principle \label{Landauer-principle}}

The past decade has provided significant insight into Landauer's principle, especially in far-from-equilibrium situations~\cite{Piechocinska2000,Sagawa2009,Esposito2011,Hilt2011,Lambson2011,Rio2011,Diana2013,Deffner2013,Zulkowski2014,Lorenzo2015,Zulkowski2015,Zulkowski2015b,Browne2014,Lorenzo2015,Goold2015,Peterson2016,Strasberg2017,Guarnieri2017,Campbell2017,Boyd2018,Klaers2019,timpanaro2020landauer,Shiraishi2019,Scandi2019thermodynamiclength,Proesmans2020,Miller2020,Proesmans2020b,Zhen2021,Van2021,Vanvu2022,Ma20022,Zhen2022,Boyd2022,Lee2022}. In what follows, we briefly comment on some of the progress relating to Landauer's principle~(see Ref.~\cite{chattopadhyay2025landauerprinciplethermodynamicscomputation} for a recent review). The theoretical tools typically used are varied, ranging from the resource-theoretic approach to quantum thermodynamics~\cite{Gour2015,lostaglio2019introductory,Czartowski2023,junior2024geometricinformationtheoreticaspectsquantum}, to notions of entropy production~\cite{landi2021irreversible}, and the framework of thermodynamic geometry~\cite{Abiuso2020,Rolandi2024}.

\subsubsection{Finite-time Landauer erasure \label{Landauer-principle-finite-time}}

The Landauer bound is optimally achieved in the quasistatic limit, where the erasure process is carried out very slowly, minimising dissipation. This naturally raises the question of what happens when the erasure process is performed within a finite time. For classical systems governed by overdamped Langevin dynamics, optimal erasure protocols have been derived for slow but finite-speed processes~\cite{Zulkowski2014,Zulkowski2015,Zulkowski2015b} and were later generalised to allow for arbitrary driving speeds~\cite{Proesmans2020,Proesmans2020b}. For quantum systems described by Markovian open quantum dynamics, the optimal erasure cost, valid for arbitrary operational times, was found~\cite{Zhen2021,Vanvu2022}. Importantly, in Markovian open quantum systems, quantum coherence has been shown to incur additional heat costs during erasure. However, this conclusion is context-dependent: recent work demonstrates that coherence can play a constructive role in optimising thermodynamic costs for certain protocols, particularly in non-Markovian settings~\cite{Rolandi2023b}. The interplay between coherence and erasure efficiency remains nuanced, as its utility or detrimentality depends on the driving regime and system-environment coupling. For slowly driven unitary quantum systems, it remains an open question whether coherence is necessary or unnecessary for minimising dissipation. The take-home message is that the minimum dissipation for erasing a classical bit has a lower bound set by the Landauer cost, plus an additional term that scales inversely with the operational time:
\begin{equation}\label{Eq:Landauer-correction-finite}
    \Delta Q_{\ms B} \geq k_B T \log 2 + \frac{\alpha}{\tau},
\end{equation}
where $\tau$ represents the total time of the process, and $\alpha \in \mathbbm{R}$ is a positive constant that depends on the specific process and the nature of the system (whether classical or quantum). 

\vspace{1cm}
\begin{mybox}{Finite-time Landauer erasure in a quantum dot~\cite{Scandi2022}}{Finite-time Landauer erasure in a quantum dot}

This experiment realises Landauer erasure with non-linear protocols that minimise dissipation beyond linear approaches.

\vspace{0.2cm}

The setup consists of three quantum dots ($D_1, D_2, D_3$) on an InAs nanowire. The main dot, $D_1$, holds the bit of information in its occupancy state, which can be 0 (unoccupied) or 1 (occupied). This state represents the bit to be erased. The energy of $D_1$ is controlled by a gate voltage, which allows for precise manipulation during the erasure process. The dot $D_2$ is prepared in a state that effectively acts as a heat bath for $D_1$. Finally, $D_3$ functions as a sensor, detecting changes in occupancy in $D_1$ and enabling real-time monitoring.

\vspace{0.2cm}
The erasure protocol begins by allowing $D_1$ to reach thermal equilibrium at an initial energy level, $E_0$, where it has a 50\% chance of being occupied. The energy is then gradually increased to $E_1$, making $D_1$ almost certainly unoccupied and thus erasing the bit. After this, the energy is quickly reset to $E_0$. Heat dissipation during this process is measured by tracking electron transitions. Unlike traditional linear approaches, non-linear protocols that minimise dissipation are employed.

\vspace{0.2cm}

Using tools from thermodynamic length, this work provides the first experimental demonstration of minimising dissipation through non-linear driving in a quantum dot.
\end{mybox}

If in the protocol, the stored information is fully erased, then $\alpha =\bar{\gamma}_{\tau}^{-1}$, where $\bar{\gamma}_{\tau}:=\tau^{-1}\int_0^{\tau}\sum_k \tr{L_k(t)\rho_t L_k(t)} \textrm{d}t$ is the time average characterising the thermal relaxation timescale, $L_k(t)$ denotes the time-dependent jump operators and $\rho_t$ the state of the system to be erased. Importantly, the aforementioned studies focused on Markovian systems that weakly interact with a heat bath. Under strong coupling conditions, however, one may question how the constant $\alpha$ in Eq.~\eqref{Eq:Landauer-correction-finite} looks like, as new effects from faster relaxation rates and non-Markovian dynamics come into play. This question was addressed by modelling a bit (a two-level system) encoded in the occupation of a single fermionic mode strongly interacting with a heat bath~\cite{Rolandi2023}. Specifically, $\alpha =  k_B T a \tau_{\text{Pl}}$, where $a \approx 2.57946$ and $\tau_{\text{Pl}} = \hbar/k_B T$ represent the Planckian time~\cite{Hartnoll2022}. In particular, this finite-time correction incorporates the product of two fundamental constants of nature--Boltzmann's constant $k_B$ and Planck’s constant $\hbar$. 

\subsubsection{Landauer’s principle at zero temperature}

Upon a closer look at the Landauer bound given by Eq.~\eqref{Eq:Landauers-principle}, we note that it becomes trivial in the zero-temperature limit. This result can be physically understood by noting that as $T \to 0$, the heat bath approaches its ground state, meaning that any physical process must satisfy $\Delta Q_{\ms B} \geq 0$. The bound is useful in showing that certain processes can occur with zero heat cost, but beyond this, it provides no further information. This motivates the question of whether it is possible to derive a modified bound that captures non-trivial information even at zero temperature. Very surprisingly, using minimal assumptions—specifically, that the environment is in a thermal state—a tighter bound was derived in Ref.~\cite{timpanaro2020landauer}. More specifically, 
\begin{equation}\label{Eq:modified-bound}
    \Delta Q_{\ms B} \geq \mathcal{Q}[\mathcal{S}^{-1}(\Delta S_{\ms S})],
\end{equation}
where $\mathcal{Q}(\beta') := \tr{H_{\ms E}[\gamma_{\ms E}(\beta') - \gamma_{\ms E}(\beta)]}$ and $\mathcal{S}(\beta') := S[\gamma_{\ms E}(\beta')] - S[\gamma_{\ms E}(\beta)]$, with $\gamma_{\ms E}(x)$ representing the thermal state of the heat bath at a given inverse temperature $x$. Observe that the tricky aspect lies in the fact that $\mathcal{Q}(\beta')$ is monotonically decreasing in $\beta'$ and thus has a unique inverse $\beta' = \mathcal{Q}^{-1}(\Delta Q_{\ms E})$. The difference now is that the bound relates less trivial functions; however, it involves only thermal equilibrium quantities, despite the process being arbitrarily far from equilibrium.

To make the bound more concrete and suitable for calculations, it is helpful to express $\mathcal{Q}$ and $\mathcal{S}$ in terms of the environment’s heat capacity. For a bath at temperature $T$ with heat capacity $C_{\ms E}(T)$, one finds~\cite{timpanaro2020landauer}:
\begin{equation}\label{Eq:Q-S-heat-capacity}
\mathcal{Q}(T') = \int_T^{T'} C_{\ms E} \mathrm{d}\tau \:\: , \:\: \mathcal{S}(T') = \int_T^{T'} \frac{C_{\ms E}(\tau)}{\tau} \mathrm{d}\tau,
\end{equation}
These expressions allow the modified bound in Eq.~\eqref{Eq:modified-bound} to be evaluated directly, provided a model for the bath is specified.

We now illustrate how Eq.~\eqref{Eq:modified-bound} can be applied by considering a simple yet instructive example: a qubit system initially in a maximally mixed state and undergoing an erasure process, while coupled to a low-temperature phonon-like environment with heat capacity $C_{\ms E}(T) = a T^3$, where $a$ is a positive constant~\cite{Ashcroft1977}. Using Eq.~\eqref{Eq:Q-S-heat-capacity}, we obtain
\begin{align}
\mathcal{S}(\beta') = \frac{a}{3}(T'^3 - T^3) \quad , \quad
\mathcal{Q}(\beta') = \frac{a}{4}(T'^4 - T^4).
\end{align}
In the zero-temperature limit $T \to 0$, the entropy change simplifies to $\mathcal{S}(\beta') = \frac{a}{3} T'^3 = \Delta S_{\ms S}$. Solving for $T'$ gives
\begin{equation}
T' = \left( \frac{3 \Delta S_{\ms S}}{a} \right)^{1/3}.
\end{equation}
Substituting this into $\mathcal{Q}$ yields the lower bound on the heat:
\begin{equation}
\Delta Q_{\ms B} \geq \mathcal{Q}[\mathcal{S}^{-1}(\Delta S_{\ms S})]
= \frac{a}{4} \left( \frac{3 \Delta S_{\ms S}}{a} \right)^{4/3}
= \frac{3^{4/3}}{4}  \frac{(\Delta S_{\ms S})^{4/3}}{a^{1/3}}.
\end{equation}

This result shows that, even at zero temperature, erasing one bit of information ($\Delta S_{\ms S} = \log 2$) requires a strictly positive amount of heat to be dissipated, provided the environment has a non-trivial heat capacity. The bound remains meaningful in regimes where the standard Landauer principle becomes uninformative.

This new bound can be used to study a variety of systems, such as the paradigmatic example of the spontaneous emission of a two-level atom into a single-mode cavity, the interaction of a system with a one-dimensional waveguide, and many other examples (see Ref.~\cite{timpanaro2020landauer}).

\subsubsection{Finite-size corrections to the Landauer erasure}

One may also ask: what are the finite-size corrections to Landauer's principle when the heat bath has a finite number of particles? More precisely, consider a reservoir consisting of $n$-particles, prepared in a Gibbs state at an inverse temperature $\beta$, and coupled to a quantum system. Can we derive bounds on the thermodynamic energetic cost in this scenario? This question naturally leads to an additional consideration of whether the bath consists of non-interacting particles or includes interactions among its components. For any thermodynamic process, with the help of an environment with a non-interacting Hamiltonian, the entropy production satisfies~\cite{PatrykMarti2024}
\begin{equation}\label{Eq:entropy-scale-n}
    \Sigma \geq \qty(\frac{\Delta S_{\ms S}}{\log d})^2\frac{1}{n}.
\end{equation}
That is, it decays at most as $\propto 1/n$ in the presence of a non-interacting environment. For a single qubit, the above result becomes $\Sigma \geq 1/3n$. Several works have analysed the decay of $\Sigma$ with $n$~\cite{reeb2014improved,Skrzypczyk2014,Bumer2019,taranto2024efficientlycoolingquantumsystems}, finding a convergence of the form $\Sigma = A/n$ for $n \gg 1$, where $A$ depends on the specific protocol. A recent work derived the ``best'' (known) protocol for a class of collisional models, finding $A \approx \pi^2/8$~\cite{taranto2024efficientlycoolingquantumsystems}.

Observe that Eq.~\eqref{Eq:entropy-scale-n} holds for thermal environments with non-interacting Hamiltonian.We can ask whether it can be violated in the case of interacting Hamiltonians. If so, this would demonstrate an advantage over non-interacting ones. To answer this question, we start by noting that for sufficiently large $n$, a finite-size correction that is valid in \emph{any} environment for thermodynamic processes that reduce the system's entropy $\Delta S_{\ms S} <0$ is given by~\cite{reeb2014improved}:
\begin{equation}\label{Eq:reebs-wolf-finite}
    \Sigma \geq \frac{2(\Delta S_{\ms S})^2}{\log^2(d-1)+4} = \mathcal{O}(n^2).
\end{equation}
Eq.~\eqref{Eq:reebs-wolf-finite} holds universally, regardless of the specific system-environment interaction and the nature of the process being implemented. The achievability of Eq.~\eqref{Eq:reebs-wolf-finite} was recently proved with an explicit process that realises Landauer erasure considering heat baths with interacting Hamiltonian~\cite{PatrykMarti2024}. Specifically, the authors considered an initial state $\rho_{\ms S} = \mathbbm{1}_{\ms S}/2$ and constructed a unitary operator and a Hamiltonian $H_{\ms B}$, which mapped $\rho_{\ms S}$ to a state $\sigma_{\ms S}$ which is $\epsilon$ close to the ground state. In doing so, they found that the entropy production in this protocol is bounded by
\begin{equation}
    \Sigma \geq 2\qty(\frac{\pi}{n})^2 + \mathcal{O}(1/n^2).
\end{equation}
Therefore, we see that entropy production decays quadratically with the size of the environment. This is in stark contrast to the case of non-interacting environments, where entropy production can decrease at most linearly with the size of the environment.

\subsubsection{Optimal cost of erasure in the single-shot scenario}

Another extension of Landauer's erasure principle considers the protocol within the so-called single-shot scenario~\cite{Gour2015,Tomamichel2016,Egloff2015,Gour2017,LipkaBartosik2018}. This setting removes the assumption of having either infinitely many independent copies of the quantum state or large numbers of identical process repetitions. Instead, it focusses on finite instances of the quantum state. In this regime, we can explore how our ability to reset a quantum system (e.g., to a known pure state) is affected when accounting for the finite-size, single-instance characteristics of the system. Single-shot tasks, such as Landauer erasure, have been studied using majorisation-based frameworks~\cite{Gour2017,Chubb2018beyondthermodynamic,Boes2022,Biswas2022,deOliveiraJunior2023}. 

For example, an erasure protocol can be considered as a process involving a unital channel\footnote{Those are channels that preserve the maximally mixed state.} that acts on the system whose state is to be erased (i.e., mapped to a fixed pure state), together with an information battery acting as a source of purity. In the simplest model, the information battery can be represented by an $n$-qubit system, where each qubit starts in a pure state and ends in a maximally mixed state. We define the work cost (in units of $k_B T \log 2$) of erasing an initial state $\rho$ as the size of the smallest information battery that allows erasure of $\rho$. This is equivalent to finding the smallest integer $n$ such that the transformation becomes possible:
\begin{equation}
    \rho \otimes\ketbra{0}{0}^{\otimes n}  \xrightarrow[\text{channel}]{\text{unital}} \ketbra{\psi}{\psi}\otimes \qty(\frac{\mathbbm{1}}{2})^{\otimes n}.
\end{equation}
Recalling that Landauer bound states that $n\geq S(\rho)$, a correction to Landauer’s bound in this setting is given by~\cite{Boes2022}:
\begin{equation}
    n \geq S(\rho) + \frac{V(\rho)}{2\sqrt{M(\rho)}},
\end{equation}
where $V(\rho):=\tr(\rho \log^2 \rho) - S(\rho)^2$ is the variance of surprisal and $M(\rho):=V(\rho)+[S(\rho)+1/\ln 2]^2$. Note that $V(\rho) = 0$ for the maximally mixed state, which is typically used as the starting point for erasure. This correction indicates that the bound increases as the initial variance of $\rho$.

Another variant of this regime, which also allows the study of finite-size corrections in Landauer erasure, is the so-called thermodynamic distillation process. This process is defined as a thermodynamic transformation from an initial system characterised by a Hamiltonian $H$ and prepared in a state $\rho$, to a target system characterised by a Hamiltonian $\tilde{H}$ and a state $\tilde{\rho}$ that is an eigenstate of $\tilde{H}$. This thermodynamic transformation can be modelled using the framework of thermal operations~\cite{Janzing2000,horodecki2013fundamental,brandao2015second}. Consequently, corrections to Landauer erasure can be studied by asking for the optimal transformation to map $N$ copies of maximally mixed states (with a trivial Hamiltonian) to $N$ copies of a target state that is $\epsilon$ close to the ground state.  The work cost can be quantified by appending a battery system, initially in an excited state, which is also transformed to the ground state during the transformation. The transformation error $\epsilon$ quantifies the quality of erasure. Recently, it was found that the erasure cost, valid for an arbitrary $N$, is given by~\cite{Biswas2022}:
\begin{equation}
    W_{\text{cost}} = \frac{N}{\beta}\Biggl[\log 2-\frac{\log(1-\epsilon)}{N}\Biggl],
\end{equation}
which recovers Landauer’s cost of erasure for when $\epsilon=0$.

Finally, other analyses of thermodynamics in the one-shot scenario, related to Landauer erasure, have been presented in~\cite{Halpern2015} and applied to various other thermodynamic tasks~\cite{Lostaglio2015b,Gemmer2015,Richens2016,Salek2017,Wang2017,LipkaBartosik2018}

\section{Conclusion}

Maxwell’s demon, once a provocative thought experiment challenging the second law of thermodynamics, has become a cornerstone of modern physics. This tutorial has traced the demon’s journey from its classical origins to its quantum ``reincarnations'', revealing how its apparent paradoxes were resolved by recognising information as a thermodynamic resource.

Stochastic and quantum thermodynamics have further expanded the demon’s legacy, inspiring research into the fundamental limits of energy, computation, and information processing. Experimental realisations across diverse platforms demonstrate that Maxwell’s demon —and related concepts such as Landauer erasure— are not merely theoretical constructs but tangible phenomena with technological implications.

No longer a paradox to be resolved, Maxwell’s demon has become a powerful tool for exploring the links between thermodynamics and information. Its story highlights how thought experiments can challenge our assumptions and lead to new physics.

\section*{Acknowledgments}

We thank Jakub Czartowski, Jake Xuereb, Albert Rico and Martí Perarnau-Llobet for valuable discussions and fruitful comments on the first version of the manuscript. The authors would like to thank the anonymous referee for their insightful comments and valuable suggestions, which significantly improved the clarity and presentation of this work. In particular, the suggestion to include Fig. 13 is especially appreciated. AOJ and JBB acknowledge financial support from the Danish National Research Foundation grant bigQ (DNRF 142) and VILLUM FONDEN through a research grant (40864). RC acknowledges the Simons Foundation (Grant Number 1023171, RC), the Brazilian National Council for Scientific and Technological Development (CNPq, Grants No.307295/2020-6 and No.403181/2024-0), the Financiadora de Estudos e Projetos (grant 1699/24 IIF-FINEP) and a guest professorship from the Otto M\o nsted Foundation.

\newpage
\bibliography{2-reference}

\end{document}